\documentclass[iop]{emulateapj}

\newcommand{\kms}{\mbox{km s$^{-1}$}}

\newcommand{\Msun}{\mbox{M$_{\odot}$}}
\newcommand{\Lsun}{\mbox{L$_{\odot}$}}
\newcommand{\hii}{\mbox{ H\,{\sc ii}}}
\newcommand{\HCO}{\mbox{HCO$^{+}$}}
\newcommand{\tHCO}{\mbox{$J=1-0$}}
\newcommand{\NdosH}{\mbox{N$_2$H$^{+}$}}
\newcommand{\tNdosHudt}{\mbox{$J=1-0,F_1=2-1,F=3-2$}}
\newcommand{\tNdosHuud}{\mbox{$J=1-0,F_1=1-1,F=2-2$}}
\newcommand{\tNdosHucu}{\mbox{$J=1-0,F_1=0-1,F=1-2$}}
\newcommand{\HNC}{\mbox{HNC}}
\newcommand{\tHNC}{\mbox{$J=1-0$}}
\newcommand{\HCN}{\mbox{HCN}}
\newcommand{\tHCNud}{\mbox{$J=1-0,F=2-1$}}
\newcommand{\tHCNuu}{\mbox{$J=1-0,F=1-1$}}
\newcommand{\tHCNuc}{\mbox{$J=1-0,F=0-1$}}
\newcommand{\CdosH}{\mbox{C$_2$H}}
\newcommand{\tCdosHudcu}{\mbox{$N=1-0,J=3/2-1/2,F=2-1$}}
\newcommand{\tCdosHuucu}{\mbox{$N=1-0,J=1/2-1/2,F=1-1$}}
\newcommand{\tCdosHuucc}{\mbox{$N=1-0,J=3/2-1/2,F=1-0$}}
\newcommand{\tCdosHuccu}{\mbox{$N=1-0,J=1/2-1/2,F=0-1$}}
\newcommand{\HtreceCO}{\mbox{H$^{13}$CO$^+$}}
\newcommand{\tHtreceCO}{\mbox{$J=1-0$}}
\newcommand{\HNtreceC}{\mbox{HN$^{13}$C}}
\newcommand{\tHNtreceC}{\mbox{$J=1-0$}}
\newcommand{\HCtresN}{\mbox{HC$_3$N}}
\newcommand{\tHCtresN}{\mbox{$J=10-9$}}
\newcommand{\HNCO}{\mbox{HNCO}}
\newcommand{\tHNCO}{\mbox{$J_{K_a,K_b}=4_{0,4}-3_{0,3}$}}
\newcommand{\SiO}{\mbox{SiO}}
\newcommand{\tSiO}{\mbox{$J=2-1$}}
\newcommand{\NHdosD}{\mbox{NH$_2$D}}
\newcommand{\tNHdosD}{\mbox{$J_{K_a,K_c}=1_{1,1}-1_{0,1}$}}
\newcommand{\SO}{\mbox{SO}}
\newcommand{\tSO}{\mbox{$N_J=2_2-1_1$}}
\newcommand{\CS}{\mbox{$^{13}$CS}}
\newcommand{\tCS}{\mbox{$J=2-1$}}
\newcommand{\HtreceCN}{\mbox{H$^{13}$CN}}
\newcommand{\tHtreceCNuu}{\mbox{$J=1-0,F=1-1$}}
\newcommand{\tHtreceCNdu}{\mbox{$J=1-0,F=2-1$}}
\newcommand{\tHtreceCNcu}{\mbox{$J=1-0,F=0-1$}}
\newcommand{\CHtresCN}{\mbox{CH$_3$CN}}
\newcommand{\tCHtresCNcc}{\mbox{$J_K=5_0-4_0$}}
\newcommand{\tCHtresCNuu}{\mbox{$J_K=5_1-4_1$}}
\newcommand{\ra}{\mbox{ $\rightarrow$ }}

\shorttitle{Chemistry in Infrared Dark Cloud Clumps}
\shortauthors{Sanhueza et al.}

\journalinfo{{\it The Astrophysical Journal}, Accepted}
\submitted{Submitted: February 18, 2012; Accepted: June 24, 2012}

\begin{document}


\title{Chemistry in Infrared Dark Cloud Clumps: a Molecular Line Survey at 3 \lowercase{mm}}

\author{Patricio Sanhueza\altaffilmark{1}, James M. Jackson\altaffilmark{1}, Jonathan B. Foster\altaffilmark{1}, Guido Garay\altaffilmark{2}, 
Andrea Silva\altaffilmark{3}, Susanna C. Finn\altaffilmark{1}}

\altaffiltext{1}{Institute for Astrophysical Research, Boston University, Boston, MA 02215, USA}
\altaffiltext{2}{Departamento de Astronom\'{\i}a, Universidad de Chile,
Casilla 36-D, Santiago, Chile}
\altaffiltext{3}{Harvard-Smithsonian Center for Astrophysics, 60 Garden Street, Cambridge, MA 02138, USA}


\begin{abstract}

We have observed 37 Infrared Dark Clouds (IRDCs), containing a total of 159
 clumps, in high-density molecular tracers at 3 mm using the 22-meter ATNF
 Mopra Telescope located in Australia. After determining 
kinematic distances, we eliminated clumps that are not located in IRDCs and 
clumps with a separation between  them of less than one Mopra beam. 
Our final sample consists of 92 IRDC clumps. The most commonly detected
 molecular lines are (detection rates 
higher than 8\%): N$_2$H$^+$, HNC, HN$^{13}$C, HCO$^+$, H$^{13}$CO$^+$, HCN,
 C$_2$H, HC$_3$N, HNCO, and SiO. We investigate the behavior of the 
different molecular tracers and look for chemical variations as a function 
of an 
evolutionary sequence based on {\it Spitzer} IRAC and MIPS emission. 
We find that the molecular tracers behave differently through the evolutionary 
sequence and some of them can be used to yield useful relative age 
information. The presence of HNC and N$_2$H$^+$ lines do 
not depend on the star formation activity. On the other hand, HC$_3$N, HNCO,
 and SiO are predominantly detected in later stages of evolution. Optical 
depth calculations show that in IRDC clumps the N$_2$H$^+$ line is 
optically thin, the C$_2$H line is moderately optically thick, and HNC and 
 HCO$^+$ are optically thick. The HCN hyperfine transitions are blended,
 and, in 
addition, show self-absorbed line profiles and extended wing emission. These 
factors combined prevent the use of HCN hyperfine transitions for 
the calculation of physical parameters. Total column densities of the different 
molecules, except C$_2$H, increase with the evolutionary stage of the 
clumps. Molecular abundances increase with the evolutionary stage for
 N$_2$H$^+$ and HCO$^+$. The N$_2$H$^+$/HCO$^+$ and N$_2$H$^+$/HNC abudance
 ratios act as chemical clocks, increasing with the evolution of the
 clumps. 

\end{abstract}


\section{Introduction}

Although far less common than low-mass stars, massive stars 
play a key role in the evolution of the energetics and 
chemistry of molecular clouds and galaxies. However, the formation of 
high-mass stars is far less clear than their low-mass counterparts for 
several reasons. Massive stars are rare and evolve quickly. In addition,
 the regions that host their early stages of formation are difficult to
 observe due to high dust extinction, and, with a few exceptions, are 
located at large distances ($\gtrsim$ 3 kpc). 
For this reason, most of the research on massive star formation has been based 
on observations of regions with current star formation  
(e.g., ``hot cores,'' \hii\ regions), which were initially easier to 
detect in surveys and to follow up in detail. In constrast, objects in the
 earlier ``prestellar'' or ``starless'' stage have been much harder to 
find and, in consequence, this stage still remains poorly understood. 

About a decade ago, however, the regions containing the earliest stages 
of massive star formation were identified. Galactic plane surveys 
revealed thousands of dark patches obscuring the bright mid-infrared 
background ({\it ISO}, \citealt{Perault96}; {\it MSX},
 \citealt{Egan98}, \citealt{Simon06a}; {\it Spitzer}, 
 \citealt{Peretto09}, \citealt{Kim10}). These dark 
silhouettes were associated with molecular and dust emission, indicating 
they consist of dense molecular gas. 
Such objects were called Infrared Dark Clouds (IRDCs). The 
first studies characterizing them suggested that they were cold ($<$25 K), 
massive ($\sim$10$^2$--10$^4$ \Msun), and dense 
($\gtrsim$10$^5$ cm$^{-3}$) molecular clouds with high column densities
 ($\gtrsim$10$^{23}$ cm$^{-2}$) \citep{Carey98,Carey00}. They also correspond to 
the densest clouds embedded within giant molecular clouds \citep{Simon06b}. 

More recent studies on IRDC clumps,\footnote{Throughout this paper, we use 
the term ``clump'' to refer to a dense object within an IRDC with a size of 
the order $\sim$1 pc and a mass $\sim$10$^2$--10$^3$ \Msun. We use the 
term ``core'' to describe a compact, dense object within a clump with a size 
$\lesssim$0.1 pc and a mass $\lesssim$50 \Msun.} 
using dust continuum emission, show that 
they have typical masses of $\sim$120 \Msun\ and sizes $\sim$0.5 pc 
\citep{Rathborne06}. Spectral energy distributions (SEDs) of 
several IRDC clumps reveal dust temperatures that range between 16 and 52 K, 
and luminosities that range from $\sim$10--10$^{5}$ \Lsun\ \citep{Rathborne10}.
 Temperatures derived using NH$_3$ 
observations are lower than dust temperatures and range from $\sim$10--20 K 
\citep{Pillai06,Sakai08,Devine11,Ragan11}. Star formation activity in IRDC 
clumps can be inferred from  high temperatures and high luminosities 
as well as the presence of UC \hii\ regions
 \citep{Battersby10}, hot cores \citep{Rathborne08}, embedded 24 $\mu$m
 sources \citep{Chambers09}, molecular outflows 
\citep{Beuther07,Sanhueza10}, and maser emission \citep{Wang06,Chambers09}.
 The high masses, densities, and column densities of the IRDC clumps,
 as well as the aforementioned signatures of star formation in them,   
indicate that IRDCs are currently forming massive stars. In addition,
 IRDCs harbor numerous candidates for the most elusive earliest phase 
of high-mass star formation, the ``prestellar'' or ``starless'' phase 
\citep{Chambers09,Rathborne10,Rygl10,Vasyunina11,Pillai11,Devine11}. 
Due to the characteristics of IRDCs and the large variety of 
evolutionary stages that they harbor, it has been suggested that they are 
the natal sites of all massive stars and stellar clusters 
\citep{Rathborne06,Rathborne10}.

With the aim of characterizing the different evolutionary stages of clumps 
found in IRDCs, \cite{Chambers09} proposed an evolutionary
sequence in which ``quiescent'' clumps evolve into 
``intermediate'', ``active'', and ``red'' clumps. This evolutionary scheme
 is based on the  {\it Spitzer}/IRAC 3-8 $\mu$m colors
 and the presence or absence of {\it Spitzer}/MIPS 24 $\mu$m point-source emission.
 A clump is called ``quiescent'' if it contains no IR-{\it Spitzer} emission (it 
is IR-dark); ``intermediate'' if it contains either an enhanced 4.5 $\mu$m 
emission, the so-called ``green fuzzies'' \citep[also known as Extended
Green Objects (EGOs);][]{Cyganowski08}, or a 24 $\mu$m source, but not both; 
``active'' if it is associated with a green fuzzy and an embedded 24
 $\mu$m source; and ``red'' if it is associated with bright 8 $\mu$m emission, 
which likely corresponds to an \hii\ region. Quiescent clumps are the best
 candidates to be in the ``prestellar'' or ``starless'' phase, and the
 massive quiescent clumps are the places where it is most probable that
 high-mass stars are in their earliest stages of evolution.  
An additional category is ``blue,'' which describes objects with bright
 3.6 $\mu$m emission that are predominantly unextincted stars. 

Although several properties of IRDCs have been determined in the last few 
years, 
there is still one that remains poorly explored, their chemistry. What is 
the chemistry in IRDC clumps? Are the evolutionary stages defined by
 \cite{Chambers09} chemically distinguishable? Presently, only a few studies 
have focused on this subject \citep{Sakai08,Sakai10,Sakai12,Vasyunina11,Miettinen11,Chen11}. 
  
In this paper, we report initial results from a program aimed at better 
understanding the chemical 
evolution of IRDC clumps, carrying out a multi-line survey at 3 mm. 
We observed several molecular lines simultaneously, which facilitates 
comparison between different lines by eliminating or reducing  
 observational errors arising from uncertainties in telescope pointing
 and calibrations. The main goals of this project are to investigate 
the behavior of the different molecular tracers and look for observable
 changes in the chemistry as a function of the evolutionary stages proposed 
by \cite{Chambers09}.

\begin{deluxetable*}{lcccccc}
\tabletypesize{\scriptsize}
\tablecaption{Summary of Observed Molecular Lines \label{tbl-obsparam}}
\tablewidth{0pt}
\tablehead{
\colhead{Molecule} &  \colhead{Transition}  & \colhead{Rest Frequency}
 & \colhead{$E_u/k$} & \colhead{$n_{\rm crit}$}& \colhead{IF} & \colhead{$T_{\rm rms}$} \\
\colhead{} &  \colhead{} & \colhead{(GHz)}&\colhead{(K)}&\colhead{(cm$^{-3}$)}&\colhead{}&\colhead{(K)}\\
}
\startdata
\NHdosD   & \tNHdosD    & 85.926260 & 20.68& $4\times 10^6$ & IF3 & 0.044 \\
\SO       & \tSO        & 86.093983 & 19.31& $2\times 10^5$ & IF3 & 0.044 \\
\HtreceCN & \tHtreceCNuu& 86.338735 & 4.14 & $2\times 10^6$ & IF3 & 0.044 \\
          & \tHtreceCNdu& 86.340167 & 4.14 & $2\times 10^6$ & IF3 & 0.044 \\
          & \tHtreceCNcu& 86.342256 & 4.14 & $2\times 10^6$ & IF3 & 0.044 \\
\HtreceCO & \tHtreceCO  & 86.754330 & 4.16 & $2\times 10^5$ & IF3 & 0.044 \\
\SiO      & \tSiO       & 86.846998 & 6.25 & $2\times 10^6$ & IF3 & 0.044 \\
\HNtreceC & \tHNtreceC  & 87.090859 & 4.18 & $3\times 10^5$ & IF3 & 0.044 \\
\CdosH    & \tCdosHudcu & 87.316925 & 4.19 & $2\times 10^5$\tablenotemark{a} & IF3 & 0.044 \\
          & \tCdosHuucc & 87.328624 & 4.19 & $2\times 10^5$\tablenotemark{a}& IF3 & 0.044 \\
          & \tCdosHuucu & 87.402004 & 4.19 & $2\times 10^5$\tablenotemark{a} & IF3 & 0.044 \\
          & \tCdosHuccu & 87.407165 & 4.19 & $2\times 10^5$\tablenotemark{a} & IF3 & 0.044 \\
\HNCO     & \tHNCO      & 87.925252 & 10.55& $1\times 10^6$ & IF2 & 0.048 \\
\HCN      & \tHCNuu     & 88.630416 & 4.25 & $3\times 10^6$ & IF2 & 0.048 \\
          & \tHCNud     & 88.631847 & 4.25 & $3\times 10^6$ & IF2 & 0.048 \\
          & \tHCNuc     & 88.633936 & 4.25 & $3\times 10^6$ & IF2 & 0.048 \\
\HCO      & \tHCO       & 89.188526 & 4.28 & $2\times 10^5$ & IF2 & 0.048 \\
\HNC      & \tHNC       & 90.663574 & 4.35 & $3\times 10^5$ & IF1 & 0.042 \\
\HCtresN  & \tHCtresN   & 90.978989 & 24.01& $5\times 10^5$ & IF1 & 0.042 \\
\CHtresCN& \tCHtresCNuu & 91.985316 & 20.39& $4\times 10^5$ & IF0 & 0.042 \\
          & \tCHtresCNcc& 91.987089 & 13.24& $5\times 10^5$ & IF0 & 0.042 \\
\CS       & \tCS        & 92.494303 & 6.00  & $3\times 10^5$ & IF0 & 0.042 \\ 
\NdosH    & \tNdosHuud  & 93.171913 & 4.47 & $3\times 10^5$ & IF0 & 0.042 \\
          & \tNdosHudt  & 93.173772 & 4.47 & $3\times 10^5$ & IF0 & 0.042 \\
          & \tNdosHucu  & 93.176261 & 4.47 & $3\times 10^5$ & IF0 & 0.042 \\
\enddata
\tablecomments{The critical density was calculated as  
$n_{\rm crit}=A_{\rm ul}/\gamma_{\rm ul}$, where $A_{\rm ul}$ is the Einstein 
coefficient and $\gamma_{\rm ul}$ is the collisional rate. Values 
of $A_{\rm ul}$ and $\gamma_{\rm ul}$ at 20 K (50 K for SO) were obtained 
for most of the molecules from the Leiden Atomic and Molecular Database 
(LAMDA) \citep{Schoier05}. Values of $A_{\rm ul}$ and $\gamma_{\rm ul}$ 
at 25 K for \NHdosD\ were obtained from \cite{Machin06}.}
\tablenotetext{a}{Critical density adopted from \cite{Lo09}.}
\end{deluxetable*}

\section{Observations}

\subsection{The Sample}
\label{sample}

We made single-pointing observations of molecular lines at 3 mm using the 
ATNF Mopra 22 m telescope\footnote{The Mopra telescope is part of the
 Australia Telescope National Facility (ATNF) which is funded by 
the Commonwealth of Australia for operation as a National Facility 
managed by CSIRO.} located in Australia. The source list comprises 37 IRDCs, 
containing a total of 159 clumps. This is a sub-sample of the 38 IRDCs 
and 190 clumps identified by \cite{Rathborne06} using 1.2 mm dust continuum
 emission at 11\arcsec\ angular resolution. 
The threshold limit for a molecular line to be considered as a detection 
was defined at the 3$\sigma$ level, where $\sigma$ is the rms noise in 
antenna temperature ($T_{\rm rms}$; see Table~\ref{tbl-obsparam}).
The clumps G015.31 MM4, G027.75 MM5, G028.37 MM5, and G030.57 MM4 present 
no molecular line emission above the 3$\sigma$ detection level; 
consequently, they are not included in any analysis 
throughout this work. We excluded 17 clumps that show signs of star
 formation at IR 
wavelengths ({\it Spitzer}) and are located at a different distance than the 
rest of the clumps within an IRDC (see Section~\ref{kinematic} for more 
details). These clumps belong to a different molecular 
cloud that is situated in the line of sight. 
Since the Mopra telescope beam 
size (38\arcsec) is sometimes comparable to the separation between
 clumps within the 
same IRDC, $\sim$30\% of the sources are contaminated by emission  
from another adjacent clump. To ensure that the observed emission is 
attributable solely to a single clump, and not a neighboring clump, each 
source with an angular separation smaller than the Mopra beam size 
from its nearest neighbor was excluded. Furthermore, blue clumps are
 not included in any analysis because they are likely 
unextincted stars. Therefore, this work will focus on the 
remaining 92 IRDC clumps unless stated otherwise. \cite{Chambers09} and
 \cite{Rathborne10} classified the initial sample of 190 IRDC
clumps using a combination of {\it Spitzer}/IRAC colors and the presence or 
absence of compact 24 $\mu$m emission. Their classifications differ in 
the method of estimating the 24 $\mu$m flux measurements. \cite{Chambers09}  
used aperture photometry, and \cite{Rathborne10} fitted a Gaussian
 to the radial profile of the emission. Due to this difference,
 in a few cases 24 $\mu$m emission is detected for a particular 
source using one 
method, but not the other. In this work, to avoid ambiguities, 
if a clump presents 24 $\mu$m emission from either of the two methods,
a 24 $\mu$m compact source will be assigned to the clump.

Table~\ref{tbl-coor} summarizes information for all 159 IRDC clumps: 
 names, coordinates, velocities, distances, dust temperatures, 
{\it Spitzer} classification, and if the source is an IRDC and if
 it is used in the present work. 

\subsection{Observing Parameters}
\label{mopra}

The observations were carried out during 2008 July--August and 2010 September. 
They were performed 
with the 3 mm Monolithic Microwave Integrated Circuits (MMIC) receiver and
 the Mopra spectrometer (MOPS)\footnote{The University of New South Wales Digital Filter Bank used for the observations with the Mopra Telescope was provided with support from the Australian Research Council.} in the 
``broad-band'' mode, resulting in a total bandwidth of 8.3 GHz split over four 
overlapping sub-bands of 2.2 GHz. Each sub-band has $2\times8096$ channels 
($1\times8096$ per each polarization) with a channel resolution of 
0.27 MHz. The velocity resolution was $\sim$0.90 \kms\ per channel.  
Both polarizations were observed and averaged to improve the 
signal to noise ratio. The observed frequencies range between 85.540 
and 93.840 GHz. This frequency range allowed us to observe 15 molecular lines 
simultaneously (see Table~\ref{tbl-obsparam}). 
The system temperatures ranged between 165 and 265 K. Typical rms noise values 
($T_{\rm rms}$) for each molecular line are shown in Table~\ref{tbl-obsparam}.
 The on-source
 integration time per object was $\sim$3 minutes. 
The angular resolution of Mopra telescope is 38\arcsec\ and its main beam 
efficiency is 0.5 at 90 GHz.\footnote{See more details
about Mopra telescope, receivers, and backends at 
http://www.narrabri.atnf.csiro.au/mopra/obsinfo.html}
All the observations were performed in position-switching mode with the
 off-position shifted from the target source by 1\arcdeg\ in Galactic 
latitude away from the Galactic midplane. The telescope pointing
 was checked by observing nearby SiO masers 
every $\sim$1 hr, and was maintained to be better than 5\arcsec.  
The uncertainty in the line intensity varies with the frequency 
from 9\% at 88 GHz to 25\% at 93 GHz (Jonathan Foster, 
private communication). 
 The initial spectral processing was done
 using the ATNF Spectral Analysis Package (ASAP) 
software.\footnote{http://www.atnf.csiro.au/computing/software/} 
The observed molecular lines, transitions, rest frequencies, upper energy 
levels, critical densities, sub-bands, and typical $T_{\rm rms}$ are listed in 
Table~\ref{tbl-obsparam}. The $T_{\rm rms}$ for each source is 
given in Table~\ref{tbl-gaussian_123_012}. 

\section{Results}

\subsection{Line Parameters}

For each molecular transition, the line center velocity, line width and 
intensity of the line were determined from Gaussian fits. Molecular lines 
that show hyperfine structure (N$_2$H$^+$ and C$_2$H) were fitted using 
a multi-Gaussian function, with a fixed frequency separation between 
the transitions. 
The Gaussian fit procedure was carried out in IDL using the MPFITFUN 
package \citep{Markwardt09}. Occasionally, two velocity components were  
detected in a source. We define the main component, i.e., the component
 associated with the IRDC clump, as the velocity component detected in
 the high-density tracers (e.g., N$_2$H$^+$ and/or H$^{13}$CO$^+$). 
 The secondary velocity 
component typically corresponds to lower-density gas located in line of sight 
and they were not used in this work. 
When two velocity components have emission from high-density tracers, 
the brightest was used as the main component. 
The Gaussian fit parameters of the main velocity component for all 
sources where emission was detected (155) are summarized
 in Table~\ref{tbl-gaussian_123_012}. No Gaussian fits were carried out for
 two distinct velocity components with separations less than 3 \kms\ or for
 self-absorbed profiles.

\subsection{Spectra and Detection Rates}

\begin{figure}[h!]
\begin{center}
\includegraphics[angle=0,scale=0.51]{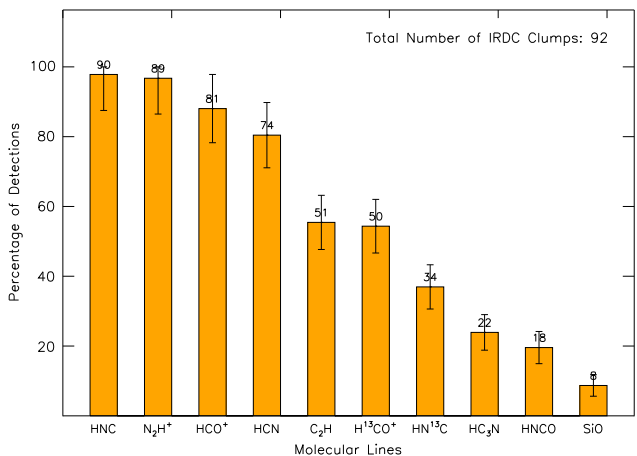}
\end{center}
\caption{Detection rates of the observed molecular lines 
toward IRDC clumps. The N$_2$H$^+$ and C$_2$H detection 
rates correspond to those of the $JF_1F=123-012$ and 
$NJF=1\,\frac{3}{2}\,2\ra0\,\frac{1}{2}\,1$ transitions,
 respectively. Error bars are the Poisson noise (i.e., the
 root square of the number of detections). Values on bars
 are number of detections for a given molecule.} 
\label{detection_rates}
\end{figure}

Figure~\ref{detection_rates} presents the detection rates, at the 3$\sigma$ 
detection level, of the 10 most frequently detected molecular species 
toward the IRDC clumps. 
For the N$_2$H$^+$ and C$_2$H lines, we show the most often detected 
transitions: $JF_1F=123\ra012$ for N$_2$H$^+$ and 
$NJF=1\,\frac{3}{2}\,2\ra0\,\frac{1}{2}\,1$ for C$_2$H. Because the HCN  
transitions are blended, and in addition, exhibit self-absorbed profiles
 and wing emission, the multi-Gaussian fit to the 
hyperfine structure was not reliable. The HCN detection rate was obtained 
by inspecting the spectrum for each source by eye, instead of comparing 
the intensity of the Gaussian fit with the $T_{\rm rms}$. 
The CH$_3$CN, $^{13}$CS, H$^{13}$CN, SO, and NH$_2$D lines 
were detected in fewer than 8 clumps; 
therefore, their detection rates are not presented in 
Figure~\ref{detection_rates}. The molecular lines most often detected were  
HNC $J=1\ra0$ and N$_2$H$^+$ $JF_1F=123\ra012$ with 90 (98\%) and 
89 (97\%) detections, respectively. On the other hand, the molecular line 
 shown in Figure~\ref{detection_rates} least often detected was the
 SiO $J=2\ra1$ with 8 (9\%) 
detections. Detection rates for all detected molecular transitions are  
given in Table~\ref{tbl-detection-rates}. Uncertainties presented in 
Figure~\ref{detection_rates} and Table~\ref{tbl-detection-rates} were 
determined assuming Poisson noise.

\LongTables
\begin{deluxetable*}{llllccccc}
\tabletypesize{\scriptsize}
\tablecaption{Clumps Properties. \label{tbl-coor}}
\tablewidth{0pt}
\tablehead{
\colhead{IRDC} &  \colhead{Clump}  & \multicolumn{2}{c}{\underline {~~~~~~Coordinates~~~~~~}}     & \colhead{$V_{lsr}$\tablenotemark{a}}  & \colhead{D} & \colhead{T$_{\rm dust}$} & \colhead{Class.\tablenotemark{b}}  & \colhead{Comment\tablenotemark{e}}\\
\colhead{Name} &  \colhead{Number} & \colhead{$\alpha$(J2000)}&\colhead{$\delta$(J2000)}& \colhead{(\kms)}  & \colhead{(kpc)}  & \colhead{(K)}  & \colhead{ } & \colhead{ } \\
}
\startdata
G015.05+00.07 &  MM1 & 18:17:50.4 & -15:53:38  &  25.2  & 2.8 & 23.5 & Q & IRDC \\
G015.05+00.07 &  MM2 & 18:17:40.0 & -15:48:55  &  29.9  & 3.1 & 30.5 & R & IRDC \\
G015.05+00.07 &  MM3 & 18:17:42.4 & -15:47:03  &  28.4  & 3.0 & 23.7 & Q & IRDC \\
G015.05+00.07 &  MM4 & 18:17:32.0 & -15:46:35  &  28.0  & 3.0 & 30.5 & Q & IRDC \\
G015.05+00.07 &  MM5 & 18:17:40.2 & -15:49:47  &  29.9  & 3.1 & 31.0 & Q & IRDC \\
G015.31-00.16 &  MM1 & 18:18:56.4 & -15:45:00  &  17.0\tablenotemark{c}  & 1.9\tablenotemark{c} & ... & R & Non-IRDC \\
G015.31-00.16 &  MM2 & 18:18:50.4 & -15:43:19  &  30.9  & 3.2 & 23.0 & I & IRDC \\
G015.31-00.16 &  MM3 & 18:18:45.3 & -15:41:58  &  31.1  & 3.2 & 23.7 & Q & IRDC \\
G015.31-00.16 &  MM4 & 18:18:48.0 & -15:44:22  &  31.1\tablenotemark{d}  & 1.9\tablenotemark{d} & ... & Q & No Detection \\
G015.31-00.16 &  MM5 & 18:18:49.1 & -15:42:47  &  31.2\tablenotemark{c}  & 3.2\tablenotemark{c} & 28.0 & Q & IRDC \\
G018.82-00.28 &  MM1 & 18:25:56.1 & -12:42:48  &  41.7  & 3.5 & ...  & I & Non-IRDC \\
G018.82-00.28 &  MM2 & 18:26:23.4 & -12:39:37  &  63.1  & 4.7 & 38.0 & R & IRDC \\
G018.82-00.28 &  MM3 & 18:25:52.6 & -12:44:37  &  44.8  & 3.7 & 34.0 & A & IRDC \\
G018.82-00.28 &  MM4 & 18:26:15.5 & -12:41:32  &  65.5  & 4.8 & 17.0 & I & IRDC \\
G018.82-00.28 &  MM6 & 18:26:18.4 & -12:41:15  &  65.8  & 4.8 & 23.7 & Q & IRDC \\
G019.27+00.07 &  MM1 & 18:25:58.5 & -12:03:59  &  26.7  & 2.4 & ...  & B & Bright Blue \\
G019.27+00.07 &  MM2 & 18:25:52.6 & -12:04:48  &  26.7  & 2.4 & 30.0 & A & IRDC \\
G022.35+00.41 &  MM1 & 18:30:24.4 & -09:10:34  &  52.7  & 3.9 & 20.0 & A & IRDC \\
G022.35+00.41 &  MM2 & 18:30:24.2 & -09:12:44  &  59.9\tablenotemark{c}  & 4.2\tablenotemark{c} & 44.0 & R & IRDC \\
G023.60+00.00 &  MM1 & 18:34:11.6 & -08:19:06  & 106.5  & 7.0 & ...  & A & Non-IRDC \\
G023.60+00.00 &  MM2 & 18:34:21.1 & -08:18:07  &  53.7  & 3.9 & ...  & A & Blended \\
G023.60+00.00 &  MM3 & 18:34:10.0 & -08:18:28  & 105.8  & 6.9 & ...  & R & Non-IRDC \\
G023.60+00.00 &  MM4 & 18:34:23.0 & -08:18:21  &  53.6  & 3.9 & ...  & R & Blended \\
G023.60+00.00 &  MM5 & 18:34:09.5 & -08:18:00  & 104.8  & 6.8 & ...  & A & Non-IRDC \\
G023.60+00.00 &  MM7 & 18:34:21.1 & -08:17:11  &  54.0  & 3.9 & 44.0 & I & IRDC \\
G023.60+00.00 &  MM9 & 18:34:22.5 & -08:16:04  &  54.3  & 3.9 & 32.5 & Q & IRDC \\
G024.08+00.04 &  MM1 & 18:34:57.0 & -07:43:26  & 114.1  & 7.8 & ...  & R & Non-IRDC \\
G024.08+00.04 &  MM2 & 18:34:51.1 & -07:45:32  & 114.1  & 7.8 & 29.0 & Q & IRDC \\
G024.08+00.04 &  MM3 & 18:35:02.2 & -07:45:25  &  51.6  & 3.7 & 29.0 & Q & IRDC \\
G024.08+00.04 &  MM4 & 18:35:02.6 & -07:45:56  &  52.2  & 3.7 & 30.0 & Q & IRDC \\
G024.33+00.11 &  MM1 & 18:35:07.9 & -07:35:04  & 114.1  & 7.8 & ...  & R & Blended \\
G024.33+00.11 &  MM2 & 18:35:34.5 & -07:37:28  & 118.2  & 7.7 & 32.5 & I & IRDC \\
G024.33+00.11 &  MM3 & 18:35:27.9 & -07:36:18  & 117.6  & 7.7 & 31.5 & R & IRDC \\
G024.33+00.11 &  MM4 & 18:35:19.4 & -07:37:17  & 115.0  & 7.8 & 31.0 & Q & IRDC \\
G024.33+00.11 &  MM5 & 18:35:33.8 & -07:36:42  & 117.2  & 7.7 & 32.0 & I & IRDC \\
G024.33+00.11 &  MM6 & 18:35:07.7 & -07:34:33  & 114.3  & 7.7 & ...  & I & Blended \\
G024.33+00.11 &  MM7 & 18:35:09.8 & -07:39:48  &  99.6  & 6.3 & 23.7 & Q & IRDC \\
G024.33+00.11 &  MM8 & 18:35:23.4 & -07:37:21  & 113.8  & 7.8 & 31.5 & Q & IRDC \\
G024.33+00.11 &  MM9 & 18:35:26.5 & -07:36:56  & 119.2  & 7.7 & 52.0 & R & IRDC \\
G024.33+00.11 & MM11 & 18:35:05.1 & -07:35:58  & 113.3  & 7.8 & 26.0 & Q & IRDC \\
G024.60+00.08 &  MM1 & 18:35:41.1 & -07:18:30  &  53.4  & 3.8 & ...  & A & Blended \\
G024.60+00.08 &  MM2 & 18:35:39.3 & -07:18:51  & 115.2  & 7.7 & 23.0 & I & IRDC \\
G024.60+00.08 &  MM3 & 18:35:40.2 & -07:18:37  &  53.8  & 3.8 & ...  & I & Blended \\
G024.60+00.08 &  MM4 & 18:35:35.7 & -07:18:09  &  53.2  & 3.8 & ...  & Q & Blended \\
G025.04-00.20 &  MM1 & 18:38:10.2 & -07:02:34  &  64.0  & 4.3 & ...  & A & Blended \\
G025.04-00.20 &  MM2 & 18:38:17.7 & -07:02:51  &  63.5  & 4.3 & 28.0 & I & IRDC \\
G025.04-00.20 &  MM3 & 18:38:10.2 & -07:02:44  &  63.7  & 4.3 & ...  & Q & Blended \\
G025.04-00.20 &  MM4 & 18:38:13.7 & -07:03:12  &  63.8  & 4.3 & 29.5 & I & IRDC \\
G025.04-00.20 &  MM5 & 18:38:12.0 & -07:02:44  &  63.9  & 4.3 & ...  & Q & Blended \\
G027.75+00.16 &  MM1 & 18:41:19.9 & -04:32:20  &  78.3  & 5.0 & ...  & A & Blended \\
G027.75+00.16 &  MM2 & 18:41:33.0 & -04:33:44  &  51.4  & 3.5 & 24.0 & Q & IRDC \\
G027.75+00.16 &  MM3 & 18:41:16.8 & -04:31:55  &  78.7  & 5.0 & ...  & I & Blended \\
G027.75+00.16 &  MM5 & 18:41:23.6 & -04:30:42  &  79.1\tablenotemark{d}  & 4.8\tablenotemark{d} & ... & Q & No Detection \\
G027.94-00.47 &  MM1 & 18:44:03.6 & -04:38:00  &  45.4  & 3.1 & 44.0 & R & IRDC \\
G027.97-00.42 &  MM1 & 18:43:52.8 & -04:36:13  &  44.6  & 3.1 & ...  & A & Blended \\
G027.97-00.42 &  MM2 & 18:43:58.0 & -04:34:24  &  19.9  & 1.5 & ...  & R & Non-IRDC \\
G027.97-00.42 &  MM3 & 18:43:54.9 & -04:36:08  &  45.9  & 3.2 & ...  & Q & Blended \\
G028.04-00.46 &  MM1 & 18:44:08.5 & -04:33:22  &  45.8  & 3.2 & 38.0 & A & IRDC \\
G028.08+00.07 &  MM1 & 18:42:20.3 & -04:16:42  &  81.4  & 5.2 & 22.0 & I & IRDC \\
G028.10-00.45 &  MM1 & 18:44:12.9 & -04:29:45  &  47.1  & 3.2 & ...  & Q & Blended \\
G028.10-00.45 &  MM2 & 18:44:14.3 & -04:29:48  &  46.9  & 3.2 & ...  & Q & Blended \\
G028.23-00.19 &  MM1 & 18:43:30.7 & -04:13:12  &  80.0  & 5.1 & 22.0 & Q & IRDC \\
G028.23-00.19 &  MM2 & 18:43:29.0 & -04:12:16  &  80.8  & 5.1 & ...  & I & Blended \\
G028.23-00.19 &  MM3 & 18:43:30.0 & -04:12:33  &  80.2  & 5.1 & ...  & Q & Blended \\
G028.28-00.34 &  MM1 & 18:44:15.0 & -04:17:54  &  48.7  & 3.3 & ...  & R & Blended \\
G028.28-00.34 &  MM2 & 18:44:21.3 & -04:17:37  &  84.9  & 5.6 & ...  & R & Non-IRDC \\
G028.28-00.34 &  MM3 & 18:44:13.4 & -04:18:05  &  49.7  & 3.4 & ...  & R & Blended \\
G028.28-00.34 &  MM4 & 18:44:11.4 & -04:17:22  &  48.8  & 3.4 & 34.5 & A & IRDC \\
G028.37+00.07 &  MM1 & 18:42:52.1 & -03:59:45  &  78.1  & 5.0 & 33.0 & A & IRDC \\
G028.37+00.07 &  MM2 & 18:42:37.6 & -04:02:05  &  80.8  & 5.1 & 39.0 & I & IRDC \\
G028.37+00.07 &  MM3 & 18:43:03.1 & -04:06:24  &  99.8  & 6.4 & ...  & R & Non-IRDC \\
G028.37+00.07 &  MM4 & 18:42:50.7 & -04:03:15  &  79.3  & 5.0 & 32.0 & A & IRDC \\
G028.37+00.07 &  MM5 & 18:42:26.8 & -04:01:30  &  78.6\tablenotemark{d}  & 4.8\tablenotemark{d} & ... & R & No Detection \\
G028.37+00.07 &  MM6 & 18:42:49.0 & -04:02:23  &  80.0  & 5.1 & 23.0 & A & IRDC \\
G028.37+00.07 &  MM7 & 18:42:56.3 & -04:07:31  &  46.6  & 3.2 & ...  & R & Non-IRDC \\
G028.37+00.07 &  MM8 & 18:42:49.7 & -04:09:54  & 107.2  & 7.5 & ...  & R & Non-IRDC \\
G028.37+00.07 &  MM9 & 18:42:46.7 & -04:04:08  &  79.4  & 5.0 & 24.5 & Q & IRDC \\
G028.37+00.07 & MM10 & 18:42:54.0 & -04:02:30  &  79.1  & 5.0 & ...  & A & Blended \\
G028.37+00.07 & MM11 & 18:42:42.7 & -04:01:44  &  80.9  & 5.1 & 36.0 & A & IRDC \\
G028.37+00.07 & MM12 & 18:43:09.9 & -04:06:52  &  34.8  & 2.5 & 23.7 & Q & IRDC \\
G028.37+00.07 & MM13 & 18:42:41.8 & -03:57:08  &  77.5  & 4.9 & 30.4 & I & IRDC \\
G028.53-00.25 &  MM1 & 18:44:18.0 & -03:59:34  &  86.8  & 5.5 & ...  & Q & Blended \\
G028.53-00.25 &  MM2 & 18:44:15.7 & -03:59:41  &  85.6  & 5.4 & ...  & A & Blended \\
G028.53-00.25 &  MM3 & 18:44:16.0 & -04:00:48  &  86.4  & 5.5 & 22.0 & Q & IRDC \\
G028.53-00.25 &  MM4 & 18:44:18.6 & -04:00:05  &  86.3  & 5.4 & ...  & I & Blended \\
G028.53-00.25 &  MM5 & 18:44:17.0 & -04:02:04  &  87.0  & 5.7 & 30.0 & I & IRDC \\
G028.53-00.25 &  MM6 & 18:44:17.8 & -04:00:05  &  86.4  & 5.5 & ...  & Q & Blended \\
G028.53-00.25 &  MM7 & 18:44:23.7 & -04:02:09  &  88.6  & 5.6 & 23.0 & Q & IRDC \\
G028.53-00.25 &  MM8 & 18:44:22.0 & -04:01:35  &  88.6  & 5.6 & 28.0 & Q & IRDC \\
G028.53-00.25 &  MM9 & 18:44:19.3 & -03:58:05  &  87.1  & 5.7 & 28.0 & I & IRDC \\
G028.53-00.25 & MM10 & 18:44:18.5 & -03:58:43  &  86.8  & 5.5 & 30.0 & Q & IRDC \\
G028.67+00.13 &  MM1 & 18:43:03.1 & -03:41:41  &  84.3\tablenotemark{c}  & 5.3\tablenotemark{c} & 40.0 & R & IRDC \\
G028.67+00.13 &  MM2 & 18:43:07.1 & -03:44:01  &  79.2  & 5.0 & 23.0 & I & IRDC \\
G028.67+00.13 &  MM3 & 18:42:58.2 & -03:48:20  & 103.9  & 6.9 & ...  & A & Non-IRDC \\
G030.14-00.06 &  MM1 & 18:46:35.7 & -02:31:03  &  86.8  & 5.5 & 27.0 & A & IRDC \\
G030.57-00.23 &  MM1 & 18:48:00.0 & -02:07:20  &  90.7  & 5.8 & 34.5 & A & IRDC \\
G030.57-00.23 &  MM2 & 18:47:58.7 & -02:15:20  & 111.3  & 7.3 & ...  & R & Non-IRDC \\
G030.57-00.23 &  MM3 & 18:47:54.5 & -02:11:15  &  95.9  & 6.3 & 23.0 & I & IRDC \\
G030.57-00.23 &  MM4 & 18:48:01.8 & -02:12:35  &  86.2\tablenotemark{d}  & 5.2\tablenotemark{d} & ... & Q & No Detection \\
G030.97-00.14 &  MM1 & 18:48:21.6 & -01:48:27  &  77.8  & 5.0 & 37.0 & A & IRDC \\
G031.02-00.10 &  MM1 & 18:48:09.9 & -01:45:17  &  76.9  & 4.9 & ...  & B & Bright Blue \\
G031.97+00.07 &  MM1 & 18:49:36.3 & -00:45:45  &  95.0  & 6.5 & ...  & A & Blended \\
G031.97+00.07 &  MM2 & 18:49:36.0 & -00:46:16  &  94.6  & 6.5 & ...  & Q & Blended \\
G031.97+00.07 &  MM3 & 18:49:32.3 & -00:47:02  &  94.1  & 6.4 & ...  & Q & Blended \\
G031.97+00.07 &  MM4 & 18:49:33.0 & -00:47:33  &  95.5  & 6.6 & ...  & I & Blended \\
G031.97+00.07 &  MM5 & 18:49:21.9 & -00:50:35  &  96.5  & 6.8 & 35.0 & I & IRDC \\
G031.97+00.07 &  MM6 & 18:49:35.0 & -00:46:44  &  94.8  & 6.5 & ...  & Q & Blended \\
G031.97+00.07 &  MM7 & 18:49:28.4 & -00:48:54  &  93.9  & 6.4 & 27.0 & Q & IRDC \\
G031.97+00.07 &  MM8 & 18:49:29.1 & -00:48:12  &  94.5  & 6.4 & 47.0 & A & IRDC \\
G033.69-00.01 &  MM1 & 18:52:58.8 &  00:42:37  & 105.7  & 7.1 & 25.0 & R & IRDC \\
G033.69-00.01 &  MM2 & 18:52:49.9 &  00:37:57  & 104.8  & 7.1 & 41.0 & R & IRDC \\
G033.69-00.01 &  MM3 & 18:52:50.8 &  00:36:43  & 103.1  & 7.1 & 43.0 & R & IRDC \\
G033.69-00.01 &  MM4 & 18:52:56.4 &  00:43:08  & 106.2  & 7.1 & 28.0 & A & IRDC \\
G033.69-00.01 &  MM5 & 18:52:47.8 &  00:36:47  & 105.5  & 7.1 & 37.0 & A & IRDC \\
G033.69-00.01 & MM11 & 18:52:56.2 &  00:41:48  & 107.0  & 7.1 & 25.0 & Q & IRDC \\
G034.43+00.24 &  MM1 & 18:53:18.0 &  01:25:24  &  57.9  & 3.8 & 38.0 & A & IRDC \\
G034.43+00.24 &  MM2 & 18:53:18.6 &  01:24:40  &  57.4  & 3.8 & ...  & R & Blended \\
G034.43+00.24 &  MM3 & 18:53:20.4 &  01:28:23  &  59.4  & 3.9 & ...  & A & Blended \\
G034.43+00.24 &  MM4 & 18:53:19.0 &  01:24:08  &  57.6  & 3.8 & ...  & A & Blended \\
G034.43+00.24 &  MM5 & 18:53:19.8 &  01:23:30  &  58.0  & 3.8 & 23.0 & A & IRDC \\
G034.43+00.24 &  MM6 & 18:53:18.6 &  01:27:48  &  58.6  & 3.8 & ...  & I & Blended \\
G034.43+00.24 &  MM7 & 18:53:18.3 &  01:27:13  &  58.1  & 3.8 & 29.5 & I & IRDC \\
G034.43+00.24 &  MM8 & 18:53:16.4 &  01:26:20  &  57.2  & 3.7 & 43.0 & A & IRDC \\
G034.43+00.24 &  MM9 & 18:53:18.4 &  01:28:14  &  58.7  & 3.8 & ...  & Q & Blended \\
G034.77-00.55 &  MM1 & 18:56:48.2 &  01:18:47  &  44.2  & 2.9 & 40.4 & R & IRDC \\
G034.77-00.55 &  MM2 & 18:56:50.3 &  01:23:16  &  42.0  & 2.8 & ...  & Q & Blended \\
G034.77-00.55 &  MM3 & 18:56:44.7 &  01:20:42  &  43.1  & 2.8 & 30.4 & I & IRDC \\
G034.77-00.55 &  MM4 & 18:56:48.9 &  01:23:34  &  42.1  & 2.8 & ...  & Q & Blended \\
G035.39-00.33 &  MM1 & 18:56:41.2 &  02:09:52  &  64.3  & 4.2 & ...  & R & Non-IRDC \\
G035.39-00.33 &  MM2 & 18:56:59.2 &  02:04:53  &  54.0  & 3.5 & ...  & A & Non-IRDC \\
G035.39-00.33 &  MM3 & 18:57:05.3 &  02:06:29  &  55.3  & 3.6 & ...  & R & Non-IRDC \\
G035.39-00.33 &  MM4 & 18:57:06.7 &  02:08:23  &  45.3  & 3.0 & ...  & A & Blended \\
G035.39-00.33 &  MM5 & 18:57:08.8 &  02:08:09  &  90.1  & 6.9 & ...  & Q & Blended \\
G035.39-00.33 &  MM6 & 18:57:08.4 &  02:09:09  &  45.5  & 3.0 & ...  & A & Blended \\
G035.39-00.33 &  MM7 & 18:57:08.1 &  02:10:50  &  45.9  & 3.0 & 34.0 & A & IRDC \\
G035.59-00.24 &  MM1 & 18:57:02.3 &  02:17:04  &  50.4  & 3.3 & 36.0 & R & IRDC \\
G035.59-00.24 &  MM2 & 18:57:07.4 &  02:16:14  &  45.3  & 3.0 & 24.0 & A & IRDC \\
G035.59-00.24 &  MM3 & 18:57:11.6 &  02:16:08  &  45.0  & 3.0 & 30.5 & A & IRDC \\
G036.67-00.11 &  MM1 & 18:58:39.6 &  03:16:16  &  53.5  & 3.5 & 20.0 & Q & IRDC \\
G036.67-00.11 &  MM2 & 18:58:35.6 &  03:15:06  &  53.3  & 3.5 & 16.0 & Q & IRDC \\
G038.95-00.47 &  MM1 & 19:04:07.4 &  05:08:48  &  42.4  & 2.8 & 16.0 & I & IRDC \\
G038.95-00.47 &  MM2 & 19:04:03.4 &  05:07:56  &  42.1  & 2.8 & 40.0 & R & IRDC \\
G038.95-00.47 &  MM3 & 19:04:07.4 &  05:09:44  &  42.4  & 2.8 & 28.0 & I & IRDC \\
G038.95-00.47 &  MM4 & 19:04:00.6 &  05:09:06  &  42.1  & 2.8 & 43.0 & A & IRDC \\
G048.65-00.29 &  MM2 & 19:21:47.6 &  13:49:22  &  33.8  & 2.5 & ...  & Q & Blended \\
G053.11+00.05 &  MM1 & 19:29:17.2 &  17:56:21  &  21.8  & 1.8 & 38.0 & R & IRDC \\
G053.11+00.05 &  MM2 & 19:29:20.2 &  17:57:06  &  22.9  & 1.9 & 36.0 & A & IRDC \\
G053.11+00.05 &  MM3 & 19:29:00.6 &  17:55:11  &   2.6  & 0.5 & ...  & R & Non-IRDC \\
G053.11+00.05 &  MM4 & 19:29:20.4 &  17:55:04  &  21.7  & 1.8 & 40.0 & A & IRDC \\
G053.11+00.05 &  MM5 & 19:29:26.3 &  17:54:53  &  21.7  & 1.8 & 45.0 & R & IRDC \\
G053.25+00.04 &  MM1 & 19:29:39.0 &  18:01:42  &  24.9  & 2.0 & 35.0 & I & IRDC \\
G053.25+00.04 &  MM2 & 19:29:33.0 &  18:01:00  &  24.3  & 2.0 & ...  & B & Bright Blue \\
G053.25+00.04 &  MM3 & 19:29:44.0 &  17:58:47  &  23.5  & 1.9 & 35.0 & I & IRDC \\
G053.25+00.04 &  MM4 & 19:29:34.5 &  18:01:39  &  24.5  & 2.0 & 37.0 & A & IRDC \\
G053.25+00.04 &  MM5 & 19:29:39.4 &  17:58:40  &  23.7  & 1.9 & 35.0 & I & IRDC \\
G053.25+00.04 &  MM6 & 19:29:31.5 &  17:59:50  &  23.8  & 1.9 & 46.0 & A & IRDC \\
G053.31+00.00 &  MM1 & 19:29:50.0 &  18:05:07  &  22.0  & 1.8 & ...  & I & Blended \\
G053.31+00.00 &  MM2 & 19:29:42.1 &  18:03:57  &  25.2  & 2.0 & 25.0 & Q & IRDC \\
G053.31+00.00 &  MM3 & 19:29:49.7 &  18:04:39  &  22.2  & 1.8 & ...  & I & Blended \\
\enddata
\tablecomments{Units of right ascension are hours, minutes and seconds,
and units of declination are degrees, arcminutes and arcseconds.}
\tablenotetext{a}{Velocity from the N$_2$H$^+$ $JF_1F=123\ra012$ transition.}
\tablenotetext{b}{Denotes the classification based on {\it Spitzer}/IRAC colors 
and the presence or absence of {\it Spitzer}/MIPS 24 $\mu$m emission.}
\tablenotetext{c}{Velocity and distance determined from the HCO$^+$ line.}
\tablenotetext{d}{Velocity and distance from \cite{Rathborne10}.}
\tablenotetext{e}{IRDC: the clump is located in an IRDC (the 92 sources of this 
study). Non-IRDC: the clump is not associated with an IRDC. Non-Detection: no
 molecular line was detected in this survey. Bright Blue: objects with bright
 3.6 $\mu$m emission that are predominantly unextincted stars. Blended: the 
angular separation between two sources is less than one Mopra beam.}
\end{deluxetable*}

\begin{figure*}
\begin{center}
\includegraphics[angle=0,scale=0.36]{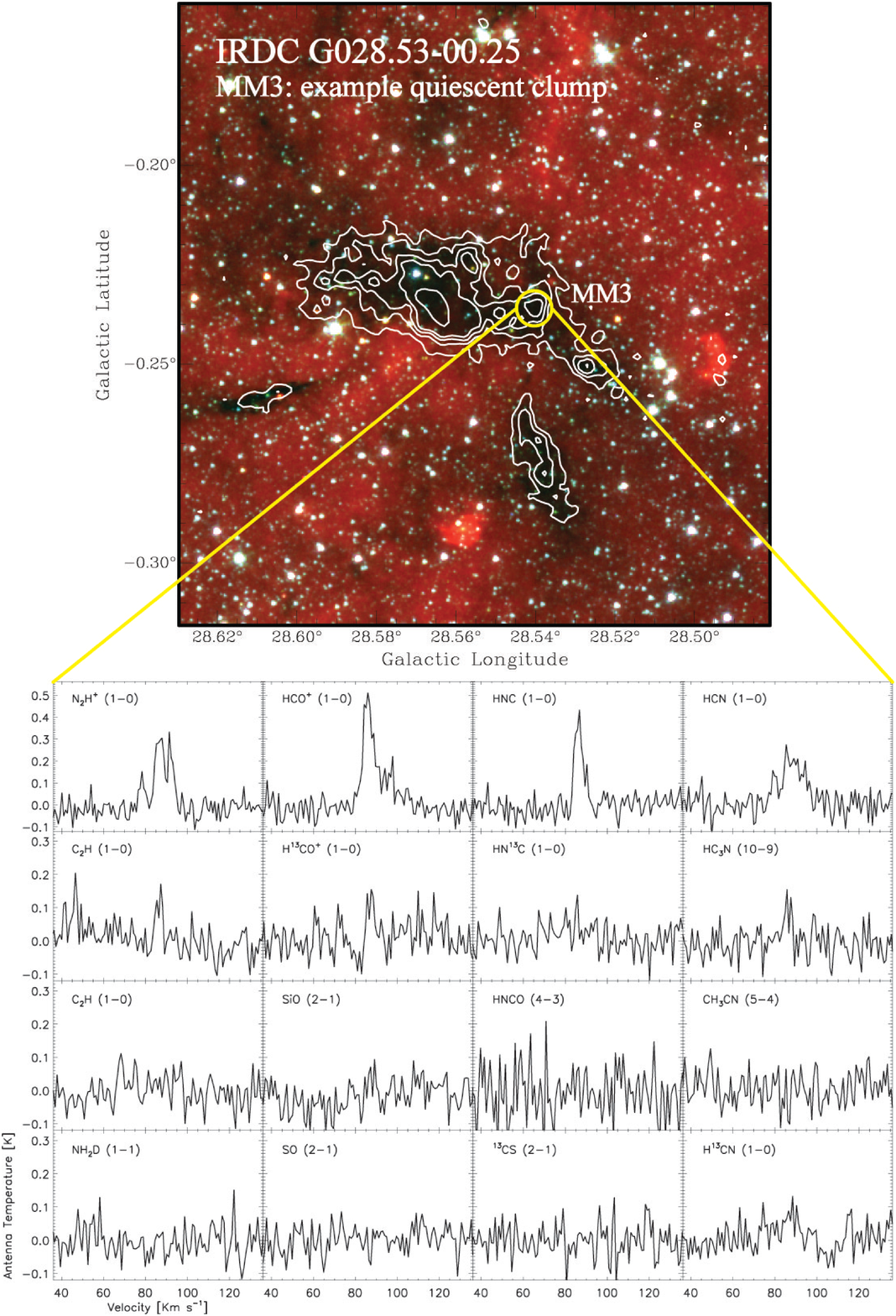}
\end{center}
\caption{ {\bf Quiescent clump example.} {\it Top:} IRAC 3-color 
(3.6 $\mu$m in blue, 4.5 $\mu$m in green and 8.0 $\mu$m in red) 
 image of the IRDC 
G028.53-00.25 overlaid with IRAM 1.2 mm continuum emissiom from
 \cite{Rathborne06}. {\it Bottom:} Molecular spectra of the quiescent MM3 clump.
Quiescent clumps contain none IR emission. The circle shows the 
telescope beam size and marks the position of the observed clump.}
\label{fig-quiescent}
\end{figure*}

\begin{figure*}
\begin{center}
\includegraphics[angle=0,scale=0.36]{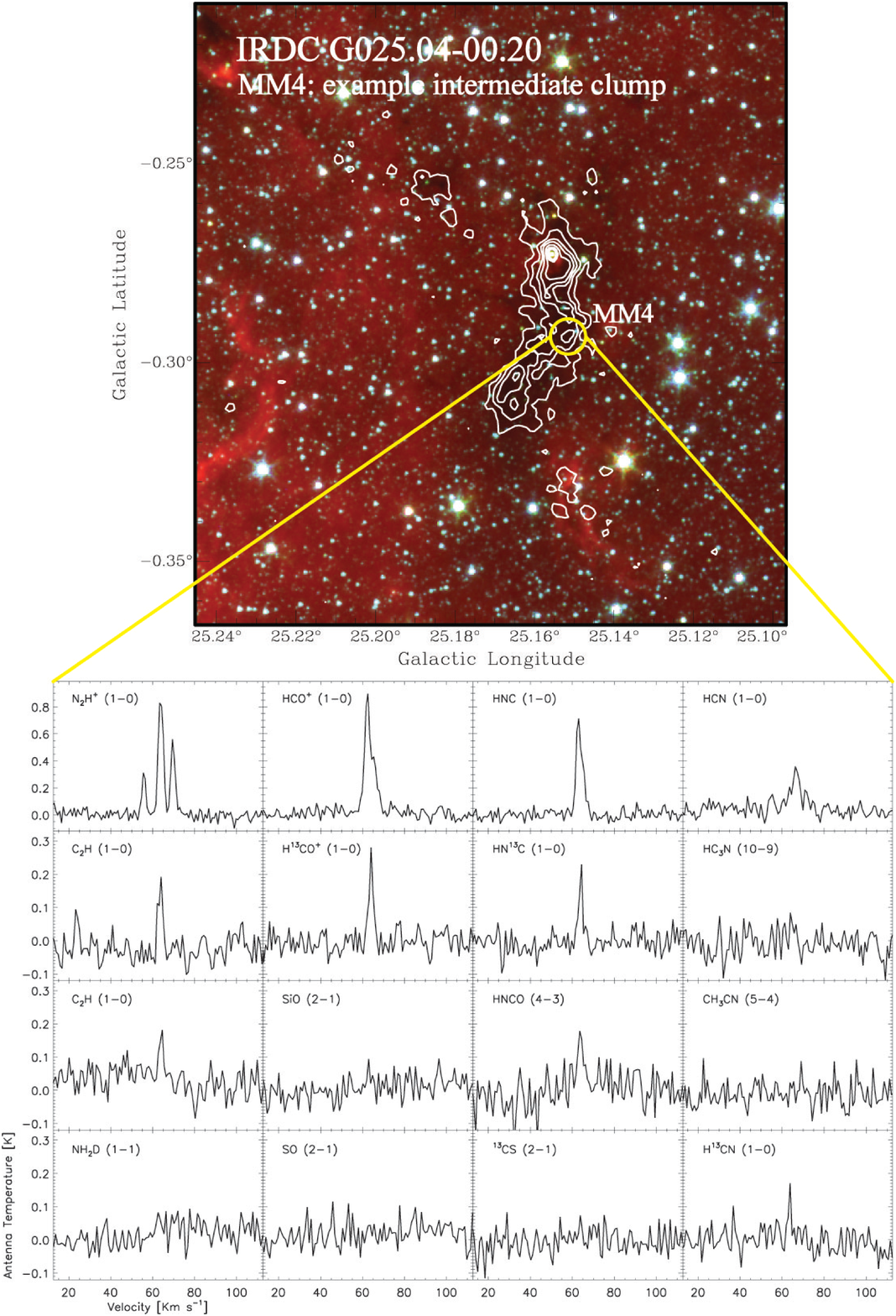}
\end{center}
\caption{ {\bf Intermediate clump example.} {\it Top:} IRAC 3-color  
(3.6 $\mu$m in blue, 4.5 $\mu$m in green and 8.0 $\mu$m in red) 
  image of the IRDC 
G025.04-00.20 overlaid with IRAM 1.2 mm continuum emissiom from
 \cite{Rathborne06}. {\it Bottom:} Molecular spectra of the intermediate MM4 clump.
Intermediate clumps contain either a green fuzzy or a 24 $\mu$m source,
but not both.  The circle shows the telescope beam size and marks the
 position of the observed clump.}
\label{fig-intermediate}
\end{figure*}

\begin{figure*}
\begin{center}
\includegraphics[angle=0,scale=0.36]{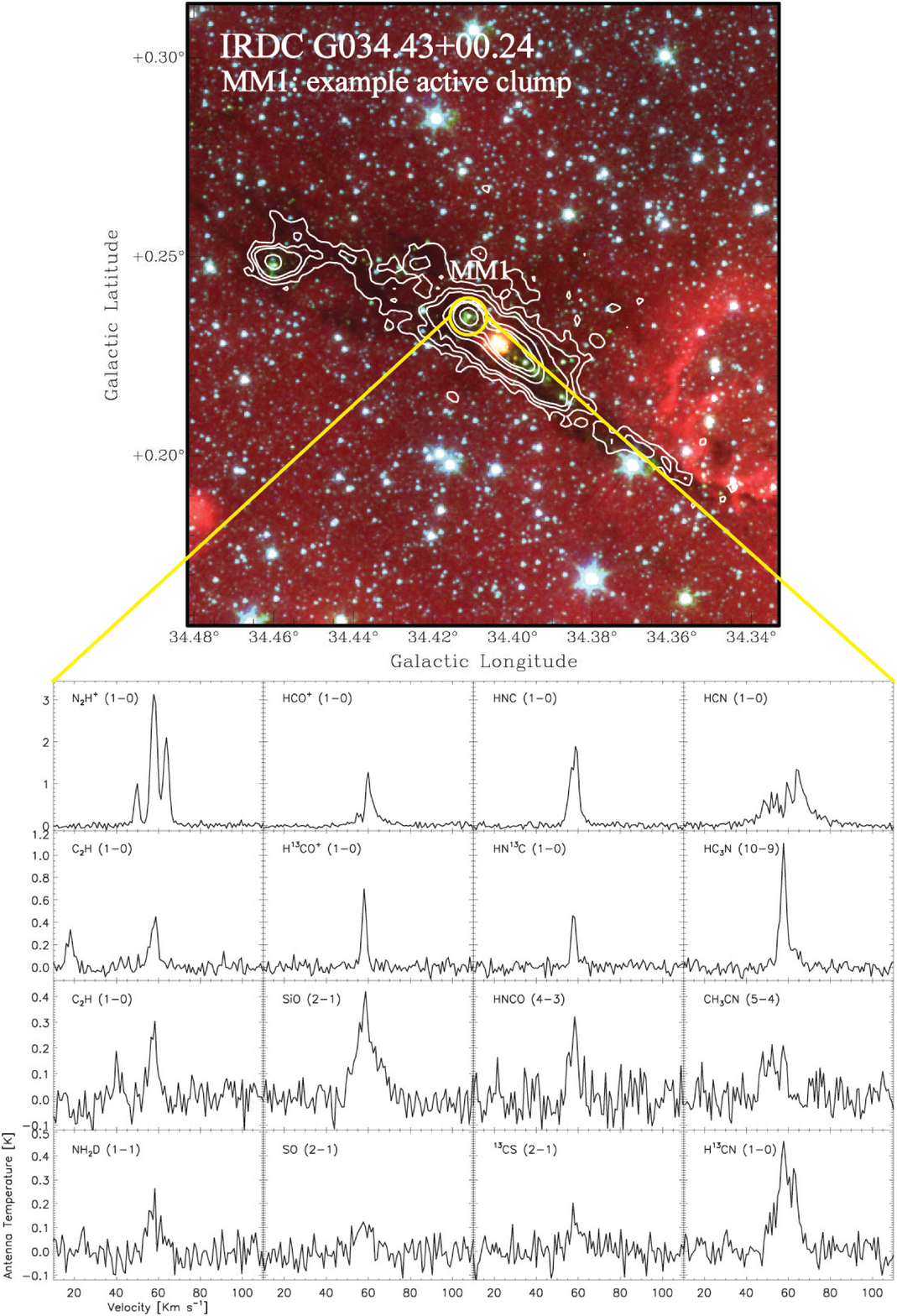}
\end{center}
\caption{ {\bf Active clump example.}  {\it Top:} IRAC 3-color 
(3.6 $\mu$m in blue, 4.5 $\mu$m in green and 8.0 $\mu$m in red) 
   image of the IRDC 
G034.43+00.24 overlaid with IRAM 1.2 mm continuum emissiom from
 \cite{Rathborne06}. {\it Bottom:} Molecular spectra of the active MM1 clump.
Active clumps are associated with enhanced 4.5 $\mu$m emission, the 
so-called green fuzzies, and an embedded 24 $\mu$m source. 
 The circle shows the telescope beam size and marks the position
 of the observed clump.}
\label{fig-active}
\end{figure*}

\begin{figure*}
\begin{center}
\includegraphics[angle=0,scale=0.36]{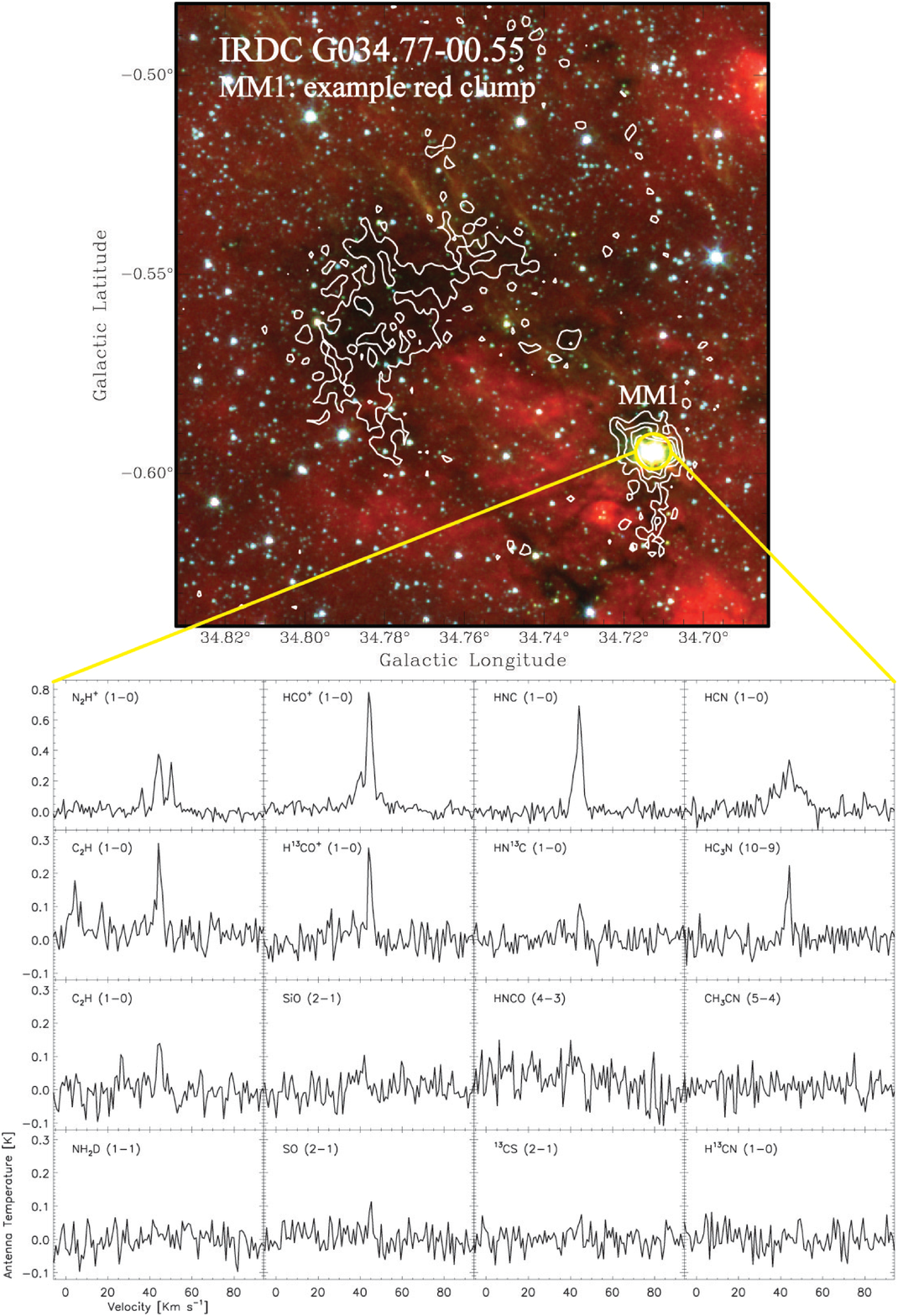}
\end{center}
\caption{ {\bf Red clump example.}  {\it Top:} IRAC 3-color 
(3.6 $\mu$m in blue, 4.5 $\mu$m in green and 8.0 $\mu$m in red) 
   image of the
 IRDC G034.77-00.55 overlaid with IRAM 1.2 mm continuum emissiom from
 \cite{Rathborne06}. {\it Bottom:} Molecular spectra of the red MM1 clump. 
Red clumps are associated with bright 8 $\mu$m emission. 
 The circle shows the telescope beam size and marks the position
 of the observed clump.}
\label{fig-red}
\end{figure*}

\begin{figure}[!h]
\begin{center}
\includegraphics[angle=0,scale=0.5]{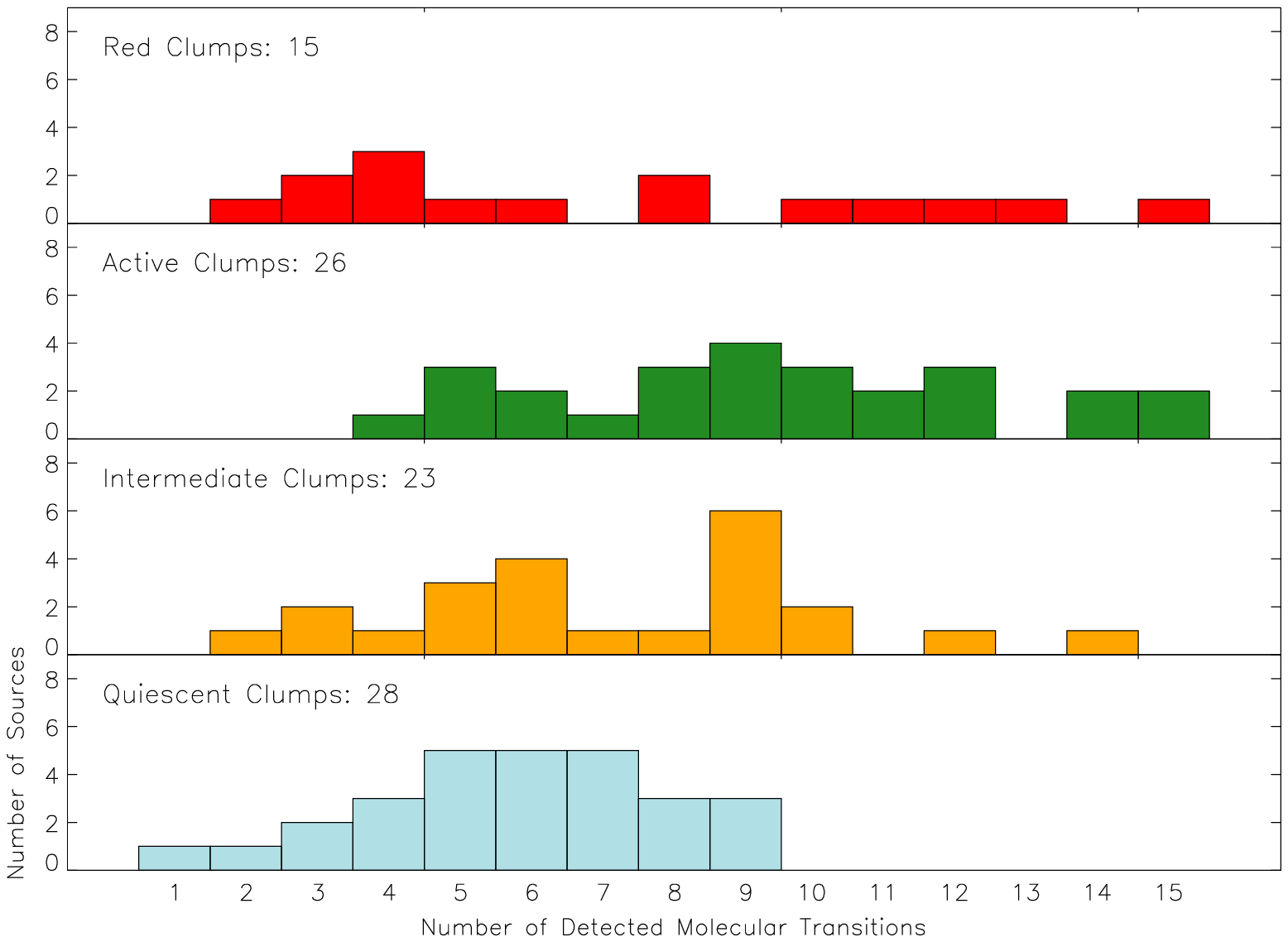}
\end{center}
\caption{Histograms with the number of detected molecular transitions per  
source for each evolutionary stage. The name of the corresponding
 evolutionary stage and the number of sources contained in each histogram are 
given on the top left side of each panel. We find, as expected, that 
more evolved clumps show a rich spectrum with numerous transition lines, 
except that the most evolved/red clumps show a population with few 
detections.}
\label{number_lines}
\end{figure}

Figures~\ref{fig-quiescent},~\ref{fig-intermediate},~\ref{fig-active}, 
and~\ref{fig-red} show representative examples of spectra for 
one clump of each proposed evolutionary stage: the quiescent G028.53 MM3,
the intermediate G025.04 MM4, the active G034.43 MM1, and the red G034.77 MM1, 
respectively. The {\it Spitzer}/IRAC image of the 
corresponding host IRDC is also displayed in each figure. 
The active clump G034.43 MM1 (Figure~\ref{fig-active}) exhibits more
 molecular line emission than the quiescent clump G028.53 MM3 
(Figure~\ref{fig-quiescent}). This pattern is typical through the sample. 
It is expected that when star-forming clumps evolve to later stages,   
they should display a rich molecular spectrum with numerous lines 
because the proto-stars heat their environment releasing more complex 
molecules from dust mantles into the gas phase;  
in contrast, in earlier stages clumps should show a 
 sparse spectrum with a smaller number of detectable molecular lines 
because many molecules are frozen out onto grain surfaces. 
However, one must be aware that although the analysis of detection 
rates may reflect the chemical composition of clumps to some extent, 
it does not give unambiguous conclusions because detection rates 
may just show the dependence on H$_2$ column density. A more  
definitive conclusion must be drawn from molecular abundances (see 
Sections~\ref{chemistry}~and~\ref{molecules}). Figure~\ref{number_lines}
 presents histograms which compare the number of detected molecular
 transitions for each group. More molecular transitions are detected 
toward clumps that show signs of star formation. 
The median number of detections for quiescent clumps is 6 molecular 
transitions. On the other hand, intermediate and active clumps have a
 median of 7 and 9 detected molecular transitions, respectively.
To quantify the 
differences between quiescent and active clumps, we have calculated the 
probability that these two distributions in Figure~\ref{number_lines} 
arise from the same underlying 
population using the Kolmogorov--Smirnov (K--S) test. The K--S test is 
a more robust method for measuring the similarity between two distributions  
 than the comparison of the median and/or mean values. Through this paper, 
we will mention that the difference between two distributions is 
statistically significant when the K-S test gives a percentage lower 
than 5\%. The 
probability that the quiescent and active clumps are derived from the 
same parent population is 0.1\%.

\begin{deluxetable*}{llcccc}
\tabletypesize{\scriptsize}
\tablecaption{N$_2$H$^+$ $JF_1F=123\ra012$ Gaussian Parameters. \label{tbl-gaussian_123_012}}
\tablewidth{0pt}
\tablehead{
\colhead{IRDC} &  \colhead{Clump}  & \colhead{$T_{\rm rms}$} & \colhead{$T_{\rm A}$} & \colhead{$V_{lsr}$} & \colhead{$\Delta$V}\\
\colhead{Name} &  \colhead{Number} & \colhead{(K)}       & \colhead{(K)}        & \colhead{(\kms)}   & \colhead{(\kms)}   \\
}
\startdata
G015.05+00.07 &  MM1  & 0.04 & 0.38 $\pm$ 0.03 &  25.21 $\pm$ 0.10 & 3.18 $\pm$ 0.31 \\
G015.05+00.07 &  MM2  & 0.04 & 0.19 $\pm$ 0.03 &  29.87 $\pm$ 0.20 & 3.64 $\pm$ 0.62 \\
G015.05+00.07 &  MM3  & 0.04 & 0.18 $\pm$ 0.03 &  28.41 $\pm$ 0.21 & 2.50 $\pm$ 0.54 \\
G015.05+00.07 &  MM4  & 0.04 & 0.44 $\pm$ 0.03 &  27.95 $\pm$ 0.08 & 2.37 $\pm$ 0.21 \\
G015.05+00.07 &  MM5  & 0.05 & 0.25 $\pm$ 0.05 &  29.93 $\pm$ 0.16 & 2.03 $\pm$ 0.46 \\
\enddata
\tablecomments{This table is available in its entirety in a machine-readable
 form in the online journal. A portion is shown here for guidance regarding
 its form and content.}
\end{deluxetable*}

 The median for red clumps is 6 detections. 
Red clumps (which show bright 8 $\mu$m emission) likely correspond to 
embedded H\,{\sc ii} regions, a more evolved stage than active clumps.
 Despite this, they have the same 
median value as quiescent clumps. In the histogram, red clumps seem to have a
bimodal distribution with a first population of 8 clumps showing a 
low number of detected molecular transitions (left side of the histogram) 
and a second one of 7 clumps presenting a higher number (right side of the 
histogram). Moreover, we note that 
the first population only has one maser detection, and the second 
population has five \citep{Chambers09}. These two different populations for 
red clumps may be the result of the evolutionary state of 
the H\,{\sc ii} region. 
We expect to detect more molecular lines from an early H\,{\sc ii} region 
compared to a late H\,{\sc ii} region because, since the early H\,{\sc ii}
 region 
has not had as much time to ionize the surrounding gas, it should have 
more molecular gas near the central heating star(s). 
 The molecular spectrum of a late stage H\,{\sc ii} region,
 where much of the gas has been ionized, should be more sparse.  

\begin{figure}
\begin{center}
\includegraphics[angle=0,scale=0.67]{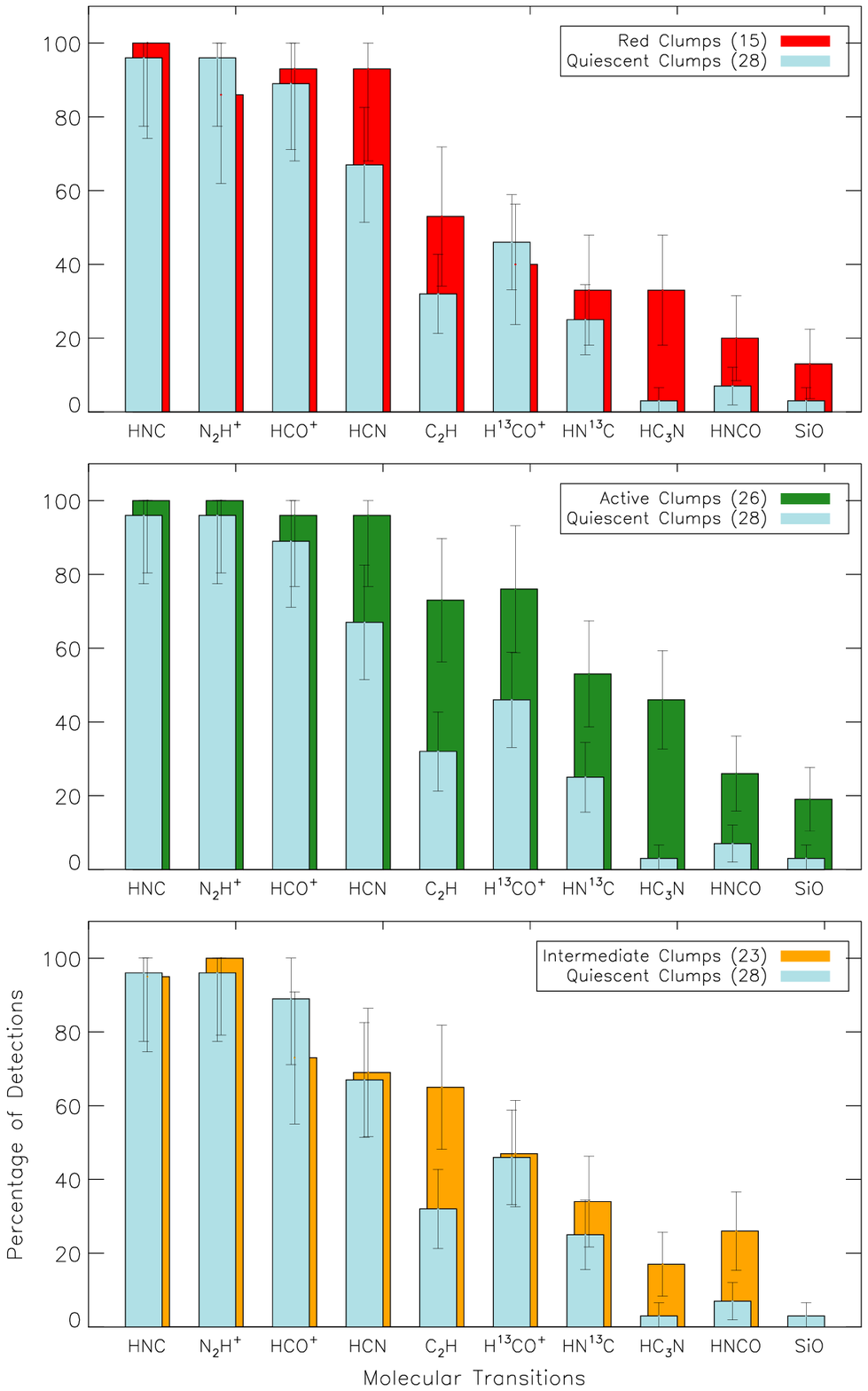}
\end{center}
\caption{ Detection rates of the observed molecular lines toward 
  red, active, and intermediate clumps compared to detections
 rates percentages in quiescent clumps. The N$_2$H$^+$ and
 C$_2$H detection rates correspond to the $JF_1F=123-012$ and  
  $NJF=1\,\frac{3}{2}\,2\ra0\,\frac{1}{2}\,1$transitions, respectively. 
The evolutionary 
sequence name and the total number of sources are given on the top right 
corner of each panel. HNC and N$_2$H$^+$ lines are detected in almost
 every IRDC clumps at every evolutionary stage. On the other hand, HC$_3$N,
 HNCO, and SiO lines are predominantly detected in later stages of
 evolution. Error bars are the Poisson noise (i.e., the root square of
 the number of detections). }
\label{detection_rates_per_category}
\end{figure}

The active clump G034.43 MM1 (Figure~\ref{fig-active}) exhibits 
a spectrum of a ``hot core.'' It shows emission in the HNCO, HC$_3$N, and 
CH$_3$CN lines, which require high densities and temperatures for excitation
 ($n_{\rm crit} \geq 5 \times 10^5$ cm$^{-3}$ and 
$E_u/k > 10$ K), and depend on the release of their 
parent molecules from dust grains to be formed. In addition, it shows 
SiO emission, which is 
usually associated with molecular outflows since SiO  
abundance is highly enhanced by shocks \citep[e.g.,][]{Schilke97,Caselli97}.
 In contrast, the quiescent clump 
G028.53 MM3 (Figure~\ref{fig-quiescent}) presents no emission of 
HNCO, HC$_3$N, CH$_3$CN, and SiO over the $3\sigma$ limit.  
The intermediate clump 
G025.04 MM4 (Figure~\ref{fig-intermediate}) seems to be in 
an intermediate chemical state, 
only exhibiting one high-excitation line (HNCO). To carry out a more 
comprehensive study of the full sample of IRDC clumps, we have split the 
histogram of detection rates (Figure~\ref{detection_rates}) into four 
 new histograms. In Figure~\ref{detection_rates_per_category}  
 the detection rates for each proposed evolutionary sequence are presented.   
The detection rates for the quiescent clumps are displayed in every panel 
for comparison. Quiescent, intermediate, active, and red clumps have
28, 23, 26, and 15 members, respectively. 
 From the histograms in Figure~\ref{detection_rates_per_category}, it is 
seen that the HNC $J=1\ra0$, N$_2$H$^+$ $JF_1F=123\ra012$, and  
HCO$^+$ $J=1\ra0$ 
  lines are present in almost every IRDC clump at any evolutionary stage.
Red and active clumps have a high detection rate ($\sim$95\%) for the  
HCN $J=1\ra0$ line, comparable with the three lines previously mentioned, 
while in less evolved clumps this value is lower ($\sim$69\%). There is a 
high enhancement in the detection of the brightest C$_2$H line, 
transition $NJF=1\,\frac{3}{2}\,2\ra0\,\frac{1}{2}\,1$, toward active 
and intermediate clumps with respect to quiescent clumps. The percentages of 
detection for the H$^{13}$CO$^+$ $J=1\ra0$ line are 
roughly the same for red, intermediate, and quiescent clumps; however, 
the value for active clumps increases by $\sim$30\% with respect to the 
other three evolutionary stages. The detection rates of 
the HN$^{13}$C $J=1\ra0$ 
line show no significant difference between red and intermediate clumps with 
respect to quiescent clumps; on the other hand, the value for active clumps 
increases by $\sim$30\%. The most remarkable difference between detection
 rates for the different evolutionary stages is found for the 
HC$_3$N $J=10\ra9$ line. The detection of this line is significantly high  
for all stages with IR signs of star formation. HC$_3$N is only detected 
in one quiescent source, and the difference between the detection rate in  
active and quiescent clumps is $\sim$40\%. While the HNCO 
$J_{K_a,K_b}=4_{0,4}\ra3_{0,3}$ line is only detected twice (7\%) in quiescent 
clumps, active and intermediate clumps have a HNCO 
$J_{K_a,K_b}=4_{0,4}\ra3_{0,3}$ detection rate of $\sim$27\%, 
indicating that this line is also observed more often toward clumps with 
current star formation. In the quiescent, intermediate, and red clumps, 
SiO $J=2-1$ emission was rarely detected. In active clumps, the detection 
rate of SiO reaches the 19\%.  

\subsection{Velocity Widths}

\begin{figure}
\begin{center}
\includegraphics[angle=0,scale=0.5]{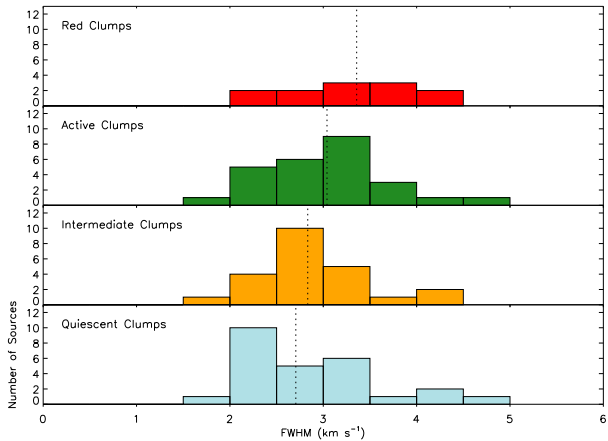}
\end{center}
\caption{Histograms of the number distributions of N$_2$H$^+$ linewidths
 ($\Delta$V) for each evolutionary sequence. The name of the evolutionary
 stage is given on the top left corner of each panel. The vertical 
dashed lines indicate the median values of the $\Delta$V for each distribution. 
Although the K-S test shows that active and quiescent clumps probably
 originate from indistinguishable parent populations, the K-S test shows that
 red and quiescent distributions are different. The median 
 values of the $\Delta$V increase with the evolution of the clumps. }
\label{FWHM-histo}
\end{figure}

The widths of molecular lines give information about the turbulence of
 clumps. Although we cannot rule out that clumps have organized velocity 
structures, such as rotation, in this paper we will assume that 
turbulence dominates the widths of the lines. 
 To investigate if turbulence is different in clumps with and without 
signs of star formation, we made histograms of the 
number distributions of full width at half maximum line widths ($\Delta$V) 
for the four evolutionary stages using  
the three most commonly detected lines: N$_2$H$^+$, HCO$^+$, and HNC. 
In Figure~\ref{FWHM-histo}, the number distributions of 
N$_2$H$^+$ $\Delta$V for each 
evolutionary sequence is presented. The range of line widths varies 
between 1.6 and 4.6 \kms\ for the full sample of IRDC clumps. 
The median values for quiescent, intermediate, active, and red clumps 
 are 2.7, 2.8, 3.0, and 3.4 \kms, respectively. Using the K--S test, we  
calculated a probability of 67\% that active and quiescent, and 6\% that  
red and quiescent distributions originate from the same parent populations.  
Although the K--S test shows that there is little difference between the 
active and quiescent clumps, it seems that there is an increase in 
$\Delta$V from clumps with no apparent star formation to active and 
red clumps. We think this trend is less clearly shown by the K--S test 
for active and quiescent distributions because of the small number of 
members of each group; in spite of this, the K--S test shows a 
difference between the red and quiescent populations because the shape of 
the number distributions are more dissimilar.
We conclude that the slightly broader widths found 
in the N$_2$H$^+$ line toward red and active clumps, compared with 
quiescent and intermediate clumps, is produced by the 
rise in the turbulence from the ongoing star formation activity. 

The number distributions of HCO$^+$ and HNC $\Delta$V show no clear trend, and 
their histograms are not presented. The range of line widths varies 
between 2.0 and 13.0 \kms\ for HCO$^+$, and between 1.0 and 
7.6 \kms\ for HNC. The $\Delta$V 
median values of HCO$^+$ for quiescent, intermediate, active, and red 
 clumps are 3.8, 4.4, 3.6, and 4.0 \kms; and for HNC are 
3.2, 3.8, 3.4, and 2.8 \kms, respectively. We identify three reasons 
that may explain why HCO$^+$ and HNC do not present a similar trend to 
that one shown by  N$_2$H$^+$. First, the HCO$^+$ and HNC lines  
 are optically thick (see Section~\ref{opacity}) and consequently, 
the line widths can be broadened by opacity 
\citep[e.g.,][]{Beltran05}. Second, the HCO$^+$ line can show broadening 
and wing emission produced by molecular outflow activity. Certainly, 
this is what is happening in G018.82 MM3, which has a $\Delta$V of 13.0 \kms\ 
in the HCO$^+$ line.  
Finally, because the Mopra beam is $\sim$38\arcsec\ (0.8 pc at the average 
 distance of 4.3 kpc), we are observing the bulk 
motions of the gas and some level of clumping may also be present.  
This also could explain why the trend of broader N$_2$H$^+$ lines toward 
more evolved objects is not as clear as we expect. However, since N$_2$H$^+$
 is thought to be a higher density tracer 
\citep[e.g.,][]{Caselli02,Pirogov03}, its emission should emanate  
mostly from the center of the clumps and thus we still can see a weak trend. 

\begin{deluxetable*}{lclllll}
\tabletypesize{\scriptsize}
\tablecaption{Detection Rates \label{tbl-detection-rates}}
\tablewidth{0pt}
\tablehead{
\colhead{} & \colhead{} & \colhead{Quiescent} &\colhead{Intermediate} &\colhead{Active} &  \colhead{Red}  &    \colhead{All} \\
\colhead{Molecule} & \colhead{Transition} & \colhead{Clumps}  & \colhead{Clumps}
 & \colhead{Clumps} & \colhead{Clumps}  & \colhead{Clumps} \\
}
\startdata
\HNC      & \tHNC        &  27 $\pm$ 5 (96\%)  &  22 $\pm$ 5 (96\%) & 26 $\pm$ 5 (100\%)  & 15 $\pm$ 4 (100\%) &   90 $\pm$ 10 (98\%)\\
 \NdosH    & \tNdosHuud    &  22 $\pm$ 5 (79\%)   & 22 $\pm$ 5 (96\%)  & 25 $\pm$ 5 (96\%)   & 9 $\pm$ 3 (60\%) &  78 $\pm$ 9 (85\%)\\
           & \tNdosHudt  &  27 $\pm$ 5 (96\%)   &23 $\pm$ 5 (100\%)  & 26 $\pm$ 5 (100\%)   & 13 $\pm$ 4 (87\%) &  89 $\pm$ 9 (97\%)\\
           & \tNdosHucu&  8 $\pm$ 3 (29\%)   &  12 $\pm$ 4 (52\%)  &  17 $\pm$ 4 (65\%)   &  5 $\pm$ 2 (33\%) &  42 $\pm$ 7 (46\%)\\
 \HCO      & \tHCO   &25 $\pm$ 5 (89\%)   &17 $\pm$ 4 (74\%)  & 25$\pm$ 5 (96\%)   &   14 $\pm$ 4 (93\%)   &     81 $\pm$ 9 (88\%)\\
\HCN      & \tHCO    &19 $\pm$ 4 (68\%)   &16 $\pm$ 4 (70\%)  &   25$\pm$ 5 (96\%)   &14 $\pm$ 4 (93\%)&    74 $\pm$ 9 (80\%)\\
\CdosH    & \tCdosHudcu & 9 $\pm$ 3 (32\%)   &15 $\pm$ 4 (65\%) & 19 $\pm$ 4 (73\%)   & 8 $\pm$ 3 (53\%)   &    51 $\pm$ 7 (55\%)\\
          & \tCdosHuucc & 1 $\pm$ 1  (4\%)  &3 $\pm$ 2 (13\%)  & 10 $\pm$ 3 (39\%)   &  4 $\pm$ 2 (27\%) &    18 $\pm$ 4 (20\%)\\
          & \tCdosHuucu & 1 $\pm$ 1 (4\%)   & 6 $\pm$ 3 (26\%)  & 9 $\pm$ 3 (35\%)   &3 $\pm$ 2 (20\%)  &    19 $\pm$ 4 (21\%)\\
          & \tCdosHuccu &  1 $\pm$ 1 (4\%)   & 1 $\pm$ 1 (4\%)  &   4 $\pm$ 2 (15\%)   &2 $\pm$ 1 (13\%)  &   8 $\pm$ 3 (9\%)\\
\HtreceCO & \tHtreceCO &13 $\pm$ 4 (46\%)   &11 $\pm$ 3 (48\%)  &   20 $\pm$ 5 (77\%)   &6 $\pm$ 3 (40\%)   &    50 $\pm$ 7 (54\%)\\
\HNtreceC & \tHNtreceC & 7 $\pm$ 3 (25\%)   &8 $\pm$ 3 (35\%)  &  14 $\pm$ 4 (54\%)   &5 $\pm$ 2 (33\%)   &    34 $\pm$ 6 (37\%)\\
\HCtresN  & \tHCtresN  & 1 $\pm$ 1 (4\%)   &4 $\pm$ 2 (17\%)  &  12 $\pm$ 4 (46\%)   & 5 $\pm$ 2 (33\%)  &   22 $\pm$ 5 (24\%)\\
\HNCO     & \tHNCO  & 2 $\pm$ 1 (7\%)   &6 $\pm$ 3 (26\%)  & 7 $\pm$ 3 (27\%)   & 3 $\pm$ 2 (20\%)   &    18 $\pm$ 4 (20\%)\\
\SiO      & \tSiO   & 1 $\pm$ 1 (4\%)   & 0 $\pm$ 0 (0\%)  & 5 $\pm$ 2 (19\%)   &2 $\pm$ 1 (13\%)  &    8 $\pm$ 3 (9\%)\\
\HtreceCN & \tHCO & 0 $\pm$ 0 (0\%) &  1 $\pm$ 1 (4\%)  &  3 $\pm$ 2 (12\%)  &2 $\pm$ 1 (13\%)   &    6 $\pm$ 2 (7\%) \\
\NHdosD & \tNHdosD & 0 $\pm$ 0 (0\%)  &  0 $\pm$ 0 (0\%)  & 3 $\pm$ 2 (12\%)  &     0 $\pm$ 0 (0\%)  &  3 $\pm$ 2 (3\%) \\
\SO   & \tSO  & 0 $\pm$ 0 (0\%)  & 0 $\pm$ 0 (0\%)  & 0 $\pm$ 0 (0\%)  &     1 $\pm$ 1 (7\%)  &  1 $\pm$ 1 (1\%) \\
\CS   & \tCS & 0 $\pm$ 0 (0\%)  & 0 $\pm$ 0 (0\%)  &  2 $\pm$ 1 (8\%)  &  0 $\pm$ 0 (0\%)  &   2 $\pm$ 1 (2\%) \\
\CHtresCN&$J=5-4$& 0 $\pm$ 0 (0\%)  & 0 $\pm$ 0 (0\%)  & 1 $\pm$ 1 (4\%)  &  0 $\pm$ 0 (0\%)  &   1 $\pm$ 1 (1\%) \\
\enddata
\end{deluxetable*}

Variations in N$_2$H$^+$ line widths for different evolutionary 
stages toward IRDC clumps have also been observed previously 
by \cite{Sakai08} and recently by \cite{Vasyunina11}.  
\cite{Sakai08} observed 22 {\it MSX} dark 
objects and 7 {\it MSX} sources located at distances less than 4.5 kpc. 
{\it MSX} dark objects are defined by \cite{Sakai08} as 
objects with 8 $\mu$m extinction features, although some of them have 
{\it Spitzer} 24 $\mu$m pointlike sources; thus, {\it MSX} dark objects 
correspond to our intermediate and quiescent clumps. The {\it MSX} sources 
correspond to our red clumps. \cite{Sakai08} found that the $\Delta$V of 
 N$_2$H$^+$ lines range between 0.6 and 4.0 \kms, and the mean values for 
{\it MSX} dark objects and {\it MSX} sources are 2.3 and 2.5 \kms,
 respectively. 
\cite{Vasyunina11} observed 13 ``quiescent'' clumps 
 (as defined by us), 11 ``middle'' clumps 
(which correspond to our intermediate and active clumps), and 13 ``active'' 
clumps (which correspond to our red clumps). They found that the range of 
$\Delta$V of N$_2$H$^+$ lines varies between 0.6 and 2.8 \kms;
 and ``quiescent,'' 
 ``middle,'' and ``active'' clumps have $\Delta$V mean values of 1.4, 1.7, and 
2.2 \kms, respectively. If the assumption that turbulence increases with 
  star formation activity producing broader molecular line widths in 
more evolved stages is correct, then the results on N$_2$H$^+$ line widths
 found by \cite{Sakai08}, \cite{Vasyunina11}, and this work support the
 evolutionary sequence proposed by \cite{Chambers09} for IRDC clumps.

\begin{figure*}
\begin{center}
\includegraphics[angle=0,scale=1.05]{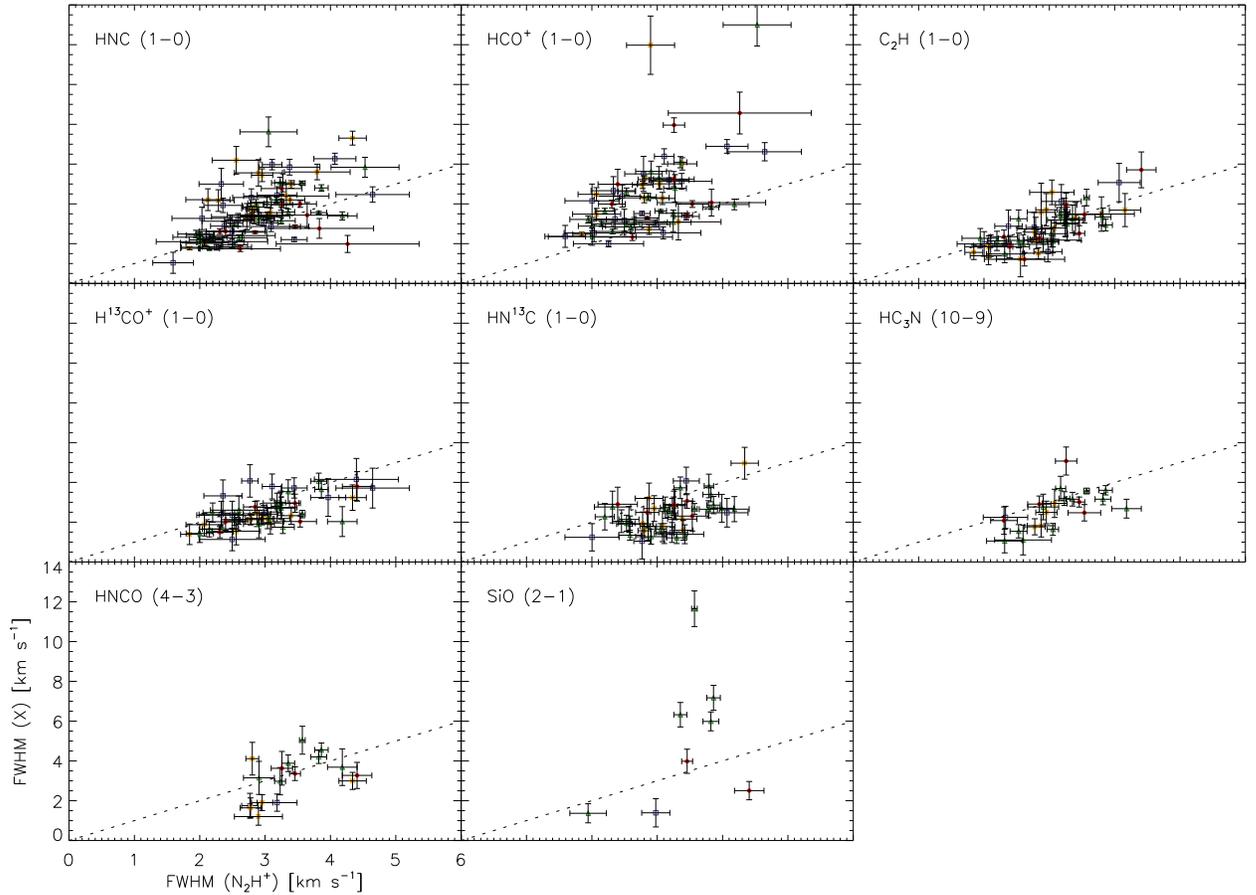}
\end{center}
\caption{Correlation plots of the N$_2$H$^+$ velocity widths against the 
velocity widths of several molecular lines. The name of the molecular lines
 are given on the top left corner of each panel. The dashed line indicates 
unity. Blue squares correspond quiescent clumps, yellow asterisks to
 intermediate clumps, green triangles to active clumps, and red circles to 
red clumps. Linewidths of HNC, HCO$^+$ and SiO show a high degree of
 scatter. C$_2$H has the best correlation with N$_2$H$^+$. H$^{13}$CO$^+$,
 HN$^{13}$C and HC$_3$N have slightly narrower line widths than N$_2$H$^+$. }
\label{FWHM-plot}
\end{figure*}

Figure~\ref{FWHM-plot} shows the $\Delta$V of the most frequently detected 
molecular lines against the widths of N$_2$H$^+$. The linewidths for 
HNC, HCO$^+$, and SiO lines show a high degree of scatter. In the case 
of HNC and HCO$^+$, this scatter is 
likely produced by the high line opacities (see Section~\ref{opacity}) and 
because these lines may also be tracing more external layers of the 
clumps. In the case of SiO, and also HCO$^+$, the scatter is probably produced 
by molecular outflow activity, which increases the $\Delta$V of the lines and 
adds extended wing emission. C$_2$H appears to present the best 
correlation with N$_2$H$^+$. H$^{13}$CO$^+$, HN$^{13}$C, and HC$_3$N have 
slightly narrower line widths than N$_2$H$^+$, indicating that they 
trace similar optically thin gas emanating from the center of the clumps.

From Figure~\ref{FWHM-plot}, it can be seen that there are four sources with 
very large linewidths in the HCO$^+$ ($>$8.0 \kms) and SiO ($>$6.0 \kms) 
lines. They are G018.82 MM3, G018.82 MM4, G035.59 MM1, and G033.69 MM2  
for HCO$^+$ and G018.82 MM3, G019.27 MM2, G022.35 MM1, and G028.04 MM1  
for SiO. Notably, the four sources that have large SiO linewidths are
 active clumps. \cite{Sakai10} mention that larger SiO linewidths and
 abundances are expected in younger sources than MSX sources (our red clumps).  
However, as they also mention, strong shocks are necessary to produce SiO in 
the gas phase; therefore, some star formation activity is needed and the 
sources should be older than quiescent clumps.  

\subsection{Kinematic Distances}
\label{kinematic}
Because the N$_2$H$^+$ $JF_1F=123\ra012$ line has a detection rate of 97\% 
and shows no self-absorbed profiles, we used the N$_2$H$^+$ systemic 
velocities of the 155 detected clumps from the initial sample to estimate 
kinematic distances. The distances were obtained 
using the \cite{Clemens85} rotation curve of the Milky Way, scaled to 
$(R_0,V_o)=(8.5~\rm{kpc},220~\rm{\kms})$ and included a small velocity 
correction (7 \kms) accounting for the measured solar peculiar motion.   
The distances of all clumps are listed in Table~\ref{tbl-coor}. 
To resolve the ambiguity between the near and far kinematic distances, 
we used the fact that IRDCs are seen as dark extinction features against 
the bright mid--infrared background emission. Hence, the ambiguity is 
resolved by assuming that all IRDCs are located at the near kinematic 
distance \cite[e.g.,][]{Simon06b,Jackson08}. The distances to the IRDC clumps 
in this work were previously established by \cite{Simon06b}. They carried out
 a morphological matching of the {\it MSX} mid--infrared extinction with 
 the $^{13}$CO $J=1\ra0$ molecular line emission from the Boston 
University-Five College Radio Astronomy Observatory (BU-FCRAO) Galactic 
Ring Survey \citep[GRS;][]{Jackson06}. However, 
because the $^{13}$CO $J=1\ra0$ emission is a low density tracer 
($n_{\rm crit}=2\times$10$^3$ cm$^{-3}$), there are typically 
 several velocity components along one line of sight. Of the initial
 155 detected IRDC clumps, the kinematic distance indicated by the high
 density tracer N$_2$H$^+$ ($n_{\rm crit}=3\times$10$^5$ cm$^{-3}$) of 
 35 (23\%) clumps differs from the distance inferred from
 $^{13}$CO by \cite{Simon06b}. 
 Of these sources, 17 show strong signatures of star formation and are
 classified as red and active clumps. Because they are isolated clumps
 located apart from the IRDC and at a different distance, they are not
 physically associated with the IRDC and were excluded from further analysis
  in this paper.  
\clearpage
\section{Discussion}

\subsection{Derivation of Physical Parameters}

Assuming local thermodynamic equilibrium (LTE) conditions, the total
 column density, $N$, can be derived from \cite[e.g.,][]{Garden91}
\begin{equation}
N=\frac{8\pi \nu^3}{c^3R}\frac{Q_{\rm rot}}{g_{\rm u} A_{\rm ul}}\frac{\exp(E_{_{l}}/kT_{\rm ex})}{[1-\exp(-h\nu/kT_{\rm ex})]}\int \tau \,dv~,  
\label{eqn-den-colum-2}
\end{equation} 
where $\tau_{\nu}$ is the optical depth of the line, $g_{\rm u}$ is the
 statistical weight of the upper level, $A_{\rm ul}$ 
is the Einstein coefficient for spontaneous emission, $E_{_{l}}$ is the 
energy of the lower state, $Q_{\rm rot}$ is the partition function, 
$\nu$ is the transition frequency, and $R$ is
 the relative intensity of the brightest hyperfine transition with respect
 to the others. $R$ is only relevant for hyperfine transitions because
 it takes into account the satellite lines correcting by their
 relative opacities. It is 5/9 for N$_2$H$^+$ and 5/12 for C$_2$H.
 It is equal to 1.0 for transitions without hyperfine structure. 

A particular case of equation~\ref{eqn-den-colum-2} is the total
 column density of a linear, rigid rotor molecule, which is given
 by \cite[e.g.,][]{Garden91}
{\small
\begin{equation}
N=\frac{3k}{8\pi^3B\mu^2R}\frac{(T_{\rm ex}+hB/3k)}{(J+1)}\frac{\exp(E_{_{J}}/kT_{\rm ex})}{[1-\exp(-h\nu/kT_{\rm ex})]}\int\tau_{\nu}\,dv~,
\label{eqn-den-colum}
\end{equation}}
where $\mu$ is the permanent dipole moment of the molecule, $J$ is
 the rotational quantum number of the lower state, $E_J=hBJ(J+1)$ is
 the energy in the level $J$, and $B$ is the rotational constant of the
 molecule. 

To calculate column densities it is necessary to
 estimate the optical depths of the lines. We calculate 
the optical depths of N$_2$H$^+$ and C$_2$H using their hyperfine structure,  
and the optical depths of HCO$^+$ and HNC using their isotopologues, 
H$^{13}$CO$^+$ and HN$^{13}$C. Under the assumption of LTE, all levels are 
populated according to the same excitation temperature ($T_{\rm ex}$).   
We assume that $T_{\rm ex}$ is equal to the dust temperature ($T_{\rm D}$) 
obtained by \cite{Rathborne10}. They determined $T_{\rm D}$ 
for 59 sources and gave lower and upper limits for 23 other sources 
in our sample of 92 IRDC clumps by fitting a graybody function to the spectral 
energy distribution (SED).
 For the sources with lower and upper limits, the average value of the 
limits was used as $T_{\rm D}$. For the remaining 10 sources, without
 SEDs, we use the median 
values found by \cite{Rathborne10} for each evolutionary sequence: 23.7, 
30.4, 34.5, and 40.4 K for quiescent, intermediate, active, and red clumps, 
respectively. When the weakest line needed to compute the optical depth was  
not detected above the 3$\sigma$ level (N$_2$H$^+$ $JF_1F=112\ra012$, 
C$_2$H $NJF=1\,\frac{3}{2}\,1\ra0\,\frac{1}{2}\,0$, H$^{13}$CO$^+$, or
 HN$^{13}$C), the intensity of the line was assumed to be 3$T_{\rm rms}$.
 This gives an upper limit of the line intensity and thus an upper limit
 in the optical depth. 

Actual optical depths were determined as follows. The main beam brightness
 temperature ($T_{\rm mb}$) is the antenna 
temperature ($T_A$) corrected by the main bean efficiency ($\eta_{\rm mb}$), 
$T_{\rm mb}=T_{A}/\eta_{\rm mb}$.  
The main beam brightness temperature of a line is related to the excitation 
temperature and the optical depth by 
\begin{equation}
T_{\rm mb} = f[J(T_{\rm ex}) - J(T_{\rm bg})](1 - e^{-\tau_{\nu}})~,
\label{Tmb-eq}
\end{equation}
where $f$ is the filling factor, $\tau_{\nu}$ is the optical depth of 
the line, $T_{\rm bg}$ is the background temperature, and $J(T)$  
is defined by
\begin{equation}
J(T) = \frac{h\nu}{k}\frac{1}{e^{h\nu/kT} - 1}~. 
\end{equation}  

Taking the ratio of Equation~\ref{Tmb-eq} evaluated for two hyperfine 
components or two isotopologues and assuming that the filling
 factors are the same, we obtain 
\begin{equation}
\frac{1-e^{-\tau_{_{1}}}}{1-e^{-\tau_{_{2}}}}=\frac{T_{\rm mb_{1}}}{T_{\rm mb_{2}}} \left[ \frac{J_{_{2}}(T_{\rm ex}) - J_{_{2}}(T_{\rm bg})}{J_{_{1}}(T_{\rm ex}) - J_{_{1}}(T_{\rm bg})}\right]~,
\label{tau-eqn2}
\end{equation}
where the subscripts ``$1$'' and ``$2$'' refer to two different   
hyperfine components or isotopologue molecules. Defining ``$r$'' as the
 ratio between the optical depths, $r=\tau_{2}/\tau_{1}$, and utilizing
 the fact that the term between brackets is $\sim$1,
 Equation~\ref{tau-eqn2} becomes
\begin{equation}  
\frac{1-e^{-\tau_{_{2}}/r}}{1-e^{-\tau_{_{2}}}}=\frac{T_{\rm mb_{1}}}{T_{\rm mb_{2}}}~.
\label{tau-eqn}
\end{equation}

Solving this Equation numerically, we obtain the optical depth for one  
of the components, and by using the relationship between opacities we can 
obtain the optical depths of the second component. 
 
\subsubsection{Optical Depths}
\label{opacity}

\vspace{0.2cm}
\hspace{3.5cm}{\it N$_2$H$^+$}
\vspace{0.2cm}

The opacities of the N$_2$H$^+$ hyperfine components are obtained from 
the ratio between the observed main beam brightness temperatures of 
the two brightest transitions: $JF_1F=123\ra012$ and $JF_1F=112\ra012$.
Rewriting Equation~\ref{tau-eqn}, we obtain 
\begin{equation}  
\frac{1-e^{-\frac{3}{5}\tau_{_{123-012}}}}{1-e^{-\tau_{_{123-012}}}}=\frac{T_{\rm mb_{112-012}}}{T_{\rm mb_{123-012}}}~,
\label{N2H+_tau-eqn}
\end{equation}
where the subscripts ``$123-012$'' and ``$112-012$'' refer to the N$_2$H$^+$ 
hyperfine components $JF_1F=123\ra012$ and $JF_1F=112\ra012$, respectively. 
We have also used the opacity ratio between these two hyperfine lines, 
$r$=$\frac{\tau_{_{123-012}}}{\tau_{_{112-012}}}$=$\frac{5}{3}$, which depends
 only on the transition moments. 

The N$_2$H$^+$ opacities were calculated for 63 IRDC clumps and the values 
for the main component are listed in Table~\ref{tbl-tau-column}. The
 main component of N$_2$H$^+$ has opacities less than 1.0 in roughly 
60\% of the sample. The other 
40\% of the sample is moderately optically thick, with a maximum
 opacity of 4.6. The median values for quiescent, intermediate, active, and 
red clumps are 0.8, 0.8, 0.5, and 1.8, respectively.  

\vspace{0.2cm}
\hspace{3.5cm}{\it C$_2$H}
\vspace{0.2cm}

The optical depths of the C$_2$H hyperfine components are obtained 
using the intensity ratio between the
 $NJF=1\,\frac{3}{2}\,2\ra0\,\frac{1}{2}\,1$ and
 $NJF=1\,\frac{3}{2}\,1\ra0\,\frac{1}{2}\,0$ transitions.
The opacity ratio between these two lines is 
$r$=$\frac{\tau_{_{12-01}}}{\tau_{_{11-00}}}$=2 \citep{Tucker74}. Thus,
 the opacity of the brightest component is given by 
\begin{equation}  
\frac{1-e^{-\frac{1}{2}\tau_{_{12-01}}}}{1-e^{-\tau_{_{12-01}}}}=\frac{T_{\rm mb_{11-00}}}{T_{\rm mb_{12-01}}}~,
\label{C2H_tau-eqn}
\end{equation}
where the subscripts ``$12-01$'' and ``$11-00$'' refer to the C$_2$H  
hyperfine components $NJF=1\,\frac{3}{2}\,2\ra0\,\frac{1}{2}\,1$ and 
$NJF=1\,\frac{3}{2}\,1\ra0\,\frac{1}{2}\,0$, respectively. 

The C$_2$H opacities were calculated for 43 IRDC clumps and the values for the 
brightest transition are listed in Table~\ref{tbl-tau-column}.   
 The emission of the brightest C$_2$H hyperfine component is mostly 
 moderately optically thick, with 86\% of the optical depths ranging from 
1.0 to 6.0. The median values of the optical depths for quiescent,
 intermediate, active, and red clumps are 2.9, 2.4, 2.2, and 1.4,
 respectively. 

\vspace{0.2cm}
\hspace{2.3cm}{\it HCO$^+$ and H$^{13}$CO$^+$}
\vspace{0.2cm}

Rewriting Equation~\ref{tau-eqn}, we obtain
\begin{equation}  
\frac{1-e^{-\tau_{_{12}}/r}}{1-e^{-\tau_{_{12}}}}=\frac{T_{\rm mb}(\rm {H^{13}CO^+})}{T_{\rm mb}(\rm {HCO^+})}~,
\label{HCO+_tau-eqn}
\end{equation} 
where ``$r$'' is the ratio between the optical depths, $r=\tau_{12}/\tau_{13}$,  
and the subscripts ``$12$'' and ``$13$'' refer to the HCO$^+$ and 
H$^{13}$CO$^+$ isotopologues, respectively. A way to determine ``$r$'' is
 to take the ratio of the expression for the 
column density (see Equation~\ref{eqn-den-colum}) evaluated for both 
isotopologues 
\begin{eqnarray}
\frac{\tau_{_{12}}}{\tau_{_{13}}} & = & \left[\frac{\rm{HCO}^+}{\rm{H^{13}CO}^+}\right]\frac{(kT_{\rm ex}/hB_{13} + 1/3)}{(kT_{\rm ex}/hB_{12} + 1/3)}\frac{\exp (E_{_{J_{13}}}/kT_{\rm ex})}{\exp(E_{_{J_{12}}}/kT_{\rm ex})}\times \nonumber\\ 
&& \frac{[1 - \exp (-h\nu_{_{12}}/kT_{\rm ex})]}{[1 - \exp(-h\nu_{_{13}}/kT_{\rm ex})]}~,
\label{r-HCO+_tau}
\end{eqnarray} 
where [HCO$^+$/H$^{13}$CO$^+$] is the isotopic abundance ratio. 
The isotopic abundance ratio of [$^{12}$C/$^{13}$C] ranges from $\sim$20 
to $\sim$70 \cite[e.g.,][]{Savage02}.   
Assuming a constant [HCO$^+$/H$^{13}$CO$^+$] abundance ratio of 50 for all 
sources,  we compute the optical depths of 62 IRDC clumps.
The values for HCO$^+$ are summarized in Table~\ref{tbl-tau-column}. 
 All sources have  
optically thick HCO$^+$ emission, with opacities ranging between 7  
and 75. On the other hand, H$^{13}$CO$^+$ is optically thin in most of 
the sources (90\%). The median values of HCO$^+$ optical depths for 
quiescent, intermediate, active, and red clumps are 30, 20, 16, and 19, 
respectively; and the values for H$^{13}$CO$^+$ are 0.6, 0.4, 0.3, and 0.4,
 respectively.

\vspace{0.2cm}
\hspace{2.3cm}{\it HNC and HN$^{13}$C }
\vspace{0.2cm}

 Rewriting Equation~\ref{tau-eqn} and following the same procedure used 
for HCO$^+$, we obtain   
\begin{equation}  
\frac{1-e^{-\tau_{_{12}}/r}}{1-e^{-\tau_{_{12}}}}=\frac{T_{\rm mb}(\rm {HN^{13}C})}{T_{\rm mb}(\rm {HNC})}~,
\label{HNC_tau-eqn}
\end{equation} 
where $r$ is the ratio between the optical depths, $r=\tau_{12}/\tau_{13}$, and 
is given by 
\begin{eqnarray}
\frac{\tau_{_{12}}}{\tau_{_{13}}} & = & \left[\frac{\rm{HNC}}{\rm{HN^{13}C}}\right]\frac{(kT_{\rm ex}/hB_{13} + 1/3)}{(kT_{\rm ex}/hB_{12} + 1/3)}\frac{\exp (E_{_{J_{13}}}/kT_{\rm ex})}{\exp(E_{_{J_{12}}}/kT_{\rm ex})}\times \nonumber\\ 
&& \frac{[1 - \exp (-h\nu_{_{12}}/kT_{\rm ex})]}{[1 - \exp(-h\nu_{_{13}}/kT_{\rm ex})]}~,
\label{r-HCO+_tau}
\end{eqnarray} 
where the subscripts ``$12$'' and ``$13$'' refer to the HNC and 
HN$^{13}$C isotopologues, respectively, and [HNC/HN$^{13}$C] is the isotopic 
abundance ratio. 

Assuming an [HNC/HN$^{13}$C] abundance ratio of 50, we compute the
 optical depths of 85 IRDC clumps. The values for HNC are summarized
 in Table~\ref{tbl-tau-column}.  
 All sources have optically thick HNC emission, whereas 
HN$^{13}$C is optically thin in most of the sources (95\%). 
The median values of HNC optical depths for quiescent, intermediate, 
active, and red clumps are 24, 22, 18, and 16, respectively; and for
 HN$^{13}$C are 0.5, 0.4, 0.3, and 0.3, respectively.

\begin{deluxetable*}{lccccccccccc}
\tabletypesize{\tiny}
\tablecaption{Optical Depths and Column Densities. \label{tbl-tau-column}}
\tablewidth{0pt}
\tablehead{
\colhead{IRDC Clump}  & \multicolumn{4}{c}{\underline {~~~~~~~~Optical Depth~~~~~~~~}} & \multicolumn{7}{c}{\underline {~~~~~~~~~~~~~~~~~~~~~~~~~~~~~~~~~~~~~~~~~~~Column Density (cm$^{-2}$)~~~~~~~~~~~~~~~~~~~~~~~~~~~~~~~~~~~~~~~~~~}}\\
\colhead{}              &\colhead{ $\tau_{\rm N_2H^+}\!\!\!\!\!$}&\colhead{$\tau_{\rm HCO^+}\!\!\!\!\!$}&
\colhead{$\tau_{\rm HNC}\!\!\!\!\!$} & \colhead{$\tau_{\rm C_2H}\!\!\!\!\!$} & \colhead{N$_2$H$^+$} 
& \colhead{HCO$^+$}       & \colhead{HNC}            & \colhead{C$_2$H}   
& \colhead{HC$_3$N}       & \colhead{HNCO}           & \colhead{SiO}\\
\colhead{}               & \colhead{}                & \colhead{}    &
 \colhead{}              & \colhead{}               &\colhead{$\times10^{13}$}&
 \colhead{$\times10^{14}$}&\colhead{$\times10^{14}$} &\colhead{$\times10^{14}$}&
 \colhead{$\times10^{13}$}&\colhead{$\times10^{13}$}&\colhead{$\times10^{12}$}\\
}
\startdata
 G015.05 MM1 & 1.00 & 29.9 & 26.9 & 2.8 &  1.56(0.12) &  2.05(0.72) &  2.27(1.09) &  2.75(3.02) &  ... &  1.67(0.69) & ...\\
 G015.05 MM2 & 0.04 &  ... & 59.4 & ... &  0.71(0.15) &  0.58(0.21) &  2.66(1.91) &     ...     &  ... &  ... &  ... \\
 G015.05 MM3 & 0.60 &  ... & 28.6 & ... &  0.36(0.10) &     ...     &  1.04(0.51) &     ...     &  ... &  ... &  ... \\
 G015.05 MM4 & ... &  ... & 16.1 & 4.5 &  1.04(0.12) &  0.58(0.20) &  1.56(0.65) &  3.06(3.29) &  ... &  ... &  ... \\
 G015.05 MM5 & 0.36 & 35.0 & 52.2 & ... &  0.52(0.16) &  1.01(0.56) &  2.40(1.55) &     ...     &  ... &  ... &  ... \\
 G015.31 MM2 & 2.89 &  ... & 69.0 & ... &  1.18(1.25) &     ...     &  2.19(1.79) &     ...     &  ... &  ... &  ... \\
 G015.31 MM3 & ... & 50.2 & 30.3 & ... &  0.31(0.08) &  0.85(0.61) &  0.51(0.51) &     ...     &  ... &  ... &  ... \\
 G015.31 MM5 & ... & 38.2 & 53.0 & ... &     ...     &  1.10(0.63) &  1.36(0.95) &     ...     &  ... &  ... &  ... \\
 G018.82 MM2 & 2.16 & 44.6 & 43.5 & ... &  2.24(2.25) &  3.20(1.94) &  3.45(2.10) &     ...     &  ... &  ... &  ... \\
 G018.82 MM3 & 2.27 & 37.7 & 27.9 & ... &  6.07(6.15) & 10.10(5.24) &  6.71(3.14) &     ...     &  ... &  ... &  ... \\
 G018.82 MM4 & 0.12 & 75.0 & 45.3 & ... &  1.06(0.18) &  8.50(6.72) &  5.02(2.85) &     ...     &  ... &  0.98(1.04) & ...\\
 G018.82 MM6 & 0.12 &  ... &  ... & ... &  1.22(0.12) &  0.52(0.19) &  0.80(0.35) &     ...     &  ... &  ... &  ... \\
 G019.27 MM2 & 0.81 &  ... & 33.8 & 5.6 &  2.18(0.15) &     ...     &  4.09(1.40) &  4.15(4.59) &  ... &  ... &  ... \\
 G022.35 MM1 & 0.31 & 14.4 & 20.9 & ... &  0.40(0.10) &  0.85(0.33) &  1.13(0.47) &     ...     &  ... &  ... &  ... \\
 G022.35 MM2 & ... &  ... & 33.2 & ... &     ...     &     ...     &  2.02(1.07) &     ...     &  ... &  ... &  ... \\
 G023.60 MM7 & 1.64 &  ... & 27.6 & ... &  4.13(4.07) &     ...     &  3.50(1.67) &     ...     &  ... &  ... &  ... \\
 G023.60 MM9 & 0.56 &  ... & 29.3 & 2.4 &  0.68(0.11) &     ...     &  1.47(0.71) &  1.51(2.16) &  ... &  ... &  ... \\
 G024.08 MM2 & 0.99 & 43.0 & 26.4 & ... &  0.70(0.12) &  1.77(0.71) &  2.51(1.12) &     ...     &  ... &  ... &  ... \\
 G024.08 MM3 & 1.27 & 31.5 & 24.1 & ... &  0.93(0.92) &  1.78(0.89) &  1.91(0.85) &     ...     &  ... &  ... &  ... \\
 G024.08 MM4 & 0.27 & 66.5 & 31.1 & ... &  0.48(0.13) &  2.43(2.20) &  1.84(0.94) &     ...     &  ... &  ... &  ... \\
 G024.33 MM2 & ... &  ... & 25.5 & ... &  3.06(0.18) &  1.37(0.36) &  6.16(1.17) &     ...     &  ... &  4.27(1.10) & ...\\
 G024.33 MM3 & ... &  ... &  ... & 1.9 &  2.65(0.17) &  1.38(0.35) &  0.85(0.29) &  3.41(4.60) &  ... &  2.81(0.90) &  3.96(1.41) \\
 G024.33 MM4 & ... &  ... &  ... & ... &  1.57(0.13) &  0.79(0.30) &  0.78(0.28) &     ...     &  ... &  ... &  ... \\
 G024.33 MM5 & 0.26 &  ... &  ... & 3.2 &  2.17(0.15) &  2.81(0.46) &  0.79(0.29) &  2.63(3.99) &  ... &  5.44(0.94) & ...\\
 G024.33 MM7 & 0.67 & 15.0 & 10.3 & 1.2 &  0.93(0.09) &  1.22(0.23) &  1.06(0.39) &  0.61(1.20) &  ... &  ... &  1.36(0.74) \\
 G024.33 MM8 & ... & 13.2 & 15.8 & ... &     ...     &  1.40(0.53) &  2.62(1.01) &     ...     &  ... &  ... &  ... \\
 G024.33 MM9 & ... & 10.9 & 11.5 & ... &     ...     &  2.30(0.89) &  3.48(1.36) &     ...     &  ... &  ... &  ... \\
G024.33 MM11 & 0.59 & 29.9 & 23.9 & ... &  1.51(0.10) &  2.82(0.62) &  4.01(1.49) &     ...     &  ... &  ... &  ... \\
 G024.60 MM2 & 0.92 & 22.4 & 16.4 & 1.1 &  1.24(0.10) &  1.51(0.49) &  1.48(0.63) &  1.16(0.40) &  0.35(0.12) &  ... & ...\\
 G025.04 MM2 & ... & 13.2 & 18.3 & ... &  0.63(0.11) &  1.26(0.51) &  1.70(0.73) &     ...     &  ... &  ... &  ... \\
 G025.04 MM4 & 0.06 &  8.7 & 24.6 & 1.0 &  2.44(0.11) &  1.51(0.54) &  4.30(1.01) &  1.09(2.22) &  ... &  3.12(0.66) & ...\\
 G027.75 MM2 & 3.17 &  ... &  ... & ... &  1.01(1.02) &     ...     &     ...     &     ...     &  ... &  ... &  ... \\
 G027.94 MM1 & 1.39 &  ... & 22.9 & 3.1 &  1.66(1.66) &     ...     &  1.97(0.68) &  1.94(3.13) &  ... &  ... &  ... \\
 G028.04 MM1 & 1.15 & 24.1 & 60.7 & 1.3 &  2.27(2.23) &  2.24(0.96) &  5.50(2.66) &  1.27(2.27) &  0.38(0.12) &  2.88(0.84) &  ... \\
 G028.08 MM1 & 1.25 &  ... & 40.7 & ... &  0.63(0.63) &     ...     &  1.92(1.02) &     ...     &  ... &  ... &  ... \\
 G028.23 MM1 & 3.62 &  ... &  ... & ... &  3.12(3.14) &  1.00(0.34) &  0.77(0.27) &     ...     &  ... &  ... &  ... \\
 G028.28 MM4 & 0.88 & 13.9 & 11.1 & 3.0 &  1.53(0.13) &  1.83(0.38) &  1.65(0.60) &  2.68(3.09) &  ... &  ... &  ... \\
 G028.37 MM1 & ... &  ... & 20.6 & 1.9 &  4.60(0.15) &  2.51(0.39) &  5.72(0.97) &  4.31(2.38) &  1.45(0.14) &  9.93(0.88) & 16.00(1.36) \\
 G028.37 MM2 & 0.46 &  ... & 24.6 & 3.0 &  2.51(0.18) &  2.42(0.45) &  6.37(1.76) &  4.54(6.44) &  0.49(0.21) &  3.61(1.50) &  ... \\
 G028.37 MM4 & 0.06 & 12.8 & 11.8 & 4.8 &  4.69(0.19) &  2.90(0.30) &  4.37(0.47) &  6.30(5.60) &  0.85(0.11) &  8.05(0.63) & 13.90(0.98) \\
 G028.37 MM6 & 0.19 & 22.2 & 20.3 & 2.3 &  2.67(0.10) &  2.72(0.52) &  2.88(0.46) &  1.80(1.89) &  0.77(0.12) &  4.30(0.51) & 10.20(0.90) \\
 G028.37 MM9 & ... & 21.9 & 15.0 & ... &  1.52(0.11) &  1.99(0.34) &  1.52(0.29) &     ...     &  ... &  ... &  ... \\
G028.37 MM11 & 0.23 &  ... & 41.8 & 3.3 &  1.05(0.20) &  0.82(0.28) &  7.21(3.80) &  2.63(4.22) &  ... &  ... &  ... \\
G028.37 MM12 & 1.37 & 65.8 & 74.0 & ... &  0.89(0.90) &  2.42(1.38) &  3.61(2.86) &     ...     &  0.31(0.15) &  ... & ...\\
G028.37 MM13 & 3.32 &  ... &  ... & ... &  1.46(1.58) &     ...     &     ...     &     ...     &  ... &  ... &  ... \\
 G028.53 MM3 & 1.73 & 19.4 & 23.7 & ... &  2.47(2.42) &  2.13(0.64) &  2.38(1.04) &     ...     &  ... &  ... &  ... \\
 G028.53 MM5 & 4.45 & 46.0 & 46.7 & ... &  2.90(3.03) &  1.50(1.02) &  3.43(2.01) &     ...     &  ... &  ... &  ... \\
 G028.53 MM7 & ... &  ... & 49.0 & ... &  1.10(0.12) &  0.91(0.26) &  3.47(1.21) &     ...     &  ... &  ... &  ... \\
 G028.53 MM8 & ... & 69.4 & 37.1 & 1.5 &  1.00(0.14) &  3.54(2.78) &  3.44(1.77) &  2.02(1.15) &  ... &  ... &  ... \\
 G028.53 MM9 & 0.81 & 13.9 & 15.0 & ... &  1.88(0.16) &  2.38(0.95) &  3.08(1.01) &     ...     &  ... &  ... &  ... \\
G028.53 MM10 & 0.10 & 13.7 & 17.2 & 2.9 &  2.04(0.20) &  2.83(1.08) &  4.90(1.46) &  5.05(6.87) &  ... &  ... &  ... \\
 G028.67 MM1 & ... & 31.0 & 31.2 & ... &     ...     &  1.64(0.82) &  2.95(1.41) &     ...     &  ... &  ... &  ... \\
 G028.67 MM2 & 0.83 &  ... & 47.7 & ... &  0.71(0.10) &     ...     &  2.64(1.57) &     ...     &  ... &  ... &  ... \\
 G030.14 MM1 & 4.58 & 37.8 & 32.6 & ... &  1.36(1.50) &  1.18(0.68) &  1.61(0.83) &     ...     &  ... &  ... &  ... \\
 G030.57 MM1 & 0.11 &  ... & 19.6 & 1.1 &  4.42(0.15) &     ...     &  3.85(1.22) &  2.64(4.13) &  0.85(0.15) &  2.86(0.98) &  ... \\
 G030.57 MM3 & 0.42 &  ... & 45.2 & 6.0 &  0.98(0.18) &     ...     &  6.33(3.59) &  2.39(3.67) &  ... &  ... &  ... \\
 G030.97 MM1 & ... & 16.4 &  9.4 & ... &  3.64(0.14) &  5.17(0.50) &  4.04(0.67) &     ...     &  0.37(0.12) &  ... & ...\\
 G031.97 MM5 & 0.04 &  ... & 15.2 & 3.5 &  2.33(0.13) &  1.25(0.29) &  3.53(1.15) &  2.93(3.80) &  ... &  ... &  ... \\
 G031.97 MM7 & 0.45 &  7.3 & 14.5 & 3.5 &  1.69(0.10) &  1.20(0.23) &  3.00(2.81) &  2.14(3.40) &  ... &  1.00(0.56) & ...\\
 G031.97 MM8 & ... &  ... &  9.5 & ... &  3.43(0.20) &     ...     &  3.41(1.18) &     ...     &  ... &  ... &  ... \\
 G033.69 MM1 & ... &  9.8 & 11.4 & ... &  3.56(0.11) &  1.70(0.25) &  2.93(0.34) &     ...     &  0.96(0.13) &  4.55(0.59) &  5.29(1.01) \\
 G033.69 MM2 & ... & 17.1 & 14.6 & 0.8 &  3.19(0.21) &  4.53(1.84) &  4.48(1.14) &  3.55(0.49) &  1.01(0.16) &  4.31(1.21) &  ... \\
 G033.69 MM3 & ... & 18.5 & 16.4 & ... &  2.40(0.28) &  3.61(1.51) &  5.18(2.09) &     ...     &  ... &  ... &  ... \\
 G033.69 MM4 & ... & 12.4 & 16.2 & ... &  3.05(0.21) &  1.76(0.64) &  3.37(0.83) &     ...     &  0.56(0.20) &  3.07(1.02) &  ... \\
 G033.69 MM5 & ... & 31.6 & 18.3 & 2.9 &  2.15(0.24) &  3.85(1.91) &  4.23(1.76) &  3.75(3.05) &  ... &  ... &  ... \\
G033.69 MM11 & ... & 12.1 & 13.9 & 4.0 &     ...     &  2.53(0.95) &  3.99(1.52) &  1.48(2.87) &  ... &  ... &  ... \\
 G034.43 MM1 & 0.42 &  ... & 18.0 & 1.7 & 13.70(0.21) &  4.77(0.48) & 13.60(1.58) &  7.75(4.71) & 3.33(0.21) &  8.59(1.43) & 34.70(2.09) \\
 G034.43 MM5 & 0.67 &  ... & 30.0 & ... &  2.98(0.09) &  1.92(0.25) &  5.23(0.79) &     ...     &  0.54(0.12) &  ... &  ... \\
 G034.43 MM7 & ... & 21.6 & 18.9 & 1.9 &  1.11(0.11) &  1.07(0.33) &  2.29(0.60) &  2.06(2.92) &  0.21(0.12) &  ... & ... \\
 G034.43 MM8 & ... & 65.7 & 19.0 & 3.9 &  1.53(0.17) &  2.71(1.21) &  2.55(1.01) &  3.56(5.98) &  ... &  ... &  ... \\
 G034.77 MM1 & 2.20 & 27.9 & 11.2 & ... &  4.09(4.04) &  4.52(0.79) &  2.44(0.80) &     ...     &  0.47(0.13) &  ... & ...\\
 G034.77 MM3 & 0.25 &  ... & 22.4 & 4.1 &  0.88(0.15) &  0.50(0.18) &  2.78(1.17) &  3.01(5.16) &  ... &  ... &  ... \\
 G035.39 MM7 & 0.18 & 27.9 & 17.6 & 2.1 &  2.41(0.17) &  6.00(0.94) &  3.81(0.82) &  4.38(4.13) &  0.32(0.15) &  ... & ...\\
 G035.59 MM1 & 4.18 & 31.0 & 40.8 & ... &  2.84(3.10) &  4.38(2.19) &  1.77(1.10) &     ...     &  ... &  ... &  ... \\
 G035.59 MM2 & ... & 21.8 & 11.6 & ... &  0.72(0.10) &  1.82(0.57) &  1.09(0.41) &     ...     &  ... &  ... &  ... \\
 G035.59 MM3 & 0.42 &  7.8 & 10.5 & 2.4 &  1.00(0.23) &  1.12(0.29) &  1.25(0.47) &  1.82(2.20) &  0.14(0.19) &  ... & ...\\
 G036.67 MM1 & 3.22 & 68.4 & 19.7 & 4.3 &  0.99(1.02) &  1.58(1.34) &  1.02(0.60) &  1.58(1.57) &  ... &  ... &  ... \\
 G036.67 MM2 & ... & 20.2 & 18.1 & ... &     ...     &  0.65(0.28) &  1.15(0.48) &     ...     &  ... &  ... &  ... \\
 G038.95 MM1 & 0.12 & 26.9 &  9.6 & 1.2 &  1.34(0.10) &  4.27(0.65) &  1.69(0.51) &  2.41(1.94) &  0.47(0.19) &  0.92(0.53) &  ... \\
 G038.95 MM2 & 0.92 & 19.7 & 11.1 & ... &  1.72(0.20) &  4.64(1.06) &  2.45(0.92) &     ...     &  0.35(0.17) &  ... & ...\\
 G038.95 MM3 & 2.44 & 18.5 & 14.0 & 0.4 &  2.98(2.97) &  3.40(0.87) &  2.35(0.92) &  2.80(0.40) &  ... &  ... &  ... \\
 G038.95 MM4 & 1.66 & 10.4 & 13.8 & 1.0 &  1.97(1.99) &  1.83(0.68) &  2.34(0.92) &  2.16(0.39) &  ... &  ... &  2.48(1.80) \\
 G053.11 MM1 & 0.56 & 10.5 &  4.5 & ... &  3.11(0.25) &  4.74(0.48) &  1.97(0.59) &     ...     &  0.80(0.14) &  ... & ...\\
 G053.11 MM2 & 1.05 & 13.6 &  9.6 & 1.1 &  2.76(2.77) &  2.80(0.56) &  1.87(0.53) &  1.62(0.62) &  ... &  ... &  ... \\
 G053.11 MM4 & 0.52 & 23.0 &  7.1 & 0.2 &  1.11(0.17) &  4.03(0.61) &  1.45(0.39) &  0.94(0.36) &  0.16(0.12) &  ... & ...\\
 G053.11 MM5 & 4.24 & 11.0 & 12.4 & 0.9 &  2.90(2.96) &  1.26(0.47) &  1.54(0.58) &  1.08(0.25) &  ... &  ... &  ... \\
 G053.25 MM1 & ... &  7.8 &  5.7 & 1.5 &  1.44(0.14) &  1.38(0.45) &  1.20(0.42) &  1.54(2.06) &  ... &  ... &  ... \\
 G053.25 MM3 & 1.79 & 15.3 & 10.4 & 4.4 &  1.37(1.44) &  1.49(0.62) &  1.35(0.52) &  3.10(5.18) &  ... &  ... &  ... \\
 G053.25 MM4 & 0.25 & 12.4 &  5.8 & ... &  1.62(0.15) &  2.44(0.56) &  1.23(0.43) &     ...     &  ... &  ... &  ... \\
 G053.25 MM5 & ... & 21.4 & 10.7 & 4.8 &     ...     &  2.55(0.88) &  1.38(0.52) &  2.05(4.63) &  ... &  ... &  ... \\
 G053.25 MM6 & 1.46 & 13.3 &  7.5 & ... &  3.84(3.88) &  2.64(0.62) &  1.65(0.62) &     ...     &  ... &  ... &  ... \\
 G053.31 MM2 & 2.69 & 11.2 & 13.3 & ... &  1.24(1.23) &  0.67(0.25) &  1.00(0.39) &     ...     &  ... &  ... &  ... \\
\enddata
\end{deluxetable*}

\clearpage
\subsubsection{Column Densities}

The column densities of N$_2$H$^+$, HCO$^+$ and HNC are calculated 
using equation~\ref{eqn-den-colum}. The SiO molecule is also
 a linear, rigid rotor. Thus, we can also use equation~\ref{eqn-den-colum} to 
estimate SiO column densities. However, since we observed only one SiO 
 isotopologue, we need to assume that the SiO 
 emission is optically thin and the filling factor is 1.0 (which gives us 
the beam-averaged column density). Therefore, replacing equation~\ref{Tmb-eq}
 in equation~\ref{eqn-den-colum}, we obtain 
\begin{eqnarray}
N & = & \frac{3k}{8\pi^3B\mu^2R}\frac{(T_{\rm ex}+hB/3k)}{(J+1)}\frac{\exp(E_{_{J}}/kT_{\rm ex})}{[1-\exp(-h\nu/kT_{\rm ex})]}\times \nonumber\\ 
&& \frac{1}{[J(T_{\rm ex}) - J(T_{\rm bg})]}\int T_{\rm mb}\,dv~.  
\label{eqn-den-colum-thin}
\end{eqnarray}

The values of the parameters used in the column density calculations 
(permanent dipole moment, rotational constant) of N$_2$H$^+$, HCO$^+$, HNC, 
and SiO are summarized in Table~\ref{tbl-param-column1}.

\begin{deluxetable}{lccl}
\tabletypesize{\scriptsize}
\tablecaption{Parameters used for N$_2$H$^+$, HCO$^+$, HNC, and
 SiO Column Density Calculations \label{tbl-param-column1}}
\tablewidth{0pt}
\tablehead{
\colhead{Molecule} &  \colhead{Dipole Moment}  & \colhead{Rotational
 Constant} & \colhead{References} \\
\colhead{} &  \colhead{$\mu$ (D)} & \colhead{$B$ (GHz)}\\
}
\startdata
\NdosH    & 3.40  & 46.586871 & 1, 2, 3\\
\HCO      & 3.89  & 44.594423 & 4, 5, 6\\
\HtreceCO & 3.89  & 43.377302 & 4, 5, 6 \\
\HNC      & 3.05  & 45.331980 & 7 ,8 \\
\HNtreceC & 3.05  & 43.545600 & 7 ,8 \\
\SiO      & 3.10  & 21.711979 & 9, 10\\
\enddata
\tablecomments{References. (1) \cite{Botschwina84}; (2) \cite{Havenith90}; 
(3) \cite{Pagani09}; (4) \cite{Botschwina93}; (5) \cite{Yamaguchi94}; 
(6) \cite{Lattanzi07}; (7) \cite{Blackman76}; (8) \cite{Tak09};
 (9) \cite{Raymonda70}; (10) \cite{Mollaaghababa91}}
\end{deluxetable}

The column densities of the remaining molecules (C$_2$H, HC$_3$N, HNCO) are 
determined using equation~\ref{eqn-den-colum-2}. Although C$_2$H is a
 linear molecule, its rotational energy levels are described by the
rotational quantum number $N$, instead of $J$. HC$_3$N is also linear
 and its partition function 
is the same used for diatomic, linear molecules \citep{Blake87}; we have not 
included the factor 3 due to the nuclear spin degeneracy because the 
hyperfine structure is not resolved. On the other hand, 
HNCO is a standard asymmetric top molecule with a different partition 
function \citep{Blake87}. For C$_2$H the column density was calculated 
using the optical depths obtained in Section~\ref{opacity}. For 
HC$_3$N and HNCO, the emission was assumed to be optically thin and 
the filling factor was used to be equal to 1.0.    
The values of the parameters used in the column density calculations 
(partition function, rotational constants, statistical weight, 
Einstein coefficient for spontaneous emission, and energy of the lower
 state) of C$_2$H, HC$_3$N and HNCO are summarized in 
Table~\ref{tbl-param-column2}. 

Upper limits to the column densities of the most commonly detected 
molecular lines (N$_2$H$^+$, HCO$^+$, HNC) were estimated when they were 
 not detected over the 3$\sigma$ level. In the case of the optically thick 
HCO$^{+}$ and HNC emission lines, the limits were obtained first for  
their isotopologues (H$^{13}$CO$^{+}$ and HN$^{13}$C). Then by using 
the isotopic abundance ratio we calculated the column densities 
for HCO$^{+}$ and HNC.  In all non-detections, the emission was assumed to 
be optically thin, with $T_{\rm mb}=3 T_{\rm rms}$, and $\Delta$V equal
 to 2.0 \kms. 

Additionally, lower limits to the column densities were estimated 
for N$_2$H$^+$ and HCO$^+$ when our normal approach was not applicable.  
For N$_2$H$^+$, the intensity ratio between the two brightest 
components, which is used to derive opacities, should not be higher than 
its value in 
the optically thin limit, 5/3. However, in some cases, this ratio exceeds 
the optically thin limit, but nevertheless remains consistent with 5/3 
within the uncertainties. To calculate a lower 
limit for the column density, the main N$_2$H$^+$ component was assumed to 
be optically thin. For HCO$^+$, in several cases the spectra show  
self-absorbed profiles, and no Gaussian fit was performed. Lower limits to 
the column densities were computed using the H$^{13}$CO$^{+}$ emission line 
and assuming that its emission is optically thin. Then, using the 
[HCO$^+$/H$^{13}$CO$^+$] isotopic abundance ratio, we obtain the HCO$^+$ 
column densities. 

Derived column densities using the optical depths obtained in
 Section~\ref{opacity} were multiplied by the filling factor to get 
the beam-averaged column density. The filling factor was calculated 
from  Equation~\ref{Tmb-eq}. Column densities for N$_2$H$^+$, HCO$^+$, 
HNC, SiO, C$_2$H, HC$_3$N, and HNCO molecules (including all limits) are listed 
in the Table~\ref{tbl-tau-column}. Median values of the column
 densities for each evolutionary sequence (excluding limits when the
 lines are not detected above the 3$\sigma$ level) are shown
 in Table~\ref{tbl-medians}. 

\begin{figure}
\begin{center}
\includegraphics[angle=0,scale=0.5]{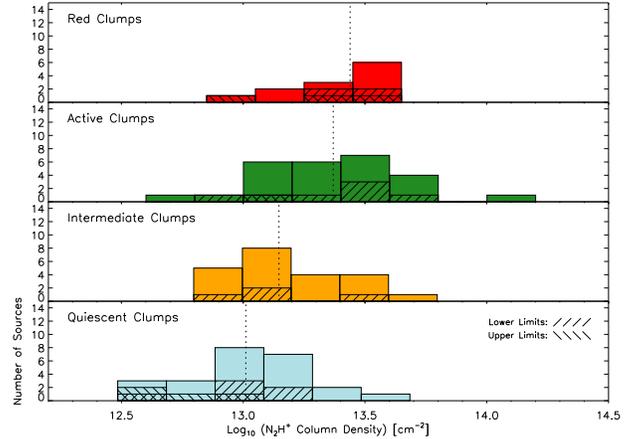}
\end{center}
\caption{Histograms of the number distributions of N$_2$H$^+$ column densities  
for each evolutionary sequence. The name of the evolutionary stage is
 given on the top left corner of each panel. 
 The vertical dashed lines indicate the median values of the column densities
 for each distribution. The diagonal patterns indicate the portions of the 
histograms that correspond to upper and lower limits (see bottom panel, 
right side). The median values and the K-S test suggest that column
 densities increase with the evolution of the clumps. Median values of
 the column densities are given in Table~\ref{tbl-medians}. }
\label{n2h+_column_density}
\end{figure}

\vspace{0.2cm}
\hspace{3.5cm}{\it N$_2$H$^+$}
\vspace{0.2cm}

Figure~\ref{n2h+_column_density} shows the number distributions 
of the N$_2$H$^+$ column densities for each evolutionary sequence.  
The K--S test gives a probability of 0.09\% that quiescent and active 
 clump distributions originate from  the same parent population. 
Figure~\ref{n2h+_column_density} shows a trend of increasing 
 column densities with the evolution of the clump, which is 
confirmed by the median values (see Table~\ref{tbl-medians}) and the
 K--S test probability. This trend does not necessary imply a 
change in the chemistry. An increase of the N$_2$H$^+$ column density 
could simply indicate that the total H$_2$ column density increases 
while maintaining constant chemical composition. In 
Sections~\ref{chemistry}~and~\ref{molecules} we discuss the 
molecular abundances, which more definitively show chemical variations.
This argument applies to all the molecules discussed in this section. 
 \cite{Sakai08} obtained N$_2$H$^+$ column densities over a sample of 11 
{\it Spitzer} and {\it MSX} dark objects (no 24 and 8 $\mu$m emission,
 respectively) that we can compare with our quiescent clumps. \cite{Pirogov03}
 obtained N$_2$H$^+$ column densities of 34 clumps, where massive stars have
 already formed, that we can compare with our active clumps.
 Table~\ref{tbl-comparison} summarizes the results of previous works,
 which are in good agreement with our results.  

\begin{figure}
\begin{center}
\includegraphics[angle=0,scale=0.5]{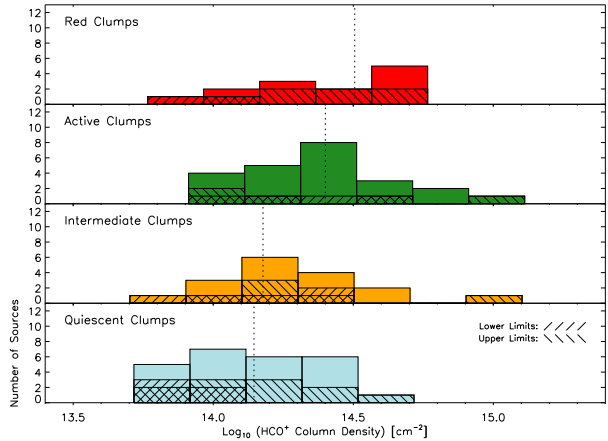}
\end{center}
\caption{Histograms of the number distributions of HCO$^+$ column densities  
for each evolutionary sequence. The name of the evolutionary stage is 
 given on the top left corner of each panel. 
 The vertical dashed lines indicate the median values of the column densities
 for each distribution. The diagonal patterns indicate the portions of the 
histograms that correspond to upper and lower limits (see bottom panel, 
right side). The median
 values and the K-S test suggest that column densities increase with the 
evolution of the clumps. Median values of the column densities are given 
in Table~\ref{tbl-medians}. }
\label{hco+_column_density}
\end{figure}

\vspace{0.2cm}
\hspace{2.3cm}{\it HCO$^+$ and H$^{13}$CO$^+$}
\vspace{0.2cm}

 Figure~\ref{hco+_column_density} shows 
the number distributions of the HCO$^+$ column densities for each 
evolutionary stage.  
 The K--S test gives us a probability of 2\% that quiescent and active 
 distributions originate from  the same parent population.
 Figure~\ref{hco+_column_density} shows a trend of increasing column
 densities with the evolution of the clump, which is supported by the
 median values (see Table~\ref{tbl-medians}) and the K--S test probability.
\cite{Sakai10} and \cite{Purcell06} obtained H$^{13}$CO$^+$ and HCO$^+$ column 
densities toward 20 and 79 massive clumps, respectively, associated with 
IR emission, methanol masers and UC H\,{\sc ii} regions. We compare this sample 
with our active and red cores in Table~\ref{tbl-comparison}. The high 
  HCO$^+$ column densities obtained by \cite{Sakai10} and \cite{Purcell06} 
confirm the trend that HCO$^+$ column densities increase with the
 evolution of the clumps, as seen in Figure~\ref{hco+_column_density}.

\vspace{0.2cm}
\hspace{2.3cm}{\it HNC and HN$^{13}$C }
\vspace{0.2cm}

\begin{figure}
\begin{center}
\includegraphics[angle=0,scale=0.5]{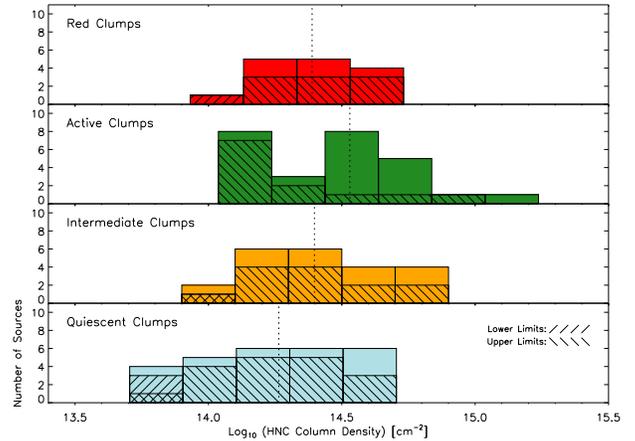}
\end{center}
\caption{Histograms of the number distributions of HNC column densities  
for each evolutionary sequence. The name of the evolutionary stage is
 given on the top left corner of each panel. 
 The vertical dashed lines indicate the median values of the column densities
 for each distribution. The diagonal patterns indicate the portions of the 
histograms that correspond to upper and lower limits (see bottom panel, 
right side). The median
 values and the K-S test suggest that column densities increase with the 
evolution of the clumps, from quiescent, to intermediate, to active clumps. 
Median values of the column densities are given in Table~\ref{tbl-medians}. }
\label{hnc_column_density}
\end{figure}

Figure~\ref{hnc_column_density} shows the number distributions of the
 HNC column densities for each evolutionary stage.
 The K--S test gives a probability of 5\% that quiescent and active 
 distributions originate from the same parent population.
 Figure~\ref{hnc_column_density} shows a weaker trend (compared with
 N$_2$H$^+$ and HCO$^+$) of increasing column densities from quiescent
 to active clumps. 
There are only a few systematic studies of HNC toward a large sample of massive 
star forming or IRDC clumps. \cite{Sakai10} observed HN$^{13}$C toward 
20 massive clumps that we can compare with our active clumps. 
Their values, showed in Table~\ref{tbl-comparison}, are in good agreement
 with our results.

\vspace{0.2cm}
\hspace{3.5cm}{\it C$_2$H}
\vspace{0.2cm}

\begin{figure}
\begin{center}
\includegraphics[angle=0,scale=0.5]{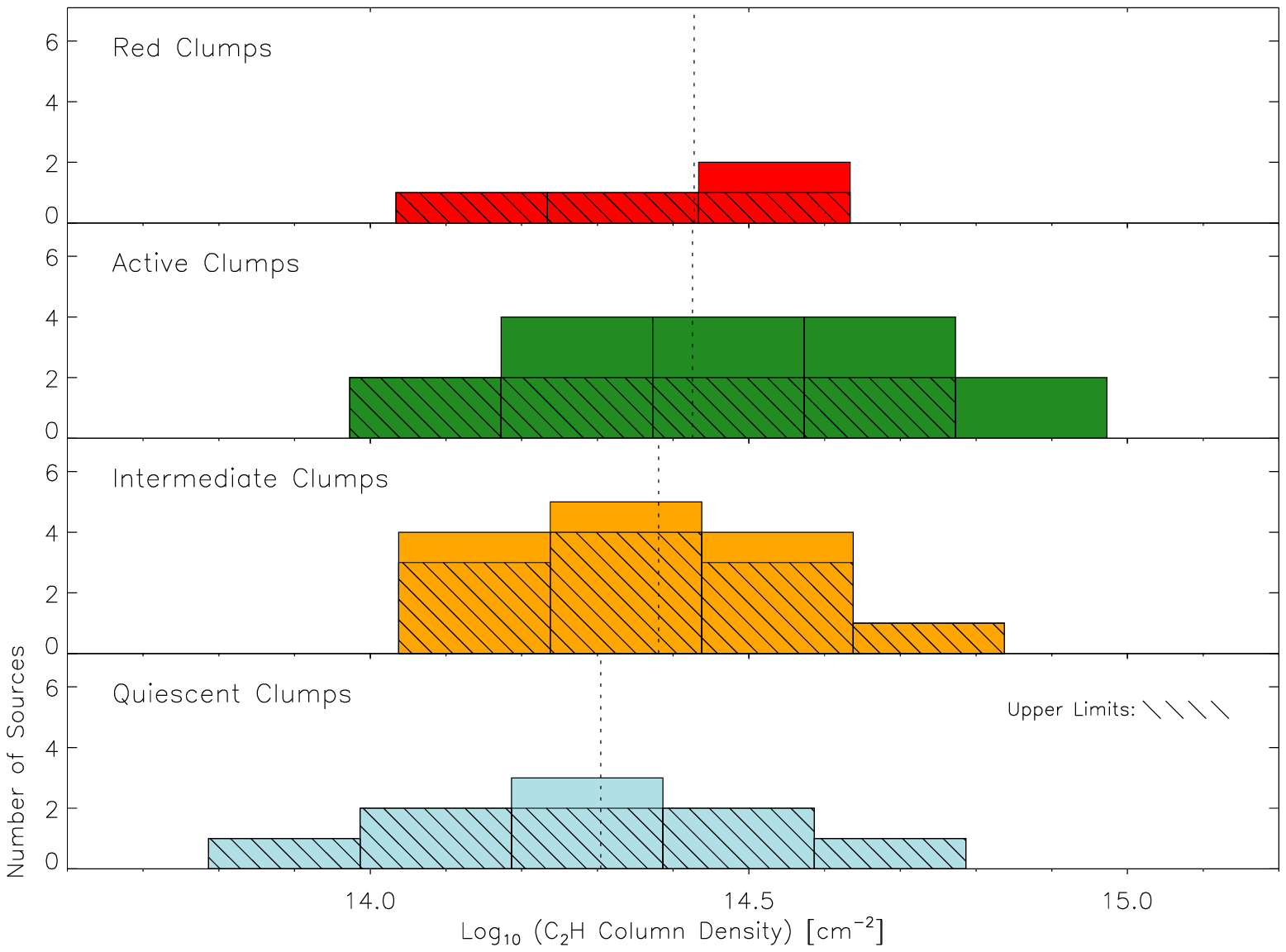}
\end{center}
\caption{Histograms of the number distributions of C$_2$H column densities  
for each evolutionary sequence. The name of the evolutionary stage is given
 on the top left corner of each panel. 
 The vertical dashed lines indicate the median values of the column densities
 for each distribution. The diagonal patterns indicate the portions of the 
histograms that correspond to upper limits. The median
 values slightly increase with the evolution of the clumps; however, the 
K-S test shows this variation is not statistically significant. Median 
values of the column densities are given in Table~\ref{tbl-medians}. }
\label{c2h_column_density}
\end{figure}

Figure~\ref{c2h_column_density} shows the number distributions of the
 C$_2$H column densities for each evolutionary sequence. 
The K--S test gives us a probability of 
38\% that active and quiescent distributions originate from  the same 
parent population. Figure~\ref{c2h_column_density} shows that the  
 trend of increasing column densities with the evolution of the clumps is 
less evident for C$_2$H than for previous molecules. 
Although the median values obtained for each 
evolutionary sequence support this trend, the K--S test probability is 
not low enough to be confident. Table~\ref{tbl-comparison} compares the
 column densities obtained for 20 massive clumps by \cite{Sakai10}
 with our active clumps, which are in a reasonable agreement. 

\vspace{0.2cm}
\hspace{3.5cm}{\it HC$_3$N}
\vspace{0.2cm}

The median values for the HC$_3$N column densities are summarized in
 Table~\ref{tbl-medians}.  HC$_3$N emission was detected only in one
 quiescent clump, G028.37 MM12, whose column density was 
$3.05\times 10^{12}$ cm$^{-2}$. 
\cite{Sakai08} obtained HC$_3$N column densities for 7 {\it Spitzer} and
 {\it MSX} dark objects (no 24 and 8 $\mu$m emission, respectively). Their 
values are comparable to that found in our one quiescent clump
 (see Table~\ref{tbl-comparison}).  

\vspace{0.2cm}
\hspace{3.5cm}{\it HNCO}
\vspace{0.2cm}

The median values for the HNCO column densities are summarized in
 Table~\ref{tbl-medians}. \cite{Zinchenko00} calculated HNCO column
 densities toward 20 massive star forming clumps using rotational diagrams.
 These sources were selected based on the presence of water masers and strong
 SiO emission, indicating they are currently forming stars. They obtained 
higher values than those found in our IRDC sample
 (Table~\ref{tbl-comparison}). This may be explained by the warmer
 rotational temperatures (median of 88 K) obtained by \cite{Zinchenko00}. 

\vspace{0.2cm}
\hspace{3.5cm}{\it SiO}
\vspace{0.2cm}

The median values for the SiO column densities are summarized in
 Table~\ref{tbl-medians}.  SiO emission was detected in no intermediate 
clumps and only in one quiescent clump, G024.33 MM7, whose column density is
$1.36\times 10^{12}$  cm$^{-2}$. \cite{Sakai10} and \cite{Miettinen06} 
obtained SiO column densities for 17 and 15 high-mass star-forming 
clumps, respectively, associated with IR and maser emission and
 UC H\,{\sc ii} regions (See Table~\ref{tbl-comparison}). Their values
 agree with those we calculated for active clumps. Red clumps, however, have
 lower column densities. 

\subsection{Chemistry}
\label{chemistry}
In order to study the chemistry toward the IRDC clumps, and investigate 
if the different evolutionary sequences proposed by \cite{Chambers09} 
are chemically distinguishable, we calculate molecular abundances using 
the 1.2 mm continuum emission obtained by \cite{Rathborne06} and
 abundance ratios of selected species.

\subsubsection{Molecular Abundances}
\label{line_abundance}

\begin{figure}
\begin{center}
\includegraphics[angle=0,scale=0.5]{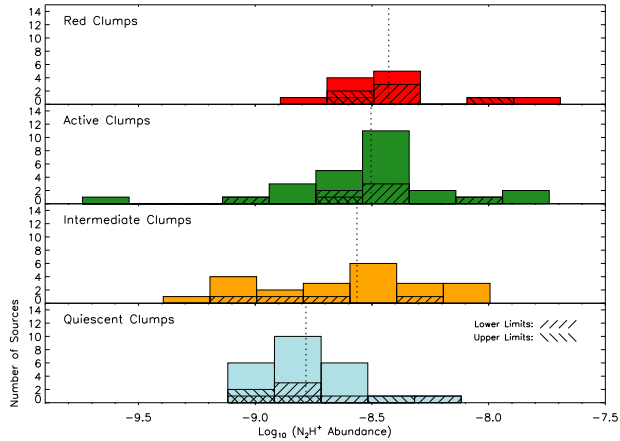}
\end{center}
\caption{Histograms of the number distributions of N$_2$H$^+$ abundances   
for each evolutionary sequence. The name of the evolutionary stage is given
 on the top left corner of each panel. 
 The vertical dashed lines indicate the median values of the abundances 
 for each distribution. The diagonal patterns indicate the portions of the 
histograms that correspond to upper and lower limits (see bottom panel, 
right side). The median
 values and the K-S test suggest that abundances increase with the 
evolution of the clumps, as it is also found by \cite{Busquet11} in 
their chemical modeling. Median values of the column densities are given in 
Table~\ref{tbl-medians}. }
\label{n2h+_abundances}
\end{figure}

\begin{figure}
\begin{center}
\includegraphics[angle=0,scale=0.5]{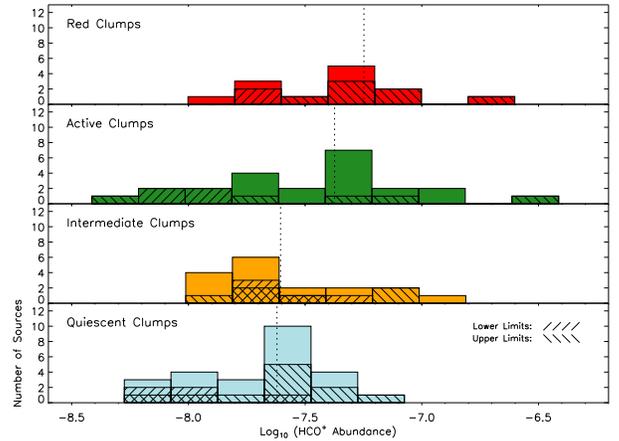}
\end{center}
\caption{Histograms of the number distributions of HCO$^+$ abundances   
for each evolutionary sequence. The name of the evolutionary stage is 
given on the top left corner of each panel. 
 The vertical dashed lines indicate the median values of the abundances 
 for each distribution. The diagonal patterns indicate the portions of the 
histograms that correspond to upper and lower limits (see bottom panel, 
right side). The median
 values and the K-S test suggest that abundances increase with the 
evolution of the clumps. Median values of the column densities are given in 
Table~\ref{tbl-medians}. }
\label{hco+_abundances}
\end{figure}

To estimate molecular abundances with respect to molecular hydrogen, 
we take the ratio between the column density of a given molecule and 
the H$_2$ column density derived from dust emission.  
Observations of 1.2 mm dust continuum emission were obtained by 
\cite{Rathborne06}  with an angular resolution of 11\arcsec. In order to 
calculate the molecular abundances, we smoothed the continuum 
emission data to the angular resolution of the molecular line data at 3 mm,
 38\arcsec. The resulting 1.2 mm peak flux, H$_2$ column densities and
 molecular abundances for the N$_2$H$^+$, HCO$^+$, HNC, C$_2$H, HC$_3$N,
 HNCO, and SiO lines (including limits for N$_2$H$^+$, HCO$^+$, and HNC)
 are listed in Table~\ref{tbl-mm-abun}. In general, the number distributions
 of abundances show few differences between 
the evolutionary sequences, with the exception of N$_2$H$^+$ and HCO$^+$ 
(Figure~\ref{n2h+_abundances}~and~\ref{hco+_abundances}). The K-S test 
gives a probability of 0.09\% and 1.2\% that quiescent and active distributions 
originate from the same parent populations for N$_2$H$^+$ and HCO$^+$, 
respectively. On the other hand, the evolutionary status of the clumps is not 
evidently distinguishable from the abundances of  HNC, C$_2$H,
 HC$_3$N, HNCO, and SiO, although their median values change 
with the evolution of the clumps. Median values of the molecular abundances
 for each evolutionary sequence (excluding the limits when the lines are
 not detected) are shown in Table~\ref{tbl-medians}. 

\cite{Vasyunina11} observed 37 IRDC clumps in the fourth quadrant
with the Mopra telescope. They also obtained molecular abundances for
 N$_2$H$^+$, HCO$^+$, HNC, SiO, C$_2$H, HC$_3$N, and HNCO. Despite 
 a couple of different assumptions (\cite{Vasyunina11} did not smooth
 their 1.2 mm 
dust continuum emission to the angular resolution of Mopra (38\arcsec) 
and they used NH$_3$ tempearatures), the abundances they find are in 
good agreement with the abundances determined in this work, except 
for HNC. They determined HNC abundances about an order 
of magnitud lower than our values. Because \cite{Vasyunina11}  did
not have the isotopologue HN$^{13}$C, they assumed that
 the emission was optically thin in order to get an estimate of the column
 density. However, as we discussed in Section~\ref{opacity}, the
 optically thin assumption is not valid for HNC in IRDCs because the
 emission is optically thick. 

\subsubsection{Abundance Ratios}
\label{abundance_ratios}

Molecular abundance ratios that can be used to estimate the age and mark 
the evolutionary stages of star-forming regions are known as ``chemical 
clocks.'' Only molecules that show differential abundances with time can 
be used to evaluate the evolutionary status of a star-forming region. 
Chemical clocks have been studied in depth in low-mass 
star-forming regions \citep[e.g.,][]{Emprechtinger09}; the concept 
 of chemical clocks has been less developed in the context of high-mass  
star-forming regions. Recent studies 
in small samples of high-mass star-forming regions show that chemical 
clocks can be extended to their high-mass counterpart 
\citep[e.g.,][]{Fontani11}. 
In this section, we explore the idea that N$_2$H$^+$/HCO$^+$ and 
N$_2$H$^+$/HNC ratios can be used as chemical clocks. We choose this set
 of lines because they reveal the clearest significant chemical variations.

\begin{deluxetable*}{lclccc}
\tabletypesize{\scriptsize}
\tablecaption{Parameters used for C$_2$H, HC$_3$N and HNCO Column
 Density Calculations \label{tbl-param-column2}}
\tablewidth{0pt}
\tablehead{
\colhead{Molecule} &  \colhead{$Q_{rot}$}  & \colhead{Rot. Const.} &
 \colhead{$g_{u}$}  & \colhead{$A_{ul}$} & \colhead{$E_{_{l}}/k$} \\
\colhead{} &  \colhead{} & \colhead{(GHz)} & \colhead{} &
 \colhead{($\times10^{-6}$ s$^{-1}$)} & \colhead{}\\
}
\startdata
C$_2$H &$kT_{\rm ex}/hB + 1/3$            &$B=43.674518^{(1)}$ &5.0 &1.52757&0.00216\\
HC$_3$N&$kT_{\rm ex}/hB + 1/3$             &$B=4.5490586^{(2)}$ &21.0&58.1300&19.6484\\
HNCO   &$[\pi(kT_{\rm ex})^3/(h^3ABC)]^{1/2}$&$A=918.417805^{(3)}$&9.0 &8.78011&6.32957\\
       &                                &$B=11.071010^{(3)}$ &    & &\\
       &                                &$C=10.910577^{(3)}$ &    & &\\
\enddata
\tablecomments{Columns are: species, partition function, rotational 
constant, statistical weight of the upper level, Einstein coefficient for 
spontaneous emission, and energy of the lower state.}
\tablecomments{References. (1) \cite{Padovani09}; (2) \cite{Thorwirth00}; (3)
 \cite{Lapinov07}. Values for $g_{\nu}$, $A_{ul}$ and $E_{_{l}}/k$ were obtained 
from The Cologne Database for Molecular Spectroscopy (CDMS)
 \citep{Muller01,Muller05}.}
\end{deluxetable*}

\vspace{0.2cm}
\hspace{2.3cm}{\it N$_2$H$^+$ and HCO$^+$}
\vspace{0.2cm}

It has been suggested that the line and abundance ratios of N$_2$H$^+$ and 
HCO$^+$ are anticorrelated, varying spatially over a given cloud and for 
different sources. 
The change in the N$_2$H$^+$/HCO$^+$ ratio was first noted 
by \cite{Turner77} and \cite{Snyder77}, and recently by \cite{Kim06} 
and \cite{Lo09}; however, it has not
 been studied in a large sample of star-forming clumps. 
Although we cannot examine spatial variations, due to our 
single-pointing observations, we can test whether the abundance ratios show 
differences for the four evolutionary stages proposed by \cite{Chambers09}. 

According to ion-molecule schemes of dense molecular clouds 
\cite[e.g.,][]{Snyder77,Jorgensen04}, N$_2$H$^+$ and HCO$^+$ are formed 
primary through reactions with H$_3^+$:  
\begin{equation}
{\rm H_3^+ +  N_2 \rightarrow  N_2H^+ + H_2}
\label{reac_N2H_form}
\end{equation} 
\begin{equation}
{\rm H_3^+ + CO \rightarrow HCO^+ + H_2 ~.}
\label{reac_HCO_form}
\end{equation} 

The main destruction mechanisms for N$_2$H$^+$ are electron recombination, 
when CO is frozen out on dust grains,
\begin{equation}
{\rm N_2H^+ + e \rightarrow N_2 + H ~,}
\label{reac_N2H_destruc1}
\end{equation} 
 and by reacting with CO,  when CO is found in the gas-phase at
 standard abundances (${\rm [CO/H_2]\sim 10^{-4}}$), 
\begin{equation}
{\rm N_2H^+ + CO \rightarrow HCO^+ + N_2 ~.}
\label{reac_N2H_destruc2}
\end{equation} 

 The dominant removal mechanism for HCO$^+$ is electron recombination
\begin{equation}
{\rm HCO^+ + e \rightarrow CO + H ~.}
\label{reac_HCO_destruc}
\end{equation} 
 
Chemical models of low-mass star-forming regions 
\cite[e.g.,][]{Lee04,Jorgensen04} and one of a massive star-forming 
region \citep{Kim06} show that the 
N$_2$H$^+$/HCO$^+$ ratio is high in early stages due to HCO$^+$ depletion  
(because most of the CO is frozen out) and is low in later stages when CO is
 evaporated from the dust grains ($T\sim$20-25 K). Because CO is both the main 
supplier of HCO$^+$ and the main destroyer of N$_2$H$^+$, the theory suggests 
that the HCO$^+$ abundance 
increases with respect to N$_2$H$^+$ as the clump evolves to a warmer phase. 
It is notable, for reasons not yet fully understood, that N$_2$H$^+$ is 
resistant to depletion in cold, dense regions. For a 
long time, it was thought that N$_2$, the main supplier of N$_2$H$^+$, 
had a lower binding strength to the 
grain surface than CO. However, recent measurements show that they are 
similar \citep{Oberg05}. Hence, the chemical network that dominates
 the formation and destruction of N$_2$H$^+$ and HCO$^+$ in 
early stages of star formation is given by 
Equations~\ref{reac_N2H_form},~\ref{reac_HCO_form},~\ref{reac_N2H_destruc1}~and~\ref{reac_HCO_destruc}, but for more evolved stages,  
Equation~\ref{reac_N2H_destruc1} becomes less important and is replaced by 
Equation~\ref{reac_N2H_destruc2}. A complete chemical network of H$_3^+$, 
 N$_2$H$^+$, and HCO$^+$, including formation and destruction rates, is 
presented in detail by \cite{Jorgensen04}. 

\begin{figure}
\begin{center}
\includegraphics[angle=0,scale=0.5]{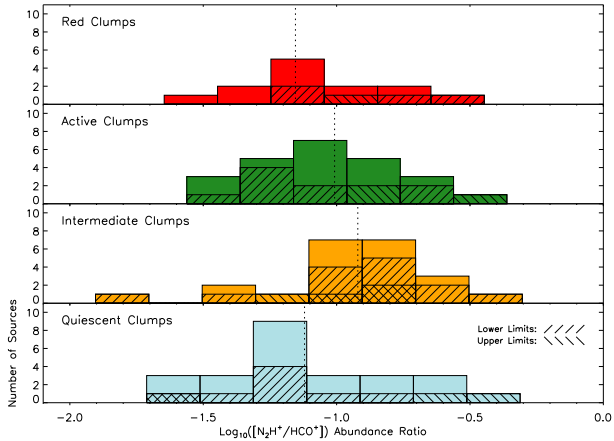}
\end{center}
\caption{ Histograms of the number distributions of N$_2$H$^+$/HCO$^+$
 abundance ratios for each evolutionary sequence. The name of the
 evolutionary stage is given on the top left corner of each panel.
 The vertical dashed lines indicate the median values of the abundance ratios  
 for each distribution. The diagonal patterns indicate the portions of the 
histograms that correspond to upper and lower limits (see bottom panel, 
right side). The median
 values and the K-S test suggest that N$_2$H$^+$/HCO$^+$ ratios act as a
 chemical clock, increasing their values with the evolution of the clumps, 
from intermediate, to active, to red clumps. It is not clear why this
 trend does not include quiescent clumps (see Section~\ref{abundance_ratios}).
 Median values of the abundance ratios are given in Table~\ref{tbl-medians}.}
\label{N2H_HCO_ratio}
\end{figure}

Figure~\ref{N2H_HCO_ratio} shows the number distributions of the
 N$_2$H$^+$/HCO$^+$ 
abundance (or column density) ratio for each evolutionary sequence. The 
median values for each distribution are 0.08, 0.12, 0.10, and 0.07 for 
quiescent, intermediate, active, and red clumps, respectively.
 As can be seen from the 
Figure~\ref{N2H_HCO_ratio}, quiescent clumps do not present the
 predicted abundance ratio to follow the trend  of decreasing
 N$_2$H$^+$/HCO$^+$ ratio with the evolution of the clumps; their
 ratios are more randomized. For this reason, 
 we use the K-S test for the intermediate and red clump distributions. 
The test shows that the probability of both populations being the 
same is low (3\%). The values of the median and the K-S test probability 
support the theoretical idea that the N$_2$H$^+$/HCO$^+$ abundance
 ratio acts as a chemical clock (at least for the
 intermediate, active, and red clumps); however, why this trend does not 
extend to quiescent clumps is unclear. We note that the relative abundance 
between N$_2$H$^+$ and HCO$^+$ acts as a 
chemical clock despite the abundances of both molecules independently 
increasing with the evolution of the clumps. We stress that a large number 
of values used in the histograms are limits and that a number of 
assumptions have gone into calculating the column densities and abundances.
A sample with a larger number of strong detections would be very helpful 
for confirming these trends.  

 The trend of decreasing N$_2$H$^+$/HCO$^+$ ratio with the 
evolution of the clumps may not extend to quiescent clumps due to one or 
a combination of the following reasons: 
(a) the sample of quiescent clumps could contain clumps with embedded star 
formation that was not detected by Spitzer, (b) CO has not
 had time to freeze out on dust grains, (c) the theoretical predictions
 at the very early stages are incorrect, and/or (d) due to the large beam of 
Mopra, we are probably detecting contaminating, diffuse emission which 
is not directly related to the star-forming process in the center of the 
clumps. Considering the last point, since the size of the Mopra beam is 
38\arcsec, we are typically observing emission on physical 
scales $\sim$0.8 pc. Typical sizes of clumps in high-mass star-forming regions 
and in IRDCs are $\sim$0.4 pc \citep{Faundez04} and $\sim$0.5 pc 
\citep{Rathborne06}, respectively. This shows that we are not only 
tracing the densest regions associated with star formation, but also the 
environment between clumps. \cite{Lee04} predict that the variation of
 the molecular abundances is much larger when we consider small scales
 inside the clumps. Further studies at higher angular resolution, using 
interferometers, are needed to understand the discrepancy shown by 
quiescent clumps. 

\vspace{0.2cm}
\hspace{2.3cm}{\it N$_2$H$^+$ and HNC}
\vspace{0.2cm}

Figure~\ref{N2H_HNC_ratio} shows the N$_2$H$^+$/HNC abundance ratio for
 each evolutionary stage. The median values for each distribution 
are 0.06, 0.06, 0.08, and 0.08 for quiescent, intermediate, active, and red 
clumps, respectively. The K-S test for the 
quiescent and active clump distributions shows that the probability of 
both populations being the same is low (5\%). The median values and 
the K-S test suggest that there is a trend of increasing N$_2$H$^+$/HNC 
abundance ratio with the evolution of the clumps. We note that N$_2$H$^+$ 
and HCO$^+$ are strongly linked because of their pathways of formation and 
destruction; on the other hand, a connection between N$_2$H$^+$ and HNC is 
less evident. The trend between these two molecules suggests that  
 HNC may be preferentially formed in cold gas. 

\vspace{0.2cm}
\hspace{2.3cm}{\it Other Abundance Ratios}
\vspace{0.2cm}

It is expected that abundances of SiO and more complex molecules, such 
as HC$_3$N and HNCO, increase in more evolved star-forming regions.
In evolved regions, these molecules or their parent molecules  
are released from the dust grains 
by shocks and sublimation due to the temperature increase. In fact, 
the abundance ratio between each one of them and N$_2$H$^+$, 
HCO$^+$, or HNC show higher ratios in later stages of evolution.
 Because the number of SiO, HC$_3$N and HNCO detections is low, 
the histograms with the number distributions of abundance ratios were not
 made. On the other hand, C$_2$H has a larger number of detections and
 shows no clear trends when abundance ratios are taken with respect to 
N$_2$H$^+$, HCO$^+$, or HNC. 

\subsection{Assumptions and limitations in calculating column densities and 
abundances}

This section discusses the major assumptions and 
limitations that affect the calculations of column densities and molecular 
abundances. Although the assumptions used in this work are widely used 
in the literature, it is important to be aware of their implications.  

{\bf Temperature.} Under the assumption of local thermodynamic 
equilibrium (LTE) conditions, 
we assume that the excitation temperature for all molecules is identical 
to the dust temperature for a given source. This assumption is reasonable 
since the critical densities of the molecules that were used to calculate
 physical parameters (between $2\times10^5$ - $2\times10^6$ cm$^{-3}$) are
 comparable to the density of the sources
 \cite[$\sim$10$^6$ cm$^{-3}$;][]{Rathborne10}, implying they are close to
 thermalization. However, the dust temperature is an average temperature 
along the line of sight, and may not necessarily reflect the temperature 
of the dust and gas in the portion of the clumps/cores which are dense 
enough to produce the emission lines we are observing. We note that the
 molecular abundance ratios 
(N$_2$H$^+$/HCO$^+$ and N$_2$H$^+$/HNC) are practically independent of 
temperature because of a similar dependence on temperature for the individual 
abundances. The abudance ratio for a given source at 15 and 50 K 
varies only $\sim$0.4\%. 

{\bf Isotopic abundance ratio.} In the calculation of optical depths and 
column densities for HCO$^+$ and HNC, we have assumed a 
constant value of 50 for the [$^{12}$C/$^{13}$C] isotopic abundance ratio for 
all sources,    
regardless of their different distances to the Galactic center.
 However, it has been shown that there is a gradient of this ratio
 increasing outwards from the Galactic center \citep{Savage02}. The
 [$^{12}$C/$^{13}$C] ratio has been measured using several methods and
 ranges from $\sim$20 to $\sim$70  \cite[and references therein]{Savage02}. 
The variation of the [$^{12}$C/$^{13}$C] ratio can give an error of a factor 
$\sim$2 in the column density estimation. 

{\bf Dust emission properties.} H$_2$ column densities were calculated 
using the smoothed 1.2 mm emission data from \cite{Rathborne06} and the same 
procedure that they used. The main assumptions are that the dust opacity 
per gram of gas, $\kappa_{1.2 \rm {mm}}$, is 1.0 cm$^2$ g$^{-1}$, and 
the gas-to-dust mass ratio is 100. The value for $\kappa_{1.2 \rm {mm}}$ was 
determined by \cite{Ossenkopf94} for protostellar cores assuming dust grains 
with thin ice mantles at gas densities of $10^6$ cm$^{-3}$. However, dust
 opacities and the gas-to-dust ratio are not well constrained. They 
could vary due to dust destruction in later stages of evolution (by shocks or 
UV radiation), or grain growth in dense, cold cores at very early 
stages. In the first scenario, the gas-to-dust ratio would be greater 
than 100; in the second scenario, the gas-to-dust ratio would be less 
than 100 and $\kappa_{1.2 \rm {mm}}$ would decrease. 

\vspace{-0.25cm}
\subsection{Molecules}
\label{molecules}

In this section, we discuss each commonly detected molecular line in depth. 
We describe the structure of the lines with hyperfine transitions (N$_2$H$^+$, 
HCN, and C$_2$H), a brief review about their formation and previous works,
 and their behavior in IRDC clumps. 
 
\begin{figure}
\begin{center}
\includegraphics[angle=0,scale=0.5]{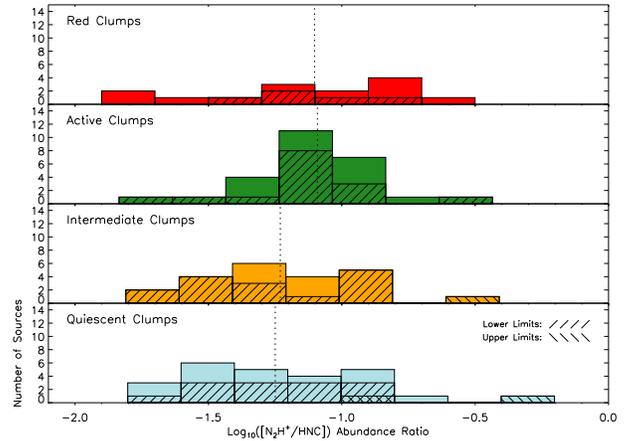}
\end{center}
\caption{Histograms of the number distributions of N$_2$H$^+$/HNC 
 abundance ratios for each evolutionary sequence. The name of the
 evolutionary stage is given on the top left corner of each panel.
 The vertical dashed lines indicate the median values of the abundance ratios  
 for each distribution. The diagonal patterns indicate the portions of the 
histograms that correspond to upper and lower limits (see bottom panel, 
right side). The median
 values and the K-S test suggest that the N$_2$H$^+$/HNC ratio may act as
 chemical clock, increasing their median values with the evolution of the
 clumps, from quiescent to red clumps. Median values of the abundance 
ratios are given in Table~\ref{tbl-medians}. }
\label{N2H_HNC_ratio}
\end{figure}
\vspace{-0.25cm}
\subsubsection{N$_2$H$^+$ (Diazenylium)} 
\label{Molecules_N2H+}
The N$_2$H$^+$ $J=1\ra0$ rotational transition exhibits a quadrupole 
hyperfine structure due to the spin of the two nitrogen nuclei. 
Although for N$_2$H$^+$ $J=1\ra0$ there are 15 allowed hyperfine 
transitions \citep{Daniel06}, only seven features can be resolved in 
low-mass star forming regions \cite[e.g.,][]{Caselli95,Crapsi05} and three in 
high-mass star forming regions \citep[e.g.,][]{Foster11}.
 In the latter case, because massive star-forming regions show turbulent 
 line widths ($>$3 \kms) much broader than thermal widths ($\sim$0.3 \kms),
 two triplets are blended and observed as one line. The relative 
intensities in the optically thin case are given by the statistical 
weights 1:3:5:7, when seven components are detected, and 1:3:5, when three 
components are detected. The relative intensities become one in the 
optically thick case. 

N$_2$H$^+$ is a good tracer of dense, cold gas because it does not 
deplete at low temperatures and high densities. 
Because the N$_2$H$^+$ spectrum show no evidence of self-absorbed profiles
 or line wings indicating outflow activity, this molecule is also a good
 tracer of dense, warm gas. However, 
some clumps have broad lines, which may suggest they are formed by more than 
one central object. Indeed, high angular resolution observations made with 
interferometers show the clumpy nature of IRDCs 
\cite[e.g.,][]{Rathborne07,Rathborne08,Zhang09}. N$_2$H$^+$ is the second  
 most frequently detected line in this survey and its detection rates do not 
depend of the star formation activity. However, the brightness temperature 
of N$_2$H$^+$ is normally weak in regions without signs of star formation. 
N$_2$H$^+$ column densities and abundances increase with the evolution
of the clumps. Although, in general, an increase of column density does not 
necessary mean an increase of the molecular abundance, we see in our 
sample of objects that column density and abundance of N$_2$H$^+$ tend to 
increase together. The increase of column density with the evolution 
of the clumps might be explained by the accretion of material. For
 instance,  \cite{Chambers09} showed that, on average, clumps with signs
 of star formation have smaller sizes and higher densities than clumps
 without apparent star formation and \cite{Zhang11} propose that embedded
 protostars and protostellar cores undergo simultaneous mass growth during
 the protostellar evolution. The increase in the abundance might be 
explained by the rise in the temperature which releases N$_2$ from the
 dust grains to the gas phase allowing the formation of more N$_2$H$^+$
 with the passage of time. However, with the increase of temperature, 
CO should also be released from the dust grains into the gas phase and 
should efficiently destroy N$_2$H$^+$ \citep{Lee04,Jorgensen04,Busquet11}. 
Why exactly N$_2$H$^+$ abundance increases with evolution is not clear. 
We suggest two possible explanations. First, the rates involving the 
formation and destruction of N$_2$, CO, N$_2$H$^+$, and HCO$^+$ might not 
be accurate, and N$_2$H$^+$ might not be destroyed as 
efficiently as is currently believed. Second, the large beam of Mopra 
is including emission from cold gas which is sorrounding the most 
compact regions directly associated with the star formation processes in 
the center of the clumps. If most of the N$_2$H$^+$ emission is 
coming from a cold envelope, in clumps that show signs of star 
formation, we are using in the abundance calculation a higher temperature 
 than the actual value. Higher temperatures would lead us to infer  
larger N$_2$H$^+$ abundances. \cite{Busquet11} performed a
 time-dependent chemical modeling of the massive protostellar cluster
 AFGL5142 based on the model of \cite{Viti04}. Their model 
consists of a prestellar and a protostellar phase. From their
 Figure~9, it can be seen that once temperature increases, from the
 prestellar to the protostellar phase, the N$_2$H$^+$ molecular
 abundance also increases, supporting what we see in our work. 
\vspace{-0.25cm}
\subsubsection{HCO$^+$ and H$^{13}$CO$^+$ (Formylium)}

HCO$^+$ has been widely used to investigate infall motions 
\cite[e.g.,][]{Fuller05,Chen10}, and, occasionally, 
its spectrum shows high-velocity wing emission which indicates outflow 
activity \cite[e.g.,][]{Cyganowski11}. This makes HCO$^+$ a good tracer of 
kinematics in star-forming regions. On the other hand, H$^{13}$CO$^+$ 
is optically thin and normally shows a Gaussian profile. The combination 
of these two lines is usually used to distinguish between an asymmetric
 ``blue profile'' 
(indicative of infall motions) and two velocity components in the line of 
sight. If the optically thin H$^{13}$CO$^+$ peaks in the self-absorption of the 
optically thick HCO$^+$, it indicates a genuine ``blue profile.'' 
In our sample of IRDC 
clumps, three intermediate clumps (G024.33 MM2, G024.33 MM5, and G031.97 MM5) 
present this characteristic profile associated with large-scale infall motions. 
As also happpens for N$_2$H$^+$, HCO$^+$ column densities and abundances
 rise with the evolution of the clumps. As we mentioned before for N$_2$H$^+$, 
 the increase in HCO$^+$ column densities might be due to the physical 
evolution of the clumps.  The rise in HCO$^+$ abundance can be plausibly
 explained by the increase in the temperature, which 
 releases CO from grain mantles into the gas phase, forming more HCO$^+$ with 
the passage of time. For example, although \cite{Busquet11} focused on the
 NH$_3$/N$_2$H$^+$ abundance ratio, they also obtained HCO$^+$ abundances
 from their chemical modeling. They showed that once temperature rises,
 from the prestellar to the protostellar phase, the HCO$^+$ molecular
 abundance rises, as well as that of N$_2$H$^+$. 

\begin{deluxetable*}{lccccc}
\tabletypesize{\scriptsize}
\tablecaption{Median Values of Derived Clump Properties for each Evolutionary Stage\label{tbl-medians}}
\tablewidth{0pt}
\tablehead{
\colhead{Molecule} & \colhead{Quiescent}  & \colhead{Intermediate} & \colhead{Active} &  \colhead{Red}  &   \colhead{All} \\
\colhead{} & \colhead{Clumps}  & \colhead{Clumps}
 & \colhead{Clumps} & \colhead{Clumps}  & \colhead{Clumps} \\
}
\startdata
          \multicolumn{6}{c}{Optical Depths ($\tau$)}     \\
\hline
\HNC     & 24  & 22  &18  &16  &   19 \\
\NdosH\tablenotemark{a}   & 0.8 &0.8 &0.5 &1.8 &    0.8\\
\HCO     & 30  & 20  &16  &19  &   21 \\
\CdosH\tablenotemark{b}   & 2.9 &2.4 &2.2 &1.4 &    2.4\\
\hline
          \multicolumn{6}{c}{Column Densities (cm$^{-2}$)}     \\
\hline
\HNC     &$1.84 \times 10^{14}$ & $2.50 \times 10^{14}$ & $3.39 \times 10^{14}$ & $2.45 \times 10^{14}$ & $2.42 \times 10^{14}$\\
\NdosH   & $1.03 \times 10^{13}$ & $1.40 \times 10^{13}$ & $2.34 \times 10^{13}$ &  $2.75 \times 10^{13}$ &  $1.60 \times 10^{13}$\\
\HCO     &$1.40 \times 10^{14}$ & $1.51 \times 10^{14}$ &  $2.51 \times 10^{14}$ &  $3.20 \times 10^{14}$ & $1.88 \times 10^{14}$\\
\CdosH   & $2.02 \times 10^{14}$ & $2.40 \times 10^{14}$ & $2.66 \times 10^{14}$ &    $2.68 \times 10^{14}$ &    $2.41 \times 10^{14}$\\
\HCtresN & $3.05 \times 10^{12}$ &  $4.10 \times 10^{12}$ & $5.47 \times 10^{12}$ &    $7.97 \times 10^{12}$ &    $4.77 \times 10^{12}$\\
\HNCO    &$1.33 \times 10^{13}$ & $3.36 \times 10^{13}$ &  $4.30 \times 10^{13}$ &    $4.31 \times 10^{13}$ &    $3.36 \times 10^{13}$\\
\SiO     & $1.36 \times 10^{12}$ & ...     & $1.39 \times 10^{13}$ &     $4.62 \times 10^{12}$ &    $7.72 \times 10^{12}$\\
\hline
          \multicolumn{6}{c}{Molecular Abundances}  \\
\hline
\HNC     & $3.52 \times 10^{-8}$ & $4.41 \times 10^{-8}$ &  $3.29 \times 10^{-8}$ &    $6.11 \times 10^{-8}$ &    $3.73 \times 10^{-8}$\\
\NdosH   & $1.65 \times 10^{-9}$ & $2.73 \times 10^{-9}$ &  $3.13 \times 10^{-9}$ &    $3.73 \times 10^{-9}$ &    $2.40 \times 10^{-9}$\\
\HCO     & $2.40 \times 10^{-8}$ & $2.48 \times 10^{-8}$ & $4.23 \times 10^{-8}$ &     $5.65 \times 10^{-8}$ &    $2.51 \times 10^{-8}$\\
\CdosH   & $3.65 \times 10^{-8}$ & $3.45 \times 10^{-8}$ &  $4.26 \times 10^{-8}$ &   $4.92 \times 10^{-8}$ &    $3.72 \times 10^{-8}$\\
\HCtresN & $3.14 \times 10^{-10}$ &$2.34 \times 10^{-10}$ & $4.97 \times 10^{-10}$ &    $4.23 \times 10^{-10}$ &    $4.23 \times 10^{-10}$\\
\HNCO    & $1.83 \times 10^{-9}$ &$3.21 \times 10^{-9}$ &$2.51 \times 10^{-9}$ &      $4.00 \times 10^{-9}$ &    $2.80 \times 10^{-9}$\\
\SiO     & $2.78 \times 10^{-10}$ & ...      &$5.95 \times 10^{-10}$ &    $4.37 \times 10^{-10}$ &    $5.79 \times 10^{-10}$\\
\hline
          \multicolumn{6}{c}{Abundance Ratios}  \\
\hline
N$_2$H$^+$/HCO$^+$ & 0.08 & 0.12 & 0.10 &  0.07 &    0.09 \\
N$_2$H$^+$/HNC     & 0.06 &0.06 & 0.08 &   0.08 &    0.07 \\
\enddata
\tablenotetext{a}{Optical depth for the brightest transition $JF_1F=123\ra012$.}
\tablenotetext{b}{Optical depth for the brightest transition $NJF=1\,\frac{3}{2}\,2\ra0\,\frac{1}{2}\,1$.}
\end{deluxetable*}

\clearpage

\begin{deluxetable*}{lccccccc}
\tabletypesize{\scriptsize}
\tablecaption{Column Density Comparison with Other Works.
 \label{tbl-comparison}}
\tablewidth{0pt}
\tablehead{
\colhead{Different Works} &  \colhead{N$_2$H$^+$}  & \colhead{HCO$^+$} &
 \colhead{HNC}  & \colhead{C$_2$H} & \colhead{HC$_3$N} & \colhead{HNCO} 
& \colhead{SiO}\\
\colhead{} &  \colhead{($\times10^{13}$)} & \colhead{($\times10^{14}$)} &
 \colhead{($\times10^{14}$)} & \colhead{($\times10^{14}$)} &
 \colhead{($\times10^{12}$)} & \colhead{($\times10^{13}$)} &
 \colhead{($\times10^{13}$)} \\
\colhead{} &  \colhead{(cm$^{-2}$)} & \colhead{(cm$^{-2}$)} &
 \colhead{(cm$^{-2}$)} & \colhead{(cm$^{-2}$)} &
 \colhead{(cm$^{-2}$)} & \colhead{(cm$^{-2}$)} &
 \colhead{(cm$^{-2}$)} \\
}
\startdata
This Work         & Q: 1.03 & A: 2.51 & A: 3.39 & A: 2.66 & Q: 3.05 & A: 4.30 & A: 1.39 \\
                  & A: 2.34 & R: 3.20 &   ...   &   ...   &   ...   & R: 4.31 & R: 0.46 \\
\hline
\cite{Sakai08}    & Q: 1.15 &   ...   &   ...   &   ...   & Q: 9.70 &   ...   &  ...    \\
\cite{Pirogov03}  & A: 2.78 &   ...   &   ...   &   ...   &  ...    &   ...   &  ...    \\
\cite{Sakai10}    &   ...   & A: 3.30 & A: 3.60 & A: 3.25 &  ...    &   ...   & A: 1.46 \\
\cite{Purcell06}  &   ...   & R: 11.7 &   ...   &   ...   &  ...    &   ...   &   ...   \\
\cite{Zinchenko00}&   ...   &   ...   &   ...   &   ...   &  ...    & A-R: 8.61 &  ...   \\
\cite{Miettinen06}&   ...   &   ...   &   ...   &   ...   &  ...    &   ...   & R: 2.00 \\
\enddata
\tablecomments{Q = Quiescent; A = Active; R = Red. \\ 
\hspace{1.4cm} This table displays median values for comparison.}
\end{deluxetable*}

\subsubsection{HNC and HN$^{13}$C (Hydrogen Isocyanide)}

HNC is the molecule with the highest detection rate in the sample and 
its presence is independent of the star formation activity. 
Because HNC is ubiquitous in the IRDC clumps, and its spectrum shows  
no evidence of line wings indicating outflows and only a few 
  self-absorbed profiles, HNC seems to be a good tracer of 
cold and warm gas. The HNC emission lines 
show, in general, broad widths (as seen in the HCO$^+$ lines as well,  
Figure~\ref{FWHM-plot}). This is probably because both lines are 
optically thick, and $\Delta$V can be broader by opacity. On the other hand,
HN$^{13}$C is optically thin and normally shows a Gaussian profile.
We note that the assumption of HNC emission being optically thin produces  
column densities of at least one order of magnitude lower. There is a 
slight increase of HNC column densities with the evolution of the clumps. 
However, there are no differences in the HNC abundances for the different 
evolutionary states. The accretion of 
material in the star-forming clumps could explain the small rise in 
the HNC column densities. From the histogram with the 
relative abundances of N$_2$H$^+$ and HNC (Figure~\ref{N2H_HNC_ratio}), it 
can be inferred that HNC may be preferentially formed in cold gas.
 Future chemical modeling is needed to clarify the behavior of HNC in
 high-mass star-forming regions. 

\subsubsection{HCN (Hydrogen Cyanide)}

The HCN $J=1\ra0$ rotational transitions show three hyperfine components 
caused by the nuclear spin of the nitrogen nucleus. In the optically 
thin case, these three components have relative intensities of 1:3:5. 
However, due to the broad linewithds found in massive star-forming regions, 
the components are blended. In addition, several sources exhibit spectra 
with extended wing emission and a wide variety of relative intensities
 other than those in the optically thin limit. The combination of these 
three factors makes it difficult to perform Gaussian fits and analyze the HCN 
line towards the IRDC clumps in this survey. For a better understanding of 
HCN hyperfine transitions, see the recent work of \cite{Loughnane}. 

\subsubsection{C$_2$H (Ethynyl)}

C$_2$H $N=1\ra0$ exhibits 6 fine and hyperfine components due to 
the presence of both an electron spin and a nuclear spin of the hydrogen 
nucleus. The spectral separation makes all of them observable in star-forming 
regions. However, two of them are relatively very weak and were not detected 
in this survey. Their relative intensities in the optically thin case are 
given by \cite{Tucker74} and \cite{Padovani09}.  

C$_2$H has been known to be a PDR tracer \cite[e.g.,][]{Fuente93}. However,
\cite{Beuther08} find that this molecule also seems to trace dense gas 
in early stages of star formation. They found that the distribution
 of C$_2$H shows a hole 
around a hot core, and suggest that C$_2$H decreases in the hot core 
phase. Based on the results of their simple chemical modeling, they suggest 
that C$_2$H may be a suitable tracer of early stages of star formation. 

So far, C$_2$H has not been systematically studied in massive star-forming 
regions or IRDCs. Although we cannot know the spatial distribution
 of C$_2$H in the IRDC clumps, we can say that there is definitely an
 increase in C$_2$H detections from quiescent to active sources (see 
Figure~\ref{detection_rates_per_category}). The $\sim$30\%  
detection rate of this line in quiescent clumps shows that its emission is not 
ubiquitous in these kinds of sources as, for example,  N$_2$H$^+$, HNC, and 
HCO$^+$ are. Certainly, higher angular resolution observations and 
mapping are needed to clarify if the C$_2$H emission comes from the external 
layers of clumps that can interact with PDR emission surrounding them, or 
if the C$_2$H emission comes from the dense, cold gas inside quiescent clumps. 
We note that C$_2$H  lines show the best $\Delta$V correlation with N$_2$H$^+$, 
suggesting that the emission originates from the same region 
(see Figure~\ref{FWHM-plot}). C$_2$H column densities and abundances show 
no clear trend of changing with evolution. There is marginal evidence for  
increasing column densities with the evolution of the clumps, but it is not 
supported by the K-S test. 

\subsubsection{HC$_3$N (Cyanoacetylene)} 

HC$_3$N is the simplest of the cyanopolyynes, molecules in the form of 
HC$_{2n+1}$N with n from 1 to 5. Its main progenitor, C$_2$H$_2$,
 exists on grains mantles and is released into the gas phase during
 the onset of heating. Thus, HC$_3$N 
is associated with warm, dense gas in regions with current
 star formation, such as hot cores \cite[e.g.,][]{Chapman09}.
This may explain the high detection rates ($\sim$50\%) for this molecule in
active clumps. The presence of HC$_3$N and the rich molecular line spectra 
found in active cores (see Figure~\ref{number_lines}), suggest
 that some active clumps may 
have embedded hot molecular cores, as, for example, \cite{Rathborne08} 
found in one IRDC clump using high angular resolution observations.
The HC$_3$N linewidths are more narrow than N$_2$H$^+$ linewidths, and 
seem to correlate better with those of H$^{13}$CO$^+$ and HN$^{13}$C 
(see Figure~\ref{FWHM-plot}). HC$_3$N column densities vary with the 
evolution of the clumps, increasing in more evolved regions as is shown 
by the median values. For HC$_3$N abundances, no real change is clear. 

\LongTables
\begin{deluxetable*}{lccccccccc}
\tabletypesize{\scriptsize}
\tablecaption{Continuum Emission Parameters and Molecular Abundances \label{tbl-mm-abun}}
\tablewidth{0pt}
\tablehead{
\colhead{IRDC Clump}  & \colhead{1.2 mm} & \colhead{N(H$_2$)} & \multicolumn{7}{c}{\underline {~~~~~~~~~~~~~~~~~~~~~~~~~~~~~~~~~~~~~~~~~~~~~~Molecular Abundance~~~~~~~~~~~~~~~~~~~~~~~~~~~~~~~~~~~~~~~~~~~~~~~~}}\\
\colhead{}              &\colhead{Flux}&\colhead{$\times10^{22}$}&
\colhead{N$_2$H$^+$}    & \colhead{HCO$^+$}       & \colhead{HNC}             &
 \colhead{C$_2$H}       & \colhead{HC$_3$N}       & \colhead{HNCO}          
 & \colhead{SiO}\\
\colhead{}               & \colhead{(mJy)}                & \colhead{(cm$^{-2}$)}    & 
\colhead{$\times10^{-9}$}& \colhead{$\times10^{-8}$}&\colhead{$\times10^{-8}$}
 &\colhead{$\times10^{-8}$}& \colhead{$\times10^{-10}$}&\colhead{$\times10^{-9}$}&
\colhead{$\times10^{-10}$}\\
}
\startdata
G015.05 MM1 &  45 &  0.86 &  1.82(0.15) &  2.39(0.85) &  2.64(1.27) &  3.20(3.52) &     ...     &  1.94(0.81) &     ...     \\
G015.05 MM2 &  25 &  0.35 &  2.05(0.45) &  1.69(0.60) &  7.70(5.53) &     ...     &     ...     &     ...     &     ...     \\
G015.05 MM3 &  18 &  0.33 &  1.08(0.31) &     ...     &  3.12(1.52) &     ...     &     ...     &     ...     &     ...     \\
G015.05 MM4 &  14 &  0.18 &  5.51(0.63) &  3.09(1.07) &  8.26(3.43) & 16.20(17.41) &     ...     &     ...     &     ...     \\
G015.05 MM5 &  24 &  0.32 &  1.63(0.51) &  3.16(1.74) &  7.50(4.84) &     ...     &     ...     &     ...     &     ...     \\
G015.31 MM2 &  22 &  0.43 &  2.73(2.89) &     ...     &  5.05(4.12) &     ...     &     ...     &     ...     &     ...     \\
G015.31 MM3 &  17 &  0.32 &  0.96(0.25) &  2.65(1.90) &  1.58(1.61) &     ...     &     ...     &     ...     &     ...     \\
G015.31 MM5 &  22 &  0.34 &     ...     &  3.22(1.83) &  3.98(2.79) &     ...     &     ...     &     ...     &     ...     \\
G018.82 MM2 &  77 &  0.82 &  2.74(2.75) &  3.91(2.38) &  4.21(2.55) &     ...     &     ...     &     ...     &     ...     \\
G018.82 MM3 &  30 &  0.36 & 16.75(16.99) & 27.86(14.46) & 18.52(8.67) &     ...     &     ...     &     ...     &     ...     \\
G018.82 MM4 &  43 &  1.26 &  0.85(0.14) &  6.73(5.33) &  3.98(2.25) &     ...     &     ...     &  0.77(0.83) &     ...     \\
G018.82 MM6 &  46 &  0.88 &  1.39(0.14) &  0.60(0.21) &  0.91(0.40) &     ...     &     ...     &     ...     &     ...     \\
G019.27 MM2 &  46 &  0.65 &  3.36(0.22) &     ...     &  6.30(2.15) &  6.40(7.08) &     ...     &     ...     &     ...     \\
G022.35 MM1 &  93 &  2.19 &  0.19(0.05) &  0.39(0.15) &  0.51(0.21) &     ...     &     ...     &     ...     &     ...     \\
G022.35 MM2 &  32 &  0.29 &     ...     &     ...     &  6.94(3.70) &     ...     &     ...     &     ...     &     ...     \\
G023.60 MM7 &  49 &  0.43 &  9.45(9.31) &     ...     &  8.00(3.82) &     ...     &     ...     &     ...     &     ...     \\
G023.60 MM9 &  32 &  0.41 &  1.67(0.27) &     ...     &  3.58(1.72) &  3.68(5.28) &     ...     &     ...     &     ...     \\
G024.08 MM2 &  48 &  0.72 &  0.98(0.16) &  2.49(1.01) &  3.52(1.57) &     ...     &     ...     &     ...     &     ...     \\
G024.08 MM3 &  33 &  0.48 &  1.94(1.92) &  3.72(1.85) &  3.98(1.77) &     ...     &     ...     &     ...     &     ...     \\
G024.08 MM4 &  45 &  0.63 &  0.76(0.21) &  3.86(3.50) &  2.92(1.49) &     ...     &     ...     &     ...     &     ...     \\
G024.33 MM2 &  44 &  0.56 &  5.44(0.32) &  2.43(0.63) & 10.94(2.07) &     ...     &     ...     &  7.58(1.95) &     ...     \\
G024.33 MM3 &  53 &  0.70 &  3.77(0.24) &  1.97(0.50) &  1.22(0.41) &  4.85(6.65) &     ...     &  4.01(1.28) &  5.64(2.02) \\
G024.33 MM4 &  53 &  0.72 &  2.19(0.19) &  1.10(0.42) &  1.10(0.40) &     ...     &     ...     &     ...     &     ...     \\
G024.33 MM5 &  51 &  0.67 &  3.24(0.22) &  4.18(0.68) &  1.18(0.42) &  3.92(5.95) &     ...     &  8.10(1.39) &     ...     \\
G024.33 MM7 &  26 &  0.49 &  1.89(0.17) &  2.49(0.47) &  2.15(0.78) &  1.24(2.46) &     ...     &     ...     &  2.78(1.51) \\
G024.33 MM8 &  42 &  0.55 &     ...     &  2.50(0.95) &  4.69(1.80) &     ...     &     ...     &     ...     &     ...     \\
G024.33 MM9 &  43 &  0.31 &     ...     &  7.24(2.81) & 10.96(4.29) &     ...     &     ...     &     ...     &     ...     \\
G024.33 MM11 &  32 &  0.55 &  2.76(0.17) &  5.16(1.14) &  7.34(2.73) &     ...     &     ...     &     ...     &     ...     \\
G024.60 MM2 &  78 &  1.54 &  0.81(0.06) &  0.98(0.32) &  0.96(0.41) &  0.75(0.26) &  2.25(0.80) &     ...     &     ...     \\
G025.04 MM2 &  59 &  0.91 &  0.68(0.12) &  1.39(0.56) &  1.87(0.80) &     ...     &     ...     &     ...     &     ...     \\
G025.04 MM4 &  62 &  0.89 &  2.74(0.12) &  1.69(0.61) &  4.82(1.13) &  1.22(2.49) &     ...     &  3.50(0.73) &     ...     \\
G027.75 MM2 &  17 &  0.32 &  3.12(3.15) &     ...     &     ...     &     ...     &     ...     &     ...     &     ...     \\
G027.94 MM1 &  41 &  0.37 &  4.44(4.46) &     ...     &  5.28(1.83) &  5.19(8.39) &     ...     &     ...     &     ...     \\
G028.04 MM1 &  44 &  0.47 &  4.87(4.79) &  4.80(2.04) & 11.79(5.69) &  2.73(4.87) &  8.07(2.62) &  6.17(1.80) &     ...     \\
G028.08 MM1 &  29 &  0.61 &  1.03(1.02) &     ...     &  3.14(1.66) &     ...     &     ...     &     ...     &     ...     \\
G028.23 MM1 &  59 &  1.23 &  2.54(2.55) &  0.81(0.27) &  0.62(0.22) &     ...     &     ...     &     ...     &     ...     \\
G028.28 MM4 &  43 &  0.51 &  2.96(0.26) &  3.55(0.73) &  3.20(1.17) &  5.19(5.99) &     ...     &     ...     &     ...     \\
G028.37 MM1 & 318 &  4.02 &  1.14(0.04) &  0.62(0.10) &  1.42(0.24) &  1.07(0.60) &  3.60(0.36) &  2.48(0.22) &  3.98(0.34) \\
G028.37 MM2 & 118 &  1.23 &  2.04(0.15) &  1.97(0.36) &  5.18(1.43) &  3.70(5.25) &  3.96(1.73) &  2.94(1.22) &     ...     \\
G028.37 MM4 & 115 &  1.51 &  3.10(0.12) &  1.92(0.20) &  2.89(0.31) &  4.17(3.73) &  5.65(0.71) &  5.33(0.42) &  9.20(0.65) \\
G028.37 MM6 &  87 &  1.71 &  1.56(0.06) &  1.59(0.30) &  1.68(0.27) &  1.06(1.38) &  4.48(0.71) &  2.51(0.30) &  5.96(0.52) \\
G028.37 MM9 &  67 &  1.21 &  1.26(0.09) &  1.64(0.29) &  1.26(0.24) &     ...     &     ...     &     ...     &     ...     \\
G028.37 MM11 &  53 &  0.60 &  1.74(0.34) &  1.36(0.47) & 11.93(6.30) &  4.36(6.98) &     ...     &     ...     &     ...     \\
G028.37 MM12 &  51 &  0.97 &  0.92(0.93) &  2.49(1.42) &  3.71(2.94) &     ...     &  3.14(1.58) &     ...     &     ...     \\
G028.37 MM13 &  20 &  0.28 &  5.16(5.59) &     ...     &     ...     &     ...     &     ...     &     ...     &     ...     \\
G028.53 MM3 &  85 &  1.78 &  1.39(1.37) &  1.19(0.36) &  1.34(0.58) &     ...     &     ...     &     ...     &     ...     \\
G028.53 MM5 &  44 &  0.63 &  4.63(4.83) &  2.39(1.62) &  5.48(3.21) &     ...     &     ...     &     ...     &     ...     \\
G028.53 MM7 &  44 &  0.86 &  1.28(0.14) &  1.06(0.31) &  4.03(1.41) &     ...     &     ...     &     ...     &     ...     \\
G028.53 MM8 &  35 &  0.54 &  1.84(0.26) &  6.53(5.13) &  6.35(3.27) &  3.73(2.12) &     ...     &     ...     &     ...     \\
G028.53 MM9 &  61 &  0.94 &  2.00(0.16) &  2.53(1.01) &  3.27(1.07) &     ...     &     ...     &     ...     &     ...     \\
G028.53 MM10 &  90 &  1.28 &  1.59(0.15) &  2.22(0.85) &  3.83(1.14) &  3.96(5.40) &     ...     &     ...     &     ...     \\
G028.67 MM1 &  27 &  0.27 &     ...     &  6.07(3.02) & 10.93(5.20) &     ...     &     ...     &     ...     &     ...     \\
G028.67 MM2 &  38 &  0.76 &  0.95(0.14) &     ...     &  3.51(2.09) &     ...     &     ...     &     ...     &     ...     \\
G030.14 MM1 &  40 &  0.64 &  2.12(2.33) &  1.83(1.05) &  2.50(1.28) &     ...     &     ...     &     ...     &     ...     \\
G030.57 MM1 & 117 &  1.40 &  3.16(0.11) &     ...     &  2.75(0.87) &  1.89(2.95) &  6.10(1.11) &  2.04(0.70) &     ...     \\
G030.57 MM3 &  17 &  0.34 &  2.92(0.55) &     ...     & 18.84(10.70) &  7.12(10.94) &     ...     &     ...     &     ...     \\
G030.97 MM1 &  98 &  1.08 &  3.38(0.14) &  4.80(0.46) &  3.75(0.62) &     ...     &  3.43(1.13) &     ...     &     ...     \\
G031.97 MM5 &  29 &  0.35 &  6.77(0.39) &  3.63(0.85) & 10.27(3.34) &  8.52(11.04) &     ...     &     ...     &     ...     \\
G031.97 MM7 &  36 &  0.59 &  2.89(0.16) &  2.05(0.40) &  5.13(4.79) &  3.66(5.80) &     ...     &  1.70(0.96) &     ...     \\
G031.97 MM8 &  38 &  0.31 & 10.84(0.62) &     ...     & 10.76(3.73) &     ...     &     ...     &     ...     &     ...     \\
G033.69 MM1 &  97 &  1.71 &  2.08(0.06) &  1.00(0.15) &  1.72(0.20) &     ...     &  5.61(0.75) &  2.66(0.35) &  3.09(0.60) \\
G033.69 MM2 &  75 &  0.73 &  4.34(0.29) &  6.17(2.51) &  6.11(1.56) &  4.84(0.67) & 13.77(2.24) &  5.87(1.66) &     ...     \\
G033.69 MM3 &  52 &  0.48 &  4.97(0.58) &  7.48(3.12) & 10.73(4.33) &     ...     &     ...     &     ...     &     ...     \\
G033.69 MM4 &  65 &  1.00 &  3.06(0.21) &  1.77(0.65) &  3.38(0.83) &     ...     &  5.59(1.97) &  3.09(1.02) &     ...     \\
G033.69 MM5 &  44 &  0.49 &  4.42(0.50) &  7.90(3.93) &  8.69(3.61) &  7.70(6.26) &     ...     &     ...     &     ...     \\
G033.69 MM11 &  37 &  0.65 &     ...     &  3.88(1.46) &  6.14(2.33) &  2.28(4.42) &     ...     &     ...     &     ...     \\
G034.43 MM1 & 548 &  5.85 &  2.34(0.04) &  0.82(0.09) &  2.33(0.27) &  1.32(0.81) &  5.70(0.36) &  1.47(0.25) &  5.94(0.36) \\
G034.43 MM5 &  89 &  1.75 &  1.70(0.05) &  1.10(0.15) &  2.99(0.45) &     ...     &  3.07(0.71) &     ...     &     ...     \\
G034.43 MM7 &  59 &  0.85 &  1.31(0.12) &  1.26(0.39) &  2.69(0.70) &  2.41(3.42) &  2.43(1.38) &     ...     &     ...     \\
G034.43 MM8 &  69 &  0.64 &  2.39(0.26) &  4.23(1.89) &  3.98(1.58) &  5.56(9.33) &     ...     &     ...     &     ...     \\
G034.77 MM1 & 111 &  1.11 &  3.68(3.65) &  4.08(0.71) &  2.20(0.72) &     ...     &  4.21(1.17) &     ...     &     ...     \\
G034.77 MM3 &  18 &  0.25 &  3.55(0.60) &  2.03(0.71) & 11.20(4.72) & 12.12(20.78) &     ...     &     ...     &     ...     \\
G035.39 MM7 &  50 &  0.61 &  3.92(0.27) &  9.77(1.54) &  6.21(1.34) &  7.13(6.73) &  5.27(2.41) &     ...     &     ...     \\
G035.59 MM1 &  22 &  0.25 & 11.25(12.28) & 17.36(8.70) &  7.02(4.36) &     ...     &     ...     &     ...     &     ...     \\
G035.59 MM2 &  19 &  0.35 &  2.03(0.29) &  5.09(1.59) &  3.05(1.14) &     ...     &     ...     &     ...     &     ...     \\
G035.59 MM3 &  30 &  0.42 &  2.40(0.55) &  2.69(0.70) &  3.00(1.13) &  4.37(5.26) &  3.32(2.99) &     ...     &     ...     \\
G036.67 MM1 &  38 &  0.90 &  1.10(1.13) &  1.75(1.49) &  1.13(0.66) &  1.75(1.74) &     ...     &     ...     &     ...     \\
G036.67 MM2 &  38 &  1.22 &     ...     &  0.54(0.24) &  0.95(0.39) &     ...     &     ...     &     ...     &     ...     \\
G038.95 MM1 & 104 &  3.33 &  0.40(0.02) &  1.28(0.20) &  0.51(0.15) &  0.72(0.58) &  1.41(0.58) &  0.27(0.16) &     ...     \\
G038.95 MM2 &  82 &  0.82 &  2.10(0.25) &  5.66(1.29) &  2.99(1.12) &     ...     &  4.23(2.10) &     ...     &     ...     \\
G038.95 MM3 &  57 &  0.88 &  3.40(3.38) &  3.88(1.00) &  2.69(1.06) &  3.20(0.46) &     ...     &     ...     &     ...     \\
G038.95 MM4 &  44 &  0.40 &  4.89(4.95) &  4.54(1.69) &  5.81(2.28) &  5.36(0.97) &     ...     &     ...     &  6.16(4.48) \\
G053.11 MM1 & 227 &  2.43 &  1.28(0.10) &  1.95(0.20) &  0.81(0.24) &     ...     &  3.28(0.57) &     ...     &     ...     \\
G053.11 MM2 &  55 &  0.63 &  4.39(4.42) &  4.47(0.90) &  2.97(0.85) &  2.58(0.98) &     ...     &     ...     &     ...     \\
G053.11 MM4 &  34 &  0.34 &  3.29(0.50) & 11.95(1.79) &  4.29(1.14) &  2.79(1.06) &  4.69(3.59) &     ...     &     ...     \\
G053.11 MM5 &  25 &  0.22 & 13.32(13.61) &  5.79(2.17) &  7.07(2.69) &  4.97(1.14) &     ...     &     ...     &     ...     \\
G053.25 MM1 &  52 &  0.61 &  2.36(0.24) &  2.28(0.75) &  1.98(0.68) &  2.54(3.40) &     ...     &     ...     &     ...     \\
G053.25 MM3 &  17 &  0.20 &  6.87(7.23) &  7.48(3.09) &  6.77(2.63) & 15.56(26.01) &     ...     &     ...     &     ...     \\
G053.25 MM4 &  37 &  0.40 &  3.99(0.37) &  6.01(1.37) &  3.04(1.07) &     ...     &     ...     &     ...     &     ...     \\
G053.25 MM5 &  17 &  0.20 &     ...     & 12.49(4.32) &  6.76(2.55) & 10.04(22.70) &     ...     &     ...     &     ...     \\
G053.25 MM6 &  34 &  0.30 & 13.05(13.20) &  8.97(2.12) &  5.61(2.10) &     ...     &     ...     &     ...     &     ...     \\
G053.31 MM2 &  34 &  0.59 &  2.09(2.07) &  1.13(0.42) &  1.69(0.65) &     ...     &     ...     &     ...     &     ...     \\
\enddata
\end{deluxetable*}

\subsubsection{HNCO (Isocyanic Acid)}

HNCO is a high density tracer that has been found in regions that span 
 a large range of temperatures 
\cite[10-500 K;][]{Jackson84,Zinchenko00,Bisschop07}. 
 The formation of HNCO is inefficient if
 only gas-phase reactions are considered \citep{Tideswell10}. In hot cores,
 the gas formation
chemistry can explain the observed HNCO abundances when reactions on the
surfaces of grains are included.  HNCO is efficiently formed on grain
mantles and processed
in more complex molecules. However, HNCO is not directly ejected from the
dust. Instead, it is formed by the dissociation of more complex molecules
(formed by HNCO on the grain mantles) once they are released to the gas phase 
\citep{Tideswell10,Rodriguez-Fernandez10}. 
\cite{Zinchenko00} found evidence that HNCO and SiO may have a common
production mechanism, presumably based on shock chemistry. 
\cite{Rodriguez-Fernandez10} found
HNCO lines that exhibit the same characteristics as other well-known shock
tracers such as CH$_3$OH and some sulfur-bearing species (SO and SO$_2$). These
authors suggest that HNCO is a shock tracer which is explained by a
combination
of grain surface and gas phase chemistry. HNCO abundances would rise in 
shock regions because this molecule can be directly ejected to the gas
phase through grain sputtering and because the efficiency of neutral-neutral 
reactions increases in the gas phase due to higher gas temperatures.

In our IRDC sample, HNCO emission is detected slightly more often in regions
 with signs of star formation. The HNCO line widths seem to correlate with 
those of N$_2$H$^+$. HNCO profiles show no evident signatures of being a tracer 
of shocks in most of the sources, except in G028.37 MM1 and G028.37 MM4 (which 
represent 10\% of the sources with HNCO detection).  
In these two sources, the HNCO spectrum presents a blue wing which is 
also observed in SiO. HNCO column densities rise with the evolution of
 the clumps, as is shown by the median values. However, HNCO abundances
 do not show this trend.

\subsubsection{SiO (Silicon Monoxide)}

It is well-known that SiO is a powerful tracer of molecular gas associated
with shocks \citep{Schilke97,Caselli97}. Its abundance is highly enhanced
 in molecular outflows, with respect to the ambient abundances. In some cases
these enhancements can be up to $10^6$ \citep{Martin-Pintado92}. The increase of
the SiO abundance is due to the silicate grain destruction, giving rise to
the injection into gas of Si atoms and/or Si-bearing species, and the
subsequent high temperature gas-phase chemistry \citep{Schilke97,Caselli97}.
 The previous SiO production mechanism explains spectral lines that
 show extended wing emission with broad line widths caused by the 
interaction of outflows and the surrounding medium.
However, an SiO component coming from quiescent gas is not well understood.
\cite{Jimenez-Serra10} detected extended narrow SiO emission not 
associated with signs of star formation in an IRDC. They suggest that this SiO
 component could be produced by the following processes: remnants of  
large-scale shocks caused by the formation process of the IRDC, decelerated 
shocked gas associated with large scale outflows from neighboring massive 
protostars, recently processed material associated with the youngest 
jets/outflows, and/or an undetected and widespread lower mass protostar 
population.    

In our IRDC sample, SiO emission is mostly detected in clumps with 
signs of star formation. The only exception is the quiescent clump 
G024.33 MM7 which shows a narrow line profile (1.4 \kms) and the lowest
 SiO abundance in the sample. The average SiO line width in the sample 
is $\sim$5 \kms. Half of the sources with SiO detection show clear 
extended wing emission indicating the presence of outflow activity.

\section{Conclusions}

We have carried out a multi-line survey at 3 mm toward 37 IRDCs, containing
159 clumps, in order to investigate the behavior of the different
molecular tracers and search for chemical variations through an
evolutionary sequence based on {\it Spitzer} IRAC and MIPS emission.
We observed N$_2$H$^+$, HNC, HN$^{13}$C, HCO$^+$, H$^{13}$CO$^+$,
HCN, C$_2$H, HC$_3$N, HNCO, and SiO lines with the Mopra 22 m telescope
located in Australia. After eliminating clumps that are not located in 
IRDCs and pairs of clumps that are placed within one Mopra beam, 
we base our study on 92 sources. 

HNC and N$_2$H$^+$ lines are detected in almost every IRDC clumps at every 
evolutionary stage, indicating that their presence does not depend on 
the star formation activity. On the other hand, HC$_3$N, HNCO, and SiO
lines are predominantly detected in later stages of evolution, as expected
from their formation paths.

The line widths of N$_2$H$^+$ slightly increase with the evolution of the
clumps, which is likely produced by the rise of turbulence due to the 
enhancement of the star formation activity at later evolutionary stages.
The increase is modest because, due to the large Mopra beam, we are also 
tracing the bulk motions of the gas, instead of just the densest
 regions associated with star formation. 

Optical depth calculations show that the N$_2$H$^+$ 
line is mostly optically thin (median of 0.8) and the C$_2$H line is moderately
optically thick (median of 2.4). HCO$^+$ and HNC lines are optically thick
(medians of 21 and 19, respectively), while their isotopologues are
optically thin (median of 0.4 for both). N$_2$H$^+$ opacities show no
variations with the evolution of the clumps, whereas C$_2$H, HCO$^+$ and
HNC show a slight decrease with the rise of star formation activity.

In general, column densities of the different molecules change for the
different evolutionary stages defined by \cite{Chambers09} and increase
with the evolution of the clumps, with the exception of C$_2$H. However,
this is not generally true for molecular abundances (i.e., after dividing 
by the total H$_2$ column density inferred from 1.2 mm continuum emission).
 Only the increases 
of N$_2$H$^+$ and HCO$^+$ abundances  are statistically significant and 
reflect chemical evolution.  
This is consistent with the results of \cite{Busquet11}, who included both 
molecules in their chemical modeling of a massive star-forming region.  
Although it is expected a rise of the HCO$^+$ abundance with the evolution of 
the clumps, it is not clear why N$_2$H$^+$ also follows this trend.  

The N$_2$H$^{+}$/HCO$^+$ abundance ratio acts as a chemical
clock, increasing its value from intermediate to active and red clumps. 
This observed trend is consistent with the theoretical predictions. The 
chemical models suggest that when clumps warm up, they release CO from
 grain mantles to the gas phase. 
This rise of CO increases the amount of HCO$^+$, because CO is its main 
supplier, with respect to N$_2$H$^+$, because CO is its main destroyer. 
However, the observed trend does not extend to quiescent clumps. This could 
be due to observational limitations or because the theoretical predictions are 
incorrect at very early stages of evolution. 
We also find that the N$_2$H$^{+}$/HNC abundance ratio increases with the 
evolution of the clumps, from quiescent to red clumps. It is not clear 
why this ratio behaves in this way, but it suggests that HNC may be
 preferentially formed in cold gas.

\acknowledgements

J.M.J. gratefully acknowledges funding support from NSF Grant
 No. AST-0808001. G.G. acknowledges support from CONICYT projects
 FONDAP No. 15010003 and BASAL PFB-06. We also thank the anonymous 
referee for helpful comments.

\end{document}